\documentclass[twocolumn,prl,superscriptaddress]{revtex4}%
\usepackage{amssymb}
\usepackage{graphicx}
\usepackage{amsmath}
\usepackage{amsfonts}
\usepackage{accents}

\newtheorem{theorem}{Theorem}
\newtheorem{lemma}[theorem]{Lemma}
\newtheorem{proposition}[theorem]{Proposition}
\newtheorem{corollary}[theorem]{Corollary}
\newtheorem{definition}[theorem]{Definition}

\begin{document}
\title{End-to-end capacities of a quantum communication network}
\author{Stefano Pirandola}
\affiliation{Department of Computer Science, University of York, York YO10 5GH, UK}
\affiliation{Research Laboratory of Electronics, Massachusetts Institute of Technology,
Cambridge, Massachusetts 02139, USA}

\begin{abstract}
In quantum mechanics, a fundamental law prevents quantum
communications to simultaneously achieve high rates and long
distances. This limitation is well known for point-to-point
protocols, where two parties are directly connected by a quantum
channel, but not yet fully understood in protocols with quantum
repeaters. Here we solve this problem bounding the ultimate rates
for transmitting quantum information, entanglement and secret keys
via quantum repeaters. We derive single-letter upper bounds for
the end-to-end capacities achievable by the most general
(adaptive) protocols of quantum and private communication, from a
single repeater chain to an arbitrarily-complex quantum network,
where systems may be routed through single or multiple paths. We
analytically establish these capacities under fundamental noise
models, including bosonic loss which is the most important for
optical communications. In this way, our results provide the
ultimate benchmarks for testing the optimal performance of
repeater-assisted quantum communications.
\end{abstract}

\maketitle

Today quantum technologies are being developed at a rapid
pace~\cite{Kimble,Rod,HybridINT,teleREVIEW}. In this scenario, quantum
communications are very advanced, with the development and implementation of a
number of point-to-point protocols of quantum key distribution
(QKD)~\cite{GisinRMP}, based on discrete variable (DV)
systems~\cite{Watrous,NielsenBook,HolevoBOOK}, such as qubits, or continuous
variable (CV) systems, such as bosonic modes~\cite{RMP,BraRMP}. Recently, we
have also witnessed the deployment of high-rate optical-based secure quantum
networks~\cite{To4,To5}. These are advantageous not only for their
multiple-user architecture but also because they may overcome the fundamental
limitations that are associated with point-to-point protocols of quantum and
private communication.

After a long series of studies that started back in 2009 with the
introduction of the reverse coherent information of a bosonic
channel~\cite{RevCohINFO,ReverseCAP}, Ref.~\cite{QKDpaper} finally
showed that the maximum rate at which two remote parties can
distribute quantum bits (qubits), entanglement bits (ebits), or
secret bits over a lossy channel (e.g., an optical fiber) is equal
to $-\log_{2}(1-\eta)$, where $\eta$ is the channel's
transmissivity. This limit is the
Pirandola-Laurenza-Ottaviani-Banchi (PLOB) bound~\cite{QKDpaper}
and cannot be surpassed even by the most powerful strategies that
exploit arbitrary local operations (LOs) assisted by two-way
classical communication (CC), also known as adaptive
LOCCs~\cite{TQC}.

To beat the PLOB bound, we need to insert a quantum repeater~\cite{Briegel} in
the communication line. In information
theory~\cite{Slepian,Schrijver,Cover&Thomas,Gamal}, a repeater or relay is any
middle node helping the communication between two end-parties. This definition
is extended to quantum information theory, where quantum repeaters are middle
nodes equipped with both classical and quantum operations, and may be arranged
to compose linear chains or more general networks. In general, they do not
need to have quantum memories (e.g., see Ref.~\cite{LoPhotonic}) even though
these are required for guaranteeing an optimal performance.

In all the ideal repeater-assisted scenarios, where we can beat the PLOB
bound, it is fundamental to determine the maximum rates that are achievable by
two end-users, i.e., to determine their end-to-end capacities for transmitting
qubits, distributing ebits, and generating secret keys. Finding these
capacities not only is important to establish the boundaries of quantum
network communications but also to benchmark practical implementations, so as
to check how far prototypes of quantum repeaters are from the ultimate
theoretical performance.

Here we address this fundamental problem. By combining methods
from quantum information
theory~\cite{Watrous,NielsenBook,HolevoBOOK,RMP,BraRMP} and
classical networks~\cite{Slepian,Schrijver,Cover&Thomas,Gamal}, we
derive tight single-letter upper bounds for the end-to-end quantum
and private capacities of repeater chains and, more generally,
quantum networks connected by arbitrary quantum channels (these
channels and the dimension of the quantum systems they transmit
may generally vary across the network). More importantly, we
establish exact formulas for these capacities under fundamental
noise models for both DV and CV systems, including dephasing,
erasure, quantum-limited amplification, and bosonic loss which is
the most important for quantum optical communications. Depending
on the routing in the quantum network (single- or multi-path),
optimal strategies are found by solving the widest
path~\cite{Pollack,MITp,newBOOK} or the maximum flow
problem~\cite{Harris,Ford,ShannonFLOW,netflow} suitably extended
to the quantum communication setting.

Our results and analytical formulas allow one to assess the rate
performance of quantum repeaters and quantum communication
networks with respect to the ultimate limits imposed by the laws
of quantum mechanics.

\section*{RESULTS}

\subsection*{Ultimate limits of repeater chains}

\begin{figure}[ptbh]
\vspace{-2.5cm}
\par
\begin{center}
\includegraphics[width=0.48\textwidth] {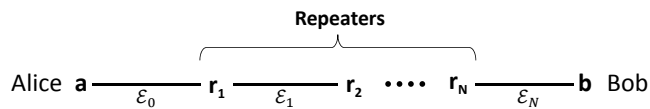} \vspace{-3.2cm}
\end{center}
\caption{Linear chain of $N$ quantum repeaters $\mathbf{r}_{1},\ldots
,\mathbf{r}_{N}$ between the two end-users, Alice $\mathbf{a}:=\mathbf{r}_{0}$
and Bob $\mathbf{b}:=\mathbf{r}_{N+1}$. The chain is connected by $N+1$
quantum channels $\{\mathcal{E}_{i}\}$.}%
\label{intropicmainn}%
\end{figure}

Consider Alice $\mathbf{a}$ and Bob $\mathbf{b}$ at the two ends of a linear
chain of $N$ quantum repeaters, labeled by $\mathbf{r}_{1},\ldots
,\mathbf{r}_{N}$. Each point has a local register of quantum systems which may
be augmented with incoming systems or depleted by outgoing ones. As also
depicted in Fig.~\ref{intropicmainn}, the chain is connected by $N+1$ quantum
channels $\{\mathcal{E}_{i}\}=\{\mathcal{E}_{0},\ldots,\mathcal{E}_{i}%
,\ldots,\mathcal{E}_{N}\}$ through which systems are sequentially
transmitted. This means that Alice transmits a system to repeater
$\mathbf{r}_{1}$, which then relays the system to repeater
$\mathbf{r}_{2}$, and so on, until Bob is reached.

Note that, in general, we may also have opposite directions for some of the
quantum channels, so that they transmit systems towards Alice; e.g., we may
have a middle relay receiving systems from both Alice and Bob. For this
reason, we generally consider the \textquotedblleft exchange\textquotedblright%
\ of a quantum system between two points by either forward or backward
transmission. Under the assistance of two-way CCs, the optimal transmission of
quantum information is related to the optimal distribution of entanglement
followed by teleportation, so that it does not depend on the physical
direction of the quantum channel but rather on the direction of the
teleportation protocol.

In a single end-to-end transmission or use of the chain, all the channels are
used exactly once. Assume that the end-points aim to share target bits, which
may be ebits or private bits~\cite{KD1,KD2}. The most general quantum
distribution protocol $\mathcal{P}_{\text{\textrm{chain}}}$ involves
transmissions which are interleaved by adaptive LOCCs among all parties, i.e.,
LOs assisted by two-way CCs among end-points and repeaters. In other words,
before and after each transmission between two nodes, there is a session of
LOCCs where all the nodes update and optimize their
registers.

After $n$ adaptive uses of the chain, the end-points share an output state
$\rho_{\mathbf{ab}}^{n}$ with $nR_{n}$ target bits. By optimizing the
asymptotic rate $\lim_{n}R_{n}$ over all protocols $\mathcal{P}%
_{\text{\textrm{chain}}}$, we define the generic two-way\ capacity
of the chain $\mathcal{C}(\{\mathcal{E}_{i}\})$. If the target are
ebits, the repeater-assisted capacity $\mathcal{C}$ is an
entanglement-distribution capacity $D_{2}$. The latter coincides
with a quantum capacity $Q_{2}$, because distributing an ebit is
equivalent to transmitting a qubit if we assume two-way CCs. If
the target are private bits, $\mathcal{C}$ is a secret-key
capacity $K\geq D_{2}$ (with the inequality holding because ebits
are specific private bits). Exact definitions and more details are
given in Supplementary Note~1.

To state our upper bound for $\mathcal{C}(\{\mathcal{E}_{i}\})$,
we introduce the notion of channel simulation, as generally
formulated by Ref.~\cite{QKDpaper} (see also
Refs.~\cite{TomPRA,RicSCIREP,RicPRA,Spyros,Cond,PBT} for
variants). Recall that any quantum channel $\mathcal{E}$ is
simulable by applying a trace-preserving LOCC $\mathcal{T}$ to the
input state $\rho$ together with some bipartite resource state
$\sigma$, so that $\mathcal{E}(\rho)=\mathcal{T}(\rho
\otimes\sigma)$. The pair $(\mathcal{T},\sigma)$ represents a
possible \textquotedblleft LOCC\ simulation\textquotedblright\ of
the channel. In particular, for channels that suitably commute
with the random unitaries of teleportation~\cite{teleREVIEW},
called \textquotedblleft
teleportation-covariant\textquotedblright\
channels~\cite{QKDpaper}, one finds that $\mathcal{T}$ is
teleportation and $\sigma$ is their Choi matrix
$\sigma_{\mathcal{E}}:=\mathcal{I}\otimes\mathcal{E}(\Phi)$, where
$\Phi$ is a maximally-entangled state. The latter is also known as
\textquotedblleft teleportation simulation\textquotedblright.

For bosonic channels, the Choi matrices are energy-unbounded, so that
simulations need to be formulated asymptotically. In general, an asymptotic
state $\sigma$ is defined as the limit of a sequence of physical states
$\sigma^{\mu}$, i.e., $\sigma:=\lim_{\mu}\sigma^{\mu}$. The simulation of a
channel $\mathcal{E}$ over an asymptotic state takes the form $||\mathcal{E}%
(\rho)-\mathcal{T}(\rho\otimes\sigma^{\mu})||_{1}\overset{\mu}{\rightarrow}0$
where the LOCC $\mathcal{T}$ may also depend on $\mu$ in the general
case~\cite{QKDpaper}. Similarly, any relevant functional on the asymptotic
state needs to be computed over the defining sequence $\sigma^{\mu}$ before
taking the limit for large $\mu$. These technicalities are fully accounted in
the Methods section.

The other notion to introduce is that of entanglement cut between Alice and
Bob. In the setting of a linear chain, a cut \textquotedblleft$i$%
\textquotedblright\ disconnects channel $\mathcal{E}_{i}$ between repeaters
$\mathbf{r}_{i}$ and $\mathbf{r}_{i+1}$. Such channel can be replaced by a
simulation with some resource state $\sigma_{i}$. After calculations (see
Methods), this allows us to write
\begin{equation}
\mathcal{C}(\{\mathcal{E}_{i}\})\leq E_{\mathrm{R}}(\sigma_{i}),
\label{chainR}%
\end{equation}
where $E_{\mathrm{R}}(\cdot)$ is the relative entropy of entanglement (REE).
Recall that the REE is defined as~\cite{RMPrelent,VedFORMm,Pleniom}%
\begin{equation}
E_{\text{\textrm{R}}}(\sigma)=\inf_{\gamma\in\text{\textrm{SEP}}}%
~S(\sigma||\gamma),
\end{equation}
where \textrm{SEP} represents the ensemble of separable bipartite
states and $S(\sigma ||\gamma):=\mathrm{Tr}\left[
\sigma(\log_{2}\sigma-\log_{2}\gamma)\right]  $ is the relative
entropy. In general, for any asymptotic state defined by the limit
$\sigma
:=\lim_{\mu}\sigma^{\mu}$, we may extend the previous definition and consider%
\begin{equation}
E_{\text{\textrm{R}}}(\sigma)=\underset{\mu}{\lim\inf}~E_{\text{\textrm{R}}%
}(\sigma^{\mu})=\inf_{\gamma^{\mu}}~\underset{\mu}{\lim\inf}~S(\sigma^{\mu
}||\gamma^{\mu}),
\end{equation}
where $\gamma^{\mu}$ is a converging sequence of separable
states~\cite{QKDpaper}.

By minimizing Eq.~(\ref{chainR}) over all cuts, we may write
\begin{equation}
\mathcal{C}(\{\mathcal{E}_{i}\})\leq\min_{i}E_{\mathrm{R}}(\sigma_{i}),
\label{chain1}%
\end{equation}
which establishes the ultimate limit for entanglement and key distribution
through a repeater chain. For a chain of teleportation-covariant channels, we
may use their teleportation simulation over Choi matrices and write
\begin{equation}
\mathcal{C}(\{\mathcal{E}_{i}\})\leq\min_{i}E_{\mathrm{R}}(\sigma
_{\mathcal{E}_{i}}). \label{teleKKK}%
\end{equation}
Note that the family of teleportation-covariant channels is large,
including Pauli channels (at any dimension)~\cite{NielsenBook} and
bosonic Gaussian channels~\cite{RMP}. Within such a family, there
are channels $\mathcal{E}$ whose generic two-way capacity
$\mathcal{C}=Q_{2}$, $D_{2}$ or $K$ satisfies
\begin{equation}
\mathcal{C}(\mathcal{E})=E_{\mathrm{R}}(\sigma_{\mathcal{E}})=D_{1}(\sigma_{\mathcal{E}}), \label{one-wayCC}%
\end{equation}
where $D_{1}(\sigma_{\mathcal{E}})$ is the one-way distillable
entanglement of the Choi matrix (defined as an asymptotic
functional in the bosonic case~\cite{QKDpaper}). These are called
\textquotedblleft distillable channels\textquotedblright\ and
include bosonic lossy channels, quantum-limited amplifiers,
dephasing and erasure channels~\cite{QKDpaper}.

For a chain of distillable channels, we therefore exactly establish the
repeater-assisted capacity as%
\begin{equation}
\mathcal{C}(\{\mathcal{E}_{i}\})=\min_{i}\mathcal{C}(\mathcal{E}_{i})=\min
_{i}E_{\mathrm{R}}(\sigma_{\mathcal{E}_{i}}). \label{DistillableCAPP}%
\end{equation}
In fact the upper bound ($\leq$) follows from Eqs.~(\ref{teleKKK})
and~(\ref{one-wayCC}). The lower bound ($\geq$) relies on the fact that an
achievable rate for end-to-end entanglement distribution consists in: (i) each
pair, $\mathbf{r}_{i}$ and $\mathbf{r}_{i+1}$, exchanging $D_{1}%
(\sigma_{\mathcal{E}_{i}})$ ebits over $\mathcal{E}_{i}$; and (ii) performing
entanglement swapping on the distilled ebits. In this way, at least $\min
_{i}D_{1}(\sigma_{\mathcal{E}_{i}})$ ebits are shared between Alice and Bob.

\subsection*{Lossy chains}

\begin{figure}[ptbh]
\begin{center}
\vspace{+0.1cm} \includegraphics[width=0.38\textwidth] {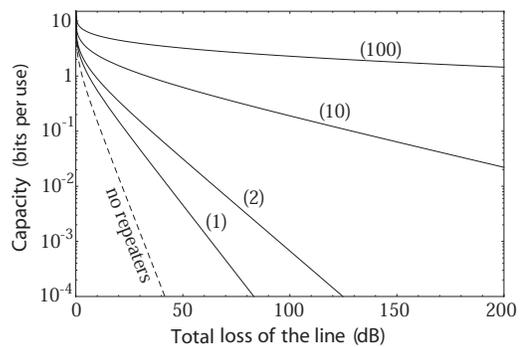}
\vspace{-0.5cm}
\end{center}
\caption{Optimal performance of lossy chains. Capacity (target bits per chain
use) versus total loss of the line (decibels, dB) for $N=1,2,10$ and $100$
equidistant repeaters. Compare the repeater-assisted capacities (solid curves)
with the point-to-point repeater-less bound~\cite{QKDpaper} (dashed curve).}%
\label{figg}%
\end{figure}

Let us specify Eq.~(\ref{DistillableCAPP}) to an important case. For a chain
of quantum repeaters connected by lossy channels with transmissivities
$\{\eta_{i}\}$, we find the capacity%
\begin{equation}
\mathcal{C}_{\text{loss}}=-\log_{2}(1-\eta_{\text{min}}),~~\eta_{\text{min}%
}:=\min_{i}\eta_{i}~. \label{EqLOSSY2}%
\end{equation}
Thus, the minimum transmissivity within the lossy chain establishes the
ultimate rate for repeater-assisted quantum/private communication between the
end-users. For instance, consider an optical fiber with transmissivity $\eta$
and insert $N$ repeaters so that the fiber is split into $N+1$ lossy channels.
The optimal configuration corresponds to equidistant repeaters, so that
$\eta_{\text{min}}=\sqrt[N+1]{\eta}$ and the maximum capacity of the lossy
chain is%
\begin{equation}
\mathcal{C}_{\text{loss}}(\eta,N)=-\log_{2}\left(  1-\sqrt[N+1]{\eta}\right)
~. \label{optLOSScap0}%
\end{equation}
This capacity is plotted in Fig.~\ref{figg} and compared with the
point-to-point PLOB bound $\mathcal{C}(\eta)=\mathcal{C}_{\text{loss}}%
(\eta,0)$. A simple calculation shows that if we want to guarantee a
performance of $1$ target bit per use of the chain, then we may tolerate at
most 3dB of loss in each individual link. This \textquotedblleft$3$dB
rule\textquotedblright\ imposes a maximum repeater-repeater distance of 15km
in standard optical fibre (at $0.2$dB/km).

\subsection*{Quantum networks under single-path routing}

A quantum communication network can be represented by an undirected finite
graph~\cite{Slepian} $\mathcal{N}=(P,E)$, where $P$ is the set of points and
$E$ the set of all edges. Each point $\mathbf{p}$\ has a local register of
quantum systems. Two points $\mathbf{p}_{i}$\textbf{ }and\textbf{ }%
$\mathbf{p}_{j}$ are connected by an edge $(\mathbf{p}_{i},\mathbf{p}_{j})\in
E$ if there is a quantum channel $\mathcal{E}_{ij}:=\mathcal{E}_{\mathbf{p}%
_{i}\mathbf{p}_{j}}$ between them. By simulating each channel $\mathcal{E}%
_{ij}$ with a resource state $\sigma_{ij}$, we simulate the entire network
$\mathcal{N}$ with a set of resource states $\sigma(\mathcal{N})=\{\sigma
_{ij}\}$. A route is an undirected path $\mathbf{a}-\mathbf{p}_{i}%
-\cdots-\mathbf{p}_{j}-\mathbf{b}$ between the two end-points,
Alice $\mathbf{a}$ and Bob $\mathbf{b}$. These are connected by an
ensemble of possible routes $\Omega=\{1,\ldots,\omega,\ldots\}$,
with the generic route $\omega$ involving the transmission through
a sequence of channels
$\{\mathcal{E}_{0}^{\omega},\ldots,\mathcal{E}_{k}^{\omega}\ldots\}$.
Finally, an entanglement cut $C$ is a bipartition
$(\mathbf{A},\mathbf{B})$ of $P$ such that
$\mathbf{a}\in\mathbf{A}$ and $\mathbf{b}\in\mathbf{B}$. Any such
cut $C$ identifies a super Alice $\mathbf{A}$ and a super Bob
$\mathbf{B}$, which are connected by the cut-set
$\tilde{C}=\{(\mathbf{x},\mathbf{y})\in
E:\mathbf{x}\in\mathbf{A},\mathbf{y}\in\mathbf{B}\}$. See the
example in Fig.~\ref{DiamondLL} and more details in Supplementary
Notes~2 and~3.
\begin{figure}[ptbh]
\vspace{-1.7cm}
\par
\begin{center}
\includegraphics[width=0.49\textwidth] {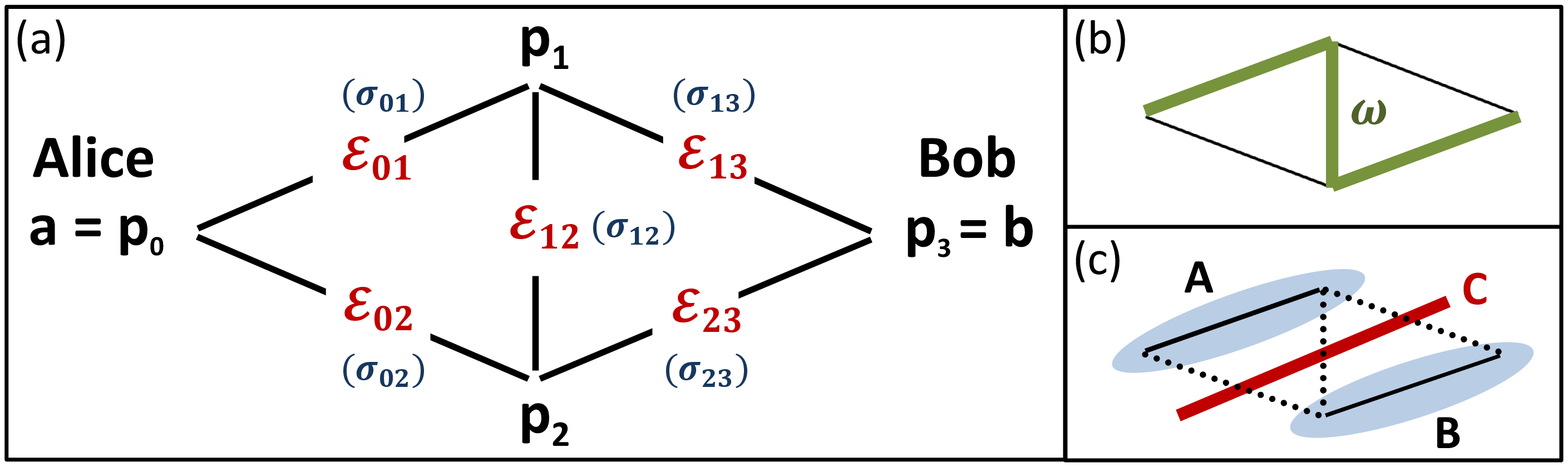} \vspace{-2.4cm}
\end{center}
\caption{Diamond quantum network $\mathcal{N}^{\diamond}$.~\textbf{(a)}~This
is a quantum network of four points $P=\{\mathbf{p}_{0},\mathbf{p}%
_{1},\mathbf{p}_{2},\mathbf{p}_{3}\}$, with end-points $\mathbf{p}%
_{0}=\mathbf{a}$ (Alice) and $\mathbf{p}_{3}=\mathbf{b}$ (Bob). Two points
$\mathbf{p}_{i}$ and $\mathbf{p}_{j}$ are connected by an edge $(\mathbf{p}%
_{i},\mathbf{p}_{j})$ if there is an associated quantum channel $\mathcal{E}%
_{ij}$. This channel has a corresponding resource state $\sigma_{ij}$ in a
simulation of the network. There are four (simple) routes: $1:\mathbf{a}%
-\mathbf{p}_{1}-\mathbf{b}$, $2:\mathbf{a}-\mathbf{p}_{2}-\mathbf{b}$,
$3:\mathbf{a}-\mathbf{p}_{2}-\mathbf{p}_{1}-\mathbf{b}$, and $4:\mathbf{a}%
-\mathbf{p}_{1}-\mathbf{p}_{2}-\mathbf{b}$. As an example, route $4$ involves
the transmission through the sequence of quantum channels $\{\mathcal{E}%
_{k}^{4}\}$ which is defined by $\mathcal{E}_{0}^{4}:=\mathcal{E}_{01}$,
$\mathcal{E}_{1}^{4}:=\mathcal{E}_{12}\ $and $\mathcal{E}_{2}^{4}%
:=\mathcal{E}_{23}$. \textbf{(b)}~We explicitly show route $\omega=4$. In a
sequential protocol, each use of the network corresponds to using a single
route $\omega$ between the two end-points, with some probability $p_{\omega}$.
\textbf{(c)}~We show an entanglement cut $C$ of the network, with super Alice
$\mathbf{A}$ and super Bob $\mathbf{B}$ made by the points in the two clouds.
These are connected by the cut-set $\tilde{C}$ composed by the dotted edges.}%
\label{DiamondLL}%
\end{figure}

Let us remark that the quantum network is here described by an undirected
graph where the physical direction of the quantum channels $\mathcal{E}_{ij}$
can be forward ($\mathbf{p}_{i}\rightarrow\mathbf{p}_{j}$) or backward
($\mathbf{p}_{j}\rightarrow\mathbf{p}_{i}$). As said before for the repeater
chains, this degree of freedom relies on the fact that we consider assistance
by two-way CC, so that the optimal transmission of qubits can always be
reduced to the distillation of ebits followed by teleportation. The logical
flow of quantum information is therefore fully determined by the LOs of the
points, not by the physical direction of the quantum channel which is used to
exchange a quantum system along an edge of the network. This study of an
undirected quantum network under two-way CC clearly departs from other investigations~\cite{Hay1,Hay2,Song}.

\begin{figure}[ptbh]
\vspace{-0.3cm}
\par
\begin{center}
\includegraphics[width=0.49\textwidth] {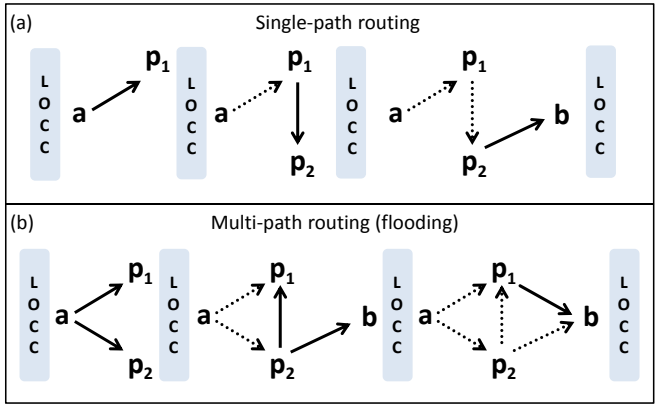} \vspace{-1.1cm}
\end{center}
\caption{Network protocols of quantum and private communication. (a)~In a
sequential protocol, systems are routed through a single path
probabilistically chosen by the points. Here it is $\mathbf{a}-\mathbf{p}%
_{1}-\mathbf{p}_{2}-\mathbf{b}$. Each transmission occurs between two adaptive
LOCCs, where all points of the network perform LOs assisted by two-way CC.
(b)~In a flooding protocol, systems are simultaneously routed from Alice to
Bob through a sequence of multipoint communications in such a way that each
edge of the network is used exactly once in an end-to-end transmission. Here
we show a possible sequence $\mathbf{a}\rightarrow\{\mathbf{p}_{1}%
,\mathbf{p}_{2}\}$, $\mathbf{p}_{2}\rightarrow\{\mathbf{p}_{1},\mathbf{b}\}$,
$\mathbf{p}_{1}\rightarrow\{\mathbf{b}\}$. Each multipoint communication
occurs between two adaptive LOCCs.}%
\label{NETproto}%
\end{figure}

In a sequential protocol $\mathcal{P}_{\text{seq}}$, the network is
initialized by a preliminary network LOCC, where all the points communicate
with each other via unlimited two-way CCs and perform adaptive LOs on their
local quantum systems. With some probability, Alice exchanges a quantum system
with repeater $\mathbf{p}_{i}$, followed by a second network LOCC; then
repeater $\mathbf{p}_{i}$ exchanges a system with repeater $\mathbf{p}_{j}$,
followed by a third network LOCC and so on, until Bob is reached through some
route in a complete sequential use of the network (see Fig.~\ref{NETproto}).
The routing is itself adaptive in the general case, with each node updating
its routing table (probability distribution) on the basis of the feedback
received by the other nodes. For large $n$ uses of the network, there is a
probability distribution associated with the ensemble $\Omega$, with the
generic route $\omega$ being used $np_{\omega}$ times. Alice and Bob's output
state $\rho_{\mathbf{ab}}^{n}$ will approximate a target state with $nR_{n}$
bits. By optimizing over $\mathcal{P}_{\text{seq}}$ and taking the limit of
large $n$, we define the sequential or single-path capacity of the network
$\mathcal{C}(\mathcal{N})$, whose nature depends on the target bits.

To state our upper bound, let us first introduce the flow of REE
through a cut. Given an entanglement cut $C$ of the network,
consider its cut-set $\tilde{C}$. For each edge
$(\mathbf{x},\mathbf{y})$ in $\tilde{C}$, we have a channel
$\mathcal{E}_{\mathbf{xy}}$ and a corresponding resource state
$\sigma_{\mathbf{xy}}$ associated with a simulation. Then we
define the
single-edge flow of REE across cut $C$ as%
\begin{equation}
E_{\mathrm{R}}(C):=\max_{(\mathbf{x},\mathbf{y})\in\tilde{C}}E_{\mathrm{R}%
}(\sigma_{\mathbf{xy}}).\label{ubseqNN0}%
\end{equation}
The minimization of this quantity over all entanglement cuts provides our
upper bound for the single-path capacity of the network, i.e.,
\begin{equation}
\mathcal{C}(\mathcal{N})\leq\min_{C}E_{\mathrm{R}}(C),\label{ubseqNN}%
\end{equation}
which is the network generalization of Eq.~(\ref{chain1}). For proof see
Methods and further details in Supplementary Note~4.

In Eq.~(\ref{ubseqNN}), the quantity $E_{\mathrm{R}}(C)$ represents the
maximum entanglement (as quantified by the REE) \textquotedblleft
flowing\textquotedblright\ through a cut. Its minimization over all the cuts
bounds the single-path capacity for quantum communication, entanglement
distribution and key generation. For a network of teleportation-covariant
channels, the resource state $\sigma_{\mathbf{xy}}$ in Eq.~(\ref{ubseqNN0}) is
the Choi matrix $\sigma_{\mathcal{E}_{\mathbf{xy}}}$ of the channel
$\mathcal{E}_{\mathbf{xy}}$. In particular, for a network of distillable
channels, we may also set
\begin{equation}
\mathcal{C}(\mathcal{E}_{\mathbf{xy}})=E_{\mathrm{R}}(\sigma_{\mathcal{E}%
_{\mathbf{xy}}})=D_{1}(\sigma_{\mathcal{E}_{\mathbf{xy}}}), \label{disNETtool}%
\end{equation}
for any edge $(\mathbf{x},\mathbf{y})$. Therefore, we may refine the previous
bound of Eq.~(\ref{ubseqNN}) into $\mathcal{C}(\mathcal{N})\leq\min
_{C}\mathcal{C}(C)$ where
\begin{equation}
\mathcal{C}(C):=\max_{(\mathbf{x},\mathbf{y})\in\tilde{C}}\mathcal{C}%
(\mathcal{E}_{\mathbf{xy}})
\end{equation}
is the maximum (single-edge) capacity of a cut.

Let us now derive a lower bound. First we prove that, for an arbitrary
network, $\min_{C}\mathcal{C}(C)=\max_{\omega}\mathcal{C}(\omega)$, where
$\mathcal{C}(\omega):=\min_{i}\mathcal{C}(\mathcal{E}_{i}^{\omega})$ is the
capacity of route $\omega$ (see Methods). Then, we observe that $\mathcal{C}%
(\omega)$ is an achievable rate. In fact, any two consecutive points on route
$\omega$ may first communicate at the rate $\mathcal{C}(\mathcal{E}%
_{i}^{\omega})$; the distributed resources are then swapped to the end-users,
e.g., via entanglement swapping or key composition at the minimum rate
$\min_{i}\mathcal{C}(\mathcal{E}_{i}^{\omega})$. For a distillable network,
this lower bound coincides with the upper bound, so that we exactly establish
the single-path capacity as%
\begin{equation}
\mathcal{C}(\mathcal{N})=\max_{\omega}\mathcal{C}(\omega)=\min_{C}%
\mathcal{C}(C)=\min_{C}E_{\mathrm{R}}(C)~. \label{distillableCCCC}%
\end{equation}

Finding the optimal route $\omega_{\ast}$ corresponds to solving the widest
path problem~\cite{Pollack} where the weights of the edges $(\mathbf{x}%
,\mathbf{y})$ are the two-way capacities $\mathcal{C}(\mathcal{E}%
_{\mathbf{xy}})$. Route $\omega_{\ast}$ can be found via modified
Dijkstra's shortest path algorithm~\cite{newBOOK}, working in time
$O(\left\vert E\right\vert \log_{2}\left\vert P\right\vert )$,
where $\left\vert E\right\vert $ is the number of edges and
$\left\vert P\right\vert $ is the number of points. Over route
$\omega_{\ast}$ a capacity-achieving protocol is non adaptive,
with point-to-point sessions of one-way entanglement distillation
followed by entanglement swapping~\cite{teleREVIEW}. In a
practical implementation, the number of distilled ebits can be
computed using the methods from Ref.~\cite{MeterPaper}. Also note
that, because the swapping is on ebits, there is no violation of
the Bellman's optimality principle~\cite{DiFranco}.

An important example is an optical lossy network $\mathcal{N}_{\text{loss}}$
where any route $\omega$ is composed of lossy channels with transmissivities
$\{\eta_{i}^{\omega}\}$. Denote by $\eta_{\omega}:=\min_{i}\eta_{i}^{\omega}$
the end-to-end transmissivity of route $\omega$. The single-path capacity is
given by the route with maximum transmissivity%
\begin{equation}
\mathcal{C}(\mathcal{N}_{\text{loss}})=-\log_{2}(1-\eta_{\mathcal{N}}%
),~~\eta_{\mathcal{N}}:=\max_{\omega\in\Omega}\eta_{\omega}.
\label{lossSINGLE}%
\end{equation}
In particular, this is the ultimate rate at which the two end-points may
generate secret bits per sequential use of the lossy network.

\subsection*{Quantum networks under multi-path routing}

In a network we may consider a more powerful routing strategy, where systems
are transmitted through a sequence of multipoint communications (interleaved
by network LOCCs). In each of these communications, a number $M$ of quantum
systems are prepared in a generally multipartite state and simultaneously
transmitted to $M$ receiving nodes. For instance, as shown in the example of
Fig.~\ref{NETproto}, Alice may simultaneously sends systems to repeaters
$\mathbf{p}_{1}$ and $\mathbf{p}_{2}$, which is denoted by $\mathbf{a}%
\rightarrow\{\mathbf{p}_{1},\mathbf{p}_{2}\}$. Then, repeater
$\mathbf{p}_{2}$ may communicate with repeater $\mathbf{p}_{1}$
and Bob $\mathbf{b}$, i.e.,
$\mathbf{p}_{2}\rightarrow\{\mathbf{p}_{1},\mathbf{b}\}$. Finally,
repeater $\mathbf{p}_{1}$ may communicate with Bob, i.e.,
$\mathbf{p}_{1}\rightarrow\mathbf{b}$. Note that each edge of the
network is used exactly once during the end-to-end transmission, a
strategy known as \textquotedblleft flooding\textquotedblright\ in
computer networks~\cite{flooding}. This is achieved by
non-overlapping multipoint communications, where the receiving
repeaters choose unused edges for the next transmissions. More
generally, each multipoint communication is assumed to be a
point-to-multipoint connection with a logical
sender-to-receiver(s) orientation but where the quantum systems
may be physically transmitted either forward or backward by the
quantum channels.

Thus, in a general quantum flooding protocol $\mathcal{P}_{\text{flood}}$, the
network is initialized by a preliminary network LOCC. Then, Alice $\mathbf{a}$
exchanges quantum systems with all her neighbor repeaters $\mathbf{a}%
\rightarrow\{\mathbf{p}_{k}\}$. This is followed by another network LOCC.
Then, each receiving repeater exchanges systems with its neighbor repeaters
through unused edges, and so on. Each multipoint communication is interleaved
by network LOCCs and may distribute multi-partite entanglement. Eventually,
Bob is reached as an end-point in the first parallel use of the network, which
is completed when all Bob's incoming edges have been used exactly once. In the
limit of many uses $n$ and optimizing over $\mathcal{P}_{\text{flood}}$, we
define the multi-path capacity of the network $\mathcal{C}^{\text{m}%
}(\mathcal{N})$.

As before, given an entanglement cut $C$, consider its cut-set $\tilde{C}$.
For each edge $(\mathbf{x},\mathbf{y})$ in $\tilde{C}$, there is a channel
$\mathcal{E}_{\mathbf{xy}}$ with a corresponding resource state $\sigma
_{\mathbf{xy}}$. We define the multi-edge flow of REE through $C$ as%
\begin{equation}
E_{\mathrm{R}}^{\text{m}}(C):=%
{\textstyle\sum\limits_{(\mathbf{x},\mathbf{y})\in\tilde{C}}}
E_{\mathrm{R}}(\sigma_{\mathbf{xy}}),
\end{equation}
which is the total entanglement (REE) flowing through a cut. The minimization
of this quantity over all entanglement cuts provides our upper bound for the
multi-path capacity of the network, i.e.,
\begin{equation}
\mathcal{C}^{\text{m}}(\mathcal{N})\leq\min_{C}E_{\mathrm{R}}^{\text{m}}(C),
\label{mpBOUND}%
\end{equation}
which is the multi-path generalization of Eq.~(\ref{ubseqNN}). For proof see
Methods and further details in Supplementary Note~5. In a
teleportation-covariant network we may simply use the Choi matrices
$\sigma_{\mathbf{xy}}=\sigma_{\mathcal{E}_{\mathbf{xy}}}$. Then, for a
distillable network, we may use $E_{\mathrm{R}}(\sigma_{\mathcal{E}%
_{\mathbf{xy}}})=\mathcal{C}(\mathcal{E}_{\mathbf{xy}})$ from
Eq.~(\ref{disNETtool}), and write the refined upper bound $\mathcal{C}%
^{\text{m}}(\mathcal{N})\leq\min_{C}\mathcal{C}^{\text{m}}(C)$, where
\begin{equation}
\mathcal{C}^{\text{m}}(C):=%
{\textstyle\sum\limits_{(\mathbf{x},\mathbf{y})\in\tilde{C}}}
\mathcal{C}(\mathcal{E}_{\mathbf{xy}})
\end{equation}
is the total (multi-edge) capacity of a cut.

To show that the upper bound is achievable for a distillable
network, we need to determine the optimal flow of qubits from
Alice to Bob. First of all, from the knowledge of the capacities
$\mathcal{C}(\mathcal{E}_{\mathbf{xy}})$, the parties solve a
classical problem of maximum
flow~\cite{Harris,Ford,ShannonFLOW,netflow} compatible with those
capacities. By using Orlin's algorithm~\cite{Orlin}, the solution
can be found in $O(|P|\times|E|)$ time. This provides an optimal
orientation for the network and the rates
$R_{\mathbf{xy}}\leq\mathcal{C}(\mathcal{E}_{\mathbf{xy}})$ to be
used. Then, any pair of neighbor points, $\mathbf{x}$ and
$\mathbf{y}$, distill $nR_{\mathbf{xy}}$ ebits via one-way CCs.
Such ebits are used to teleport $nR_{\mathbf{xy}}$ qubits from
$\mathbf{x}$ to $\mathbf{y}$ according to the optimal orientation.
In this way, a number $nR$ of qubits are teleported from Alice to
Bob, flowing as quantum information through the network. Using the
max-flow min-cut
theorem~\cite{Harris,Ford,ShannonFLOW,netflow,Karp,Dinic,Alon,Ahuja,Cheriyan,King,Orlin}%
, we have that the maximum flow is $n\mathcal{C}^{\text{m}}(C_{\text{min}})$
where $C_{\text{min}}$ is the minimum cut, i.e., $\mathcal{C}^{\text{m}%
}(C_{\text{min}})=\min_{C}\mathcal{C}^{\text{m}}(C)$. Thus, that for a
distillable $\mathcal{N}$, we find the multi-path capacity%
\begin{equation}
\mathcal{C}^{\text{m}}(\mathcal{N})=\min_{C}\mathcal{C}^{\text{m}}(C)=\min
_{C}E_{\mathrm{R}}^{\text{m}}(C), \label{multiCC}%
\end{equation}
which is the multi-path version of Eq.~(\ref{distillableCCCC}). This is
achievable by using a non adaptive protocol where the optimal routing is given
by Orlin's algorithm~\cite{Orlin}.

As an example, consider again a lossy optical network $\mathcal{N}%
_{\text{loss}}$ whose generic edge $(\mathbf{x},\mathbf{y})$ has
transmissivity $\eta_{\mathbf{xy}}$. Given a cut $C$, consider its loss
$L_{C}:=%
{\textstyle\prod\nolimits_{(\mathbf{x},\mathbf{y})\in\tilde{C}}}
(1-\eta_{\mathbf{xy}})$ and define the total loss of the network as the
maximization $L_{\mathcal{N}}:=\max_{C}L_{C}$. We find that the multi-path
capacity is just given by%
\begin{equation}
\mathcal{C}^{\text{m}}(\mathcal{N}_{\text{loss}})=-\log_{2}L_{\mathcal{N}}.
\end{equation}
It is interesting to make a direct comparison between the performance of
single- and multi-path strategies. For this purpose, consider a diamond
network $\mathcal{N}_{\text{loss}}^{\Diamond}$ whose links are lossy channels
with the same transmissivity $\eta$. In this case, we easily see that the
multi-path capacity doubles the single-path capacity of the network, i.e.,%
\begin{equation}
\mathcal{C}^{\text{m}}(\mathcal{N}_{\text{loss}}^{\Diamond})=2\mathcal{C}%
(\mathcal{N}_{\text{loss}}^{\Diamond})=-2\log_{2}(1-\eta).
\end{equation}
As expected the parallel use of the quantum network is more powerful than the
sequential use.

\subsection{Formulas for distillable chains and networks}

Here we provide explicit analytical formulas for the end-to-end capacities of
distillable chains and networks, beyond the lossy case already studied above.
In fact, examples of distillable channels are not only lossy channels but also
quantum-limited amplifiers, dephasing and erasure channels. First let us
recall their explicit definitions and their two-way capacities.

A lossy (pure-loss) channel with transmissivity $\eta\in(0,1)$
corresponds to a specific phase-insensitive Gaussian channel which
transforms input quadratures
$\mathbf{\hat{x}}=(\hat{q},\hat{p})^{T}$ as
$\mathbf{\hat{x}}\rightarrow
\sqrt{\eta}\mathbf{\hat{x}}+\sqrt{1-\eta}\mathbf{\hat{x}}_{E}$,
where $E$ is the environment in the vacuum state~\cite{RMP}. Its
two-way capacities ($Q_{2}$, $D_{2}$ and $K$) all coincide and are
given by the PLOB bound~\cite{QKDpaper}
\begin{equation}
\mathcal{C}(\eta)=-\log_{2}(1-\eta). \label{eq1}%
\end{equation}
A quantum-limited amplifier with an associated gain $g>1$ is
another phase-insensitive
Gaussian channel but realizing the transformation $\mathbf{\hat{x}}%
\rightarrow\sqrt{g}\mathbf{\hat{x}}+\sqrt{g-1}\mathbf{\hat{x}}_{E}$, where the
environment $E$ is in the vacuum state~\cite{RMP}. Its two-way capacities all
coincide and are given by~\cite{QKDpaper}
\begin{equation}
\mathcal{C}(g)=-\log_{2}(1-g^{-1}). \label{eq2}%
\end{equation}

A dephasing channel with probability $p\leq1/2$ is a Pauli channel of the form
$\rho\rightarrow(1-p)\rho+pZ\rho Z$, where $Z$ is the phase-flip Pauli
operator~\cite{NielsenBook}. Its two-way capacities all coincide and are given
by~\cite{QKDpaper}
\begin{equation}
\mathcal{C}(p)=1-H_{2}(p), \label{eq3}%
\end{equation}
where $H_{2}(p):=-p\log_{2}p-(1-p)\log_{2}(1-p)$ is the binary Shannon
entropy. Finally, an erasure channel with probability $p\leq1/2$ is a channel
of the form $\rho\rightarrow(1-p)\rho+p\left\vert e\right\rangle \left\langle
e\right\vert $, where $\left\vert e\right\rangle \left\langle e\right\vert $
is an orthogonal state living in an extra dimension~\cite{NielsenBook}. Its
two-way capacities all coincide to~\cite{QKDpaper,GEW,Erasure}
\begin{equation}
\mathcal{C}(p)=1-p. \label{eq5}%
\end{equation}

Consider now a repeater chain $\{\mathcal{E}_{i}\}$, where the
channels $\mathcal{E}_{i}$ are distillable of the same type (e.g.,
all quantum-limited amplifiers with different gains $g_{i}$). The
repeater-assisted capacity can be computed by combining
Eq.~(\ref{DistillableCAPP}) with one of the
Eqs.~(\ref{eq1})-(\ref{eq5}). The final formulas are shown in the
first column of Table~I. Then consider a quantum network
$\mathcal{N}=(P,E)$, where each edge $(\mathbf{x},\mathbf{y})\in
E$ is described by a distillable channel
$\mathcal{E}_{\mathbf{xy}}$ of the same type. For network
$\mathcal{N}$, we may consider both a generic route
$\omega\in\Omega$, with sequence of channels
$\mathcal{E}_{i}^{\omega}$, and a entanglement cut $C$, with
corresponding cut-set $\tilde{C}$. By combining
Eqs.~(\ref{distillableCCCC}) and~(\ref{multiCC}) with
Eqs.~(\ref{eq1})-(\ref{eq5}), we derive explicit formulas for the
single-path and multi-path capacities. These are given in the
second and third columns of Table~I where we set%
\begin{align}
\eta_{\mathcal{N}}  &  :=\max_{\omega\in\Omega}\min_{i}\eta_{i}^{\omega}%
=\min_{C}\max_{(\mathbf{x},\mathbf{y})\in\tilde{C}}\eta_{\mathbf{xy}},\\
g_{\mathcal{N}}  &  :=\min_{\omega\in\Omega}\max_{i}g_{i}^{\omega}=\max
_{C}\min_{(\mathbf{x},\mathbf{y})\in\tilde{C}}g_{\mathbf{xy}},\\
p_{\mathcal{N}}  &  :=\min_{\omega\in\Omega}\max_{i}p_{i}^{\omega}=\max
_{C}\min_{(\mathbf{x},\mathbf{y})\in\tilde{C}}p_{\mathbf{xy}},\\
L_{\mathcal{N}}  &  :=\max\limits_{C}{\prod\limits_{(\mathbf{x},\mathbf{y}%
)\in\tilde{C}}}(1-\eta_{\mathbf{xy}}),\\
G_{\mathcal{N}}  &  :=\max\limits_{C}{\prod\limits_{(\mathbf{x},\mathbf{y}%
)\in\tilde{C}}}(1-g_{\mathbf{xy}}^{-1}).
\end{align}

\begin{table*}[ptbh]%
\[%
\begin{tabular}
[c]{c|ccc}
& ~~%
\begin{tabular}
[c]{c}%
Chain $\{\mathcal{E}_{i}\}$\\
Repeater capacity $\mathcal{C}(\{\mathcal{E}_{i}\})$%
\end{tabular}
~~ & ~~%
\begin{tabular}
[c]{c}%
Network $\mathcal{N}$\\
Single-path capacity $\mathcal{C}(\mathcal{N})$%
\end{tabular}
~~ & ~~%
\begin{tabular}
[c]{c}%
Network $\mathcal{N}$\\
Multi-path capacity $\mathcal{C}^{\mathrm{m}}(\mathcal{N})$%
\end{tabular}
\\\hline%
\begin{tabular}
[c]{c}%
~\\
Lossy channels\\
(transmissivity $\eta$)\\
~
\end{tabular}
&
\begin{tabular}
[c]{c}%
~\\
$-\log_{2}(1-\min\limits_{i}\eta_{i})$\\
~~
\end{tabular}
~ &
\begin{tabular}
[c]{c}%
~\\
$-\log_{2}(1-\eta_{\mathcal{N}})$\\
~
\end{tabular}
&
\begin{tabular}
[c]{c}%
~\\
$-\log_{2}L_{\mathcal{N}}$\\
~
\end{tabular}
\\%
\begin{tabular}
[c]{c}%
Q-limited amplifiers\\
(gain $g$)\\
~
\end{tabular}
&
\begin{tabular}
[c]{c}%
$-\log_{2}\left[  1-(\max\limits_{i}g_{i})^{-1}\right]  $\\
~
\end{tabular}
~~ &
\begin{tabular}
[c]{c}%
$-\log_{2}(1-g_{\mathcal{N}}^{-1})$\\
~
\end{tabular}
& $%
\begin{tabular}
[c]{c}%
$-\log_{2}G_{\mathcal{N}}$\\
~
\end{tabular}
$\\%
\begin{tabular}
[c]{c}%
Dephasing channels\\
(probability $p$)\\
~
\end{tabular}
&
\begin{tabular}
[c]{c}%
$1-H_{2}(\max\limits_{i}p_{i})$\\
~
\end{tabular}
~ &
\begin{tabular}
[c]{c}%
$1-H_{2}(p_{\mathcal{N}})$\\
~
\end{tabular}
& $%
\begin{tabular}
[c]{c}%
$\min\limits_{C}{\sum\limits_{(\mathbf{x},\mathbf{y})\in\tilde{C}}}\left[
1-H_{2}(p_{\mathbf{xy}})\right]  $\\
~
\end{tabular}
$\\%
\begin{tabular}
[c]{c}%
Erasure channels\\
(probability $p$)\\
~
\end{tabular}
&
\begin{tabular}
[c]{c}%
$1-\max\limits_{i}p_{i}$\\
~
\end{tabular}
~ &
\begin{tabular}
[c]{c}%
$1-p_{\mathcal{N}}$\\
~
\end{tabular}
& $%
\begin{tabular}
[c]{c}%
$\min\limits_{C}{\sum\limits_{(\mathbf{x},\mathbf{y})\in\tilde{C}}}\left(
1-p_{\mathbf{xy}}\right)  $\\
~
\end{tabular}
$%
\end{tabular}
\ \ \ \
\]
\caption{Analytical formulas for the end-to-end capacities of distillable
chains and networks.}%
\end{table*}

Let us note that the formulas for dephasing and erasure channels can be easily
extended to arbitrary dimension $d$. In fact, a qudit erasure channel is
formally defined as before and its two-way capacities
are~\cite{QKDpaper,GEW,Erasure}
\begin{equation}
\mathcal{C}(p)=(1-p)\log_{2}d.
\end{equation}
Therefore, it is sufficient to multiply by $\log_{2}d$ the corresponding
expressions in Table~I. Then, in arbitrary dimension $d$, the dephasing
channel is defined as
\begin{equation}
\rho\rightarrow\sum_{k=0}^{d-1}p_{k}Z_{d}^{k}\rho(Z_{d}^{\dagger})^{k},
\end{equation}
where $p_{k}$\ is the probability of $k$ phase flips and $Z_{d}^{k}\left\vert
i\right\rangle =\exp(2\pi ikd^{-1})\left\vert i\right\rangle $. Its generic
two-way capacity is~\cite{QKDpaper}
\begin{equation}
\mathcal{C}(p,d)=\log_{2}d-H(\{p_{k}\}), \label{eq4}%
\end{equation}
where $H(\{p_{k}\}):=-\sum_{k}p_{k}\log_{2}p_{k}$ is the Shannon entropy. Here
the generalization is also simple. For instance, in a chain $\{\mathcal{E}%
_{i}\}$ of such $d$-dimensional dephasing channels, we would have $N+1$
distributions $\{p_{k}^{i}\}$. We then compute the most entropic distribution,
i.e., we take the maximization $\max_{i}H(\{p_{k}^{i}\})$. This is the
bottleneck that determines the repeater capacity, so that
\begin{equation}
\mathcal{C}(\{p_{k}^{i}\})=\log_{2}d-\max_{i}H(\{p_{k}^{i}\}).
\end{equation}
Generalization to dimension $d$ is also immediate for the two network
capacities $\mathcal{C}$ and $\mathcal{C}^{\mathrm{m}}$.

\section*{DISCUSSION}

This work establishes the ultimate boundaries of quantum and
private communications assisted by repeaters, from the case of a
single repeater chain to an arbitrary quantum network under
single- or multi-path routing. Assuming arbitrary quantum channels
between the nodes, we have shown that the end-to-end capacities
are bounded by single-letter quantities based on the relative
entropy of entanglement. These upper bounds are very general and
also apply to chains and networks with untrusted nodes (i.e., run
by an eavesdropper). Our theory is formulated in a general
information-theoretic fashion which also applies to other
entanglement measures, as discussed in our Methods section. The
upper bounds are particularly important because they set the
tightest upper limits on the performance of quantum repeaters in
various network configurations. For instance, our benchmarks may
be used to evaluate performances in relay-assisted QKD\ protocols
such as MDI-QKD\ and variants~\cite{MDI1,MDI2,CVMDIQKD}. Related
literature and other
developments~\cite{ref1,ref2,ref3,ref4,ref5,TomKenn,Marco,Ma} are
discussed in Supplementary Note~6.

For the lower bounds, we have employed classical composition methods of the
capacities, either based on the widest path problem or the maximum flow,
depending on the type of routing. In general, these simple and classical lower
bounds do not coincide with the quantum upper bounds. However this is
remarkably the case for distillable networks, for which the ultimate quantum
communication performance can be completely reduced to the resolution of
classical problems of network information theory. For these networks, widest
path and maximum flow determine the quantum performance in terms of secret key
generation, entanglement distribution and transmission of quantum information.
In this way, we have been able to exactly establish the various end-to-end
capacities of distillable chains and networks where the quantum systems are
affected by the most fundamental noise models, including bosonic loss, which
is the most important for optical and telecom communications, quantum-limited
amplification, dephasing and erasure. In particular, our results also showed
how the parallel or \textquotedblleft broadband\textquotedblright\ use of a
lossy quantum network via multi-path routing may greatly improve the
end-to-end rates.

\section{Methods}

We present the main techniques that are needed to prove the
results of our main text. These methods are here provided for a
more general entanglement measure $E_{\text{\textrm{M}}}$, and
specifically apply to the REE. We consider a quantum network
$\mathcal{N}$ under single- or multi-path routing. In particular,
a chain of quantum repeaters can be treated as a single-route
quantum network.

For the upper bounds, our methodology can be broken down in the
following steps: (i) Derivation of a general weak converse upper
bound in terms of a suitable entanglement measure (in particular,
the REE); (ii) Simulation of the quantum network, so that quantum
channels are replaced by resource states; (iii) Stretching of the
network with respect to an entanglement cut, so that Alice and
Bob's shared state has a simple decomposition in terms of resource
states; (iv) Data processing, subadditivity over tensor-products,
and minimization over entanglement cuts. These steps provide
entanglement-based upper bounds for the end-to-end capacities. For
the lower bounds, we perform a suitable composition of the
point-to-point capacities of the single-link channels by means of
the widest path and the maximum flow, depending on the routing.
For the case of distillable quantum networks (and chains), these
lower bounds coincide with the upper bounds expressed in terms of
the REE.

\subsection{General (weak converse) upper bound}

This closely follows the derivation of the corresponding
point-to-point upper bound first given in the second 2015 arXiv
version of Ref.~\cite{QKDpaper} and later reported as Theorem~2 in
Ref.~\cite{TQC}. Consider an arbitrary end-to-end
$(n,R_{n}^{\varepsilon},\varepsilon)$ network protocol
$\mathcal{P}$ (single- or multi-path). This outputs a shared state
$\rho_{\mathbf{ab}}^{n}$ for Alice and Bob after $n$ uses, which
is $\varepsilon$-close to a target private state~\cite{KD1,KD2}
$\phi^{n}$ having $nR_{n}^{\varepsilon}$ secret bits, i.e., in
trace norm we have $||\rho
_{\mathbf{ab}}^{n}-\phi^{n}||_{1}\leq\varepsilon$. Consider now an
entanglement measure $E_{\text{\textrm{M}}}$ which is normalized
on the target state, i.e.,
\begin{equation}
E_{\text{\textrm{M}}}(\phi^{n})\geq nR_{n}^{\varepsilon}.
\end{equation}
Assume that $E_{\text{\textrm{M}}}$ is continuous. This means
that, for $d$-dimensional states $\rho$ and $\sigma$ that are
close in trace norm as $\left\Vert \rho -\sigma\right\Vert
_{1}\leq\varepsilon$, we may write
\begin{equation}
\left\vert E_{\text{\textrm{M}}}(\rho)-E_{\text{\textrm{M}}}(\sigma
)\right\vert \leq g(\varepsilon)\log_{2}d+h(\varepsilon),
\end{equation}
with the functions $g$ and $h$ converging to zero in $\varepsilon$. Assume
also that $E_{\text{\textrm{M}}}$ is monotonic under trace-preserving LOCCs
$\bar{\Lambda}$, so that%
\begin{equation}
E_{\text{\textrm{M}}}[\bar{\Lambda}(\rho)]\leq E_{\text{\textrm{M}}}(\rho),
\end{equation}
a property which is also known as data processing inequality. Finally, assume
that $E_{\text{\textrm{M}}}$ is subadditive over tensor products, i.e.,
\begin{equation}
E_{\text{\textrm{M}}}(\rho^{\otimes n})\leq nE_{\text{\textrm{M}}}(\rho).
\end{equation}
All these properties are certainly satisfied by the REE $E_{\mathrm{R}}$ and
the squashed entanglement (SQ) $E_{\mathrm{SQ}}$, with specific expressions
for $g$ and $h$ (e.g., these expressions are explicitly reported in
Sec.~VIII.A of Ref.~\cite{TQC}).

Using the first two properties (normalization and continuity), we may write
\begin{equation}
R_{n}^{\varepsilon}\leq\frac{E_{\text{\textrm{M}}}(\rho_{\mathbf{ab}}%
^{n})+g(\varepsilon)\log_{2}d+h(\varepsilon)}{n}, \label{main}%
\end{equation}
where $d$ is the dimension of the target private state. We know that this
dimension is at most exponential in the number of uses, i.e., $\log_{2}%
d\leq\alpha nR_{n}^{\varepsilon}$ for constant $\alpha$ (e.g., see
Ref.~\cite{QKDpaper} or Lemma~1 in Ref.~\cite{TQC}). By replacing this
dimensional bound in Eq.~(\ref{main}), taking the limit for large $n$ and
small $\varepsilon$ (weak converse), we derive%
\begin{equation}
\lim_{\varepsilon}\lim_{n}R_{n}^{\varepsilon}\leq\lim_{n}\frac
{E_{\text{\textrm{M}}}(\rho_{\mathbf{ab}}^{n})}{n}.
\end{equation}
Finally, we take the supremum over all protocols $\mathcal{P}$ so
that we can write our general upper bound for the end-to-end
secret key capacity (SKC) of the network
\begin{equation}
E_{\text{\textrm{M}}}^{\star}(\mathcal{N}):=\sup_{\mathcal{P}}\lim_{n}%
\frac{E_{\text{\textrm{M}}}(\rho_{\mathbf{ab}}^{n})}{n}. \label{mention}%
\end{equation}
In particular, this is an upper bound to the single-path SKC\ $\mathcal{K}$ if
$\mathcal{P}$ are single-path protocols, and to the multi-path SKC
$\mathcal{K}^{m}$ if $\mathcal{P}$ are multi-path (flooding) protocols.

In the case of an infinite-dimensional state $\rho_{\mathbf{ab}}^{n}$, the
proof can be repeated by introducing a truncation trace-preserving
LOCC\ $\boldsymbol{T}$, so that $\delta_{\mathbf{ab}}^{n}=\boldsymbol{T}%
(\rho_{\mathbf{ab}}^{n})$ is a finite-dimensional state. The proof is repeated
for $\delta_{\mathbf{ab}}^{n}$ and finally we use the data processing
$E_{\text{\textrm{M}}}(\delta_{\mathbf{ab}}^{n})\leq E_{\text{\textrm{M}}%
}(\rho_{\mathbf{ab}}^{n})$ to write the same upper bound as in
Eq.~(\ref{mention}). This follows the same steps of the proof given in the
second 2015 arXiv version of Ref.~\cite{QKDpaper} and later reported as
Theorem~2 in Ref.~\cite{TQC}. It is worth mentioning that Eq.~(\ref{mention})
can equivalently be proven without using the exponential growth of the private
state, i.e., using the steps of the third proof given in the Supplementary
Note~3 of Ref.~\cite{QKDpaper}.

\subsection{Network simulation}

Given a network $\mathcal{N}=(P,E)$ with generic point $\mathbf{x}\in P$ and
edge $(\mathbf{x},\mathbf{y})\in E$, replace the generic channel
$\mathcal{E}_{\mathbf{xy}}$ with a simulation over a resource state
$\sigma_{\mathbf{xy}}$. This means to write $\mathcal{E}_{\mathbf{xy}}%
(\rho)=\mathcal{T}_{\mathbf{xy}}(\rho\otimes\sigma_{\mathbf{xy}})$ for any
input state $\rho$, by resorting to a suitable trace-preserving LOCC
$\mathcal{T}_{\mathbf{xy}}$ (this is always possible for any quantum
channel~\cite{QKDpaper}). If we perform this operation for all the edges, we
then define the simulation of the network $\sigma(\mathcal{N})=\{\sigma
_{\mathbf{xy}}\}_{(\mathbf{x},\mathbf{y})\in E}$ where each channel is
replaced by a corresponding resource state.\ If the channels are bosonic, then
the simulation is typically asymptotic of the type $\mathcal{E}_{\mathbf{xy}%
}(\rho)=\lim_{\mu}\mathcal{E}_{\mathbf{xy}}^{\mu}(\rho)$ where $\mathcal{E}%
_{\mathbf{xy}}^{\mu}(\rho)=\mathcal{T}_{\mathbf{xy}}^{\mu}(\rho\otimes
\sigma_{\mathbf{xy}}^{\mu})$ for some sequence of simulating LOCCs
$\mathcal{T}_{\mathbf{xy}}^{\mu}$ and sequence of resource states
$\sigma_{\mathbf{xy}}^{\mu}$.

Here the parameter $\mu$ is usually connected with the energy of the resource
state. For instance, if $\mathcal{E}_{\mathbf{xy}}$ is a
teleportation-covariant bosonic channel, then the resource state
$\sigma_{\mathbf{xy}}^{\mu}$ is its quasi-Choi matrix $\sigma_{\mathcal{E}%
_{\mathbf{xy}}}^{\mu}:=\mathcal{I}\otimes\mathcal{E}_{\mathbf{xy}}(\Phi^{\mu
})$, with $\Phi^{\mu}$ being a two-mode squeezed vacuum state
(TMSV) state~\cite{RMP} whose parameter $\mu=\bar{n}+1/2$ is
related to the mean number $\bar{n}$ of thermal photons.
Similarly, the simulating LOCC $\mathcal{T}_{\mathbf{xy}}^{\mu}$
is a Braunstein-Kimble protocol~\cite{BK,BK2} where the ideal Bell
detection is replaced by the finite-energy projection onto
$\alpha$-displaced TMSV states $D(\alpha)\Phi^{\mu}D(-\alpha)$,
with $D$ being the phase-space displacement operator~\cite{RMP}.

Given an asymptotic simulation of a quantum channel, the
associated simulation error is correctly quantified by employing
the energy-constrained diamond distance~\cite{QKDpaper}, which
must go to zero in the limit, i.e.,
\begin{equation}
\left\Vert \mathcal{E}_{\mathbf{xy}}-\mathcal{E}_{\mathbf{xy}}^{\mu
}\right\Vert _{\diamond\bar{N}}\overset{\mu}{\rightarrow}0\text{~for any
finite }\bar{N}\text{.} \label{deltaN}%
\end{equation}
Recall that, for any two bosonic channels $\mathcal{E}$ and $\mathcal{E}%
^{\prime}$, this quantity is defined as
\begin{equation}
\left\Vert \mathcal{E}-\mathcal{E}^{\prime}\right\Vert _{\diamond\bar{N}%
}:=\sup_{\rho_{AB}\in D_{\bar{N}}}\left\Vert \mathcal{I}_{A}\otimes
\mathcal{E}(\rho_{AB})-\mathcal{I}_{A}\otimes\mathcal{E}^{\prime}(\rho
_{AB})\right\Vert _{1},
\end{equation}
where $D_{\bar{N}}$ is the compact set of bipartite bosonic states with
$\bar{N}$ mean number of photons (see Ref.~\cite{Shirokov} for a later and
slightly different definition, where the constraint is only on the $B$ part).
Thus, in general, if the network has bosonic channels, we may write the
asymptotic simulation $\sigma(\mathcal{N})=\lim_{\mu}\sigma^{\mu}%
(\mathcal{N})$ where $\sigma^{\mu}(\mathcal{N}):=\{\sigma_{\mathbf{xy}}^{\mu
}\}_{(\mathbf{x},\mathbf{y})\in E}$.

\subsection{Stretching of the network}

Once we simulate a network, the next step is its stretching, which is the
complete adaptive-to-block simplification of its output state (for the exact
details of this procedure see Supplementary Note~3). As a result of
stretching, the $n$-use output state of the generic network protocol can be
decomposed as
\begin{equation}
\rho_{\mathbf{ab}}^{n}=\bar{\Lambda}_{\mathbf{ab}}\left[
{\textstyle\bigotimes\limits_{(\mathbf{x},\mathbf{y})\in E}}
\sigma_{\mathbf{xy}}^{\otimes n_{\mathbf{xy}}}\right]  ~, \label{stretN}%
\end{equation}
where $\bar{\Lambda}$ represents a trace-preserving LOCC (which is
local with respect to Alice and Bob). The LOCC $\bar{\Lambda}$
includes all the adaptive LOCCs from the original protocol besides
the simulating LOCCs. In Eq.~(\ref{stretN}), the parameter
$n_{\mathbf{xy}}$ is the number of uses of the edge
$(\mathbf{x},\mathbf{y})$, that we may always approximate to an
integer for large $n$. We have $n_{\mathbf{xy}}\leq n$ for
single-path routing, and $n_{\mathbf{xy}}=n$ for flooding
protocols in multi-path routing.

In the presence of bosonic channels and asymptotic simulations, we modify
Eq.~(\ref{stretN}) into the approximate stretching%
\begin{equation}
\rho_{\mathbf{ab}}^{n,\mu}=\bar{\Lambda}_{\mathbf{ab}}^{\mu}\left[
{\textstyle\bigotimes\limits_{(\mathbf{x},\mathbf{y})\in E}}
\sigma_{\mathbf{xy}}^{\mu\otimes n_{\mathbf{xy}}}\right]  ,
\end{equation}
which tends to the actual output $\rho_{\mathbf{ab}}^{n}$ for large $\mu$. In
fact, using a \textquotedblleft peeling\textquotedblright%
\ technique~\cite{QKDpaper,TQC} which exploits the triangle inequality and the
monotonicity of the trace distance under completely-positive trace-preserving
maps, we may write the following bound
\begin{equation}
\left\Vert \rho_{\mathbf{ab}}^{n}-\rho_{\mathbf{ab}}^{n,\mu}\right\Vert
_{1}\leq\varepsilon^{\mu}:=\sum_{(\mathbf{x},\mathbf{y})\in E}n_{\mathbf{xy}%
}\left\Vert \mathcal{E}_{\mathbf{xy}}-\mathcal{E}_{\mathbf{xy}}^{\mu
}\right\Vert _{\diamond\bar{N}},
\end{equation}
which goes to zero in $\mu$ for any finite input energy $\bar{N}$, finite
number of uses $n$ of the protocol, and finite number of edges $|E|$ in the
network (the explicit steps of the proof can be found in Supplementary Note~3).

\subsection{Stretching with respect to entanglement cuts}

The decomposition of the output state can be greatly simplified by introducing
cuts in the network. In particular, we may drastically reduce the number of
resource states in its representation. Given a cut $C$ of $\mathcal{N}$ with
cut-set $\tilde{C}$, we may in fact stretch the network with respect to that
specific cut (see again Supplementary Note~3 for exact details of the
procedure). In this way, we may write
\begin{equation}
\rho_{\mathbf{ab}}^{n}(C)=\bar{\Lambda}_{\mathbf{ab}}\left[
{\textstyle\bigotimes\limits_{(\mathbf{x},\mathbf{y})\in\tilde{C}}}
\sigma_{\mathbf{xy}}^{\otimes n_{\mathbf{xy}}}\right]  ~, \label{streC1}%
\end{equation}
where $\bar{\Lambda}_{\mathbf{ab}}$ is a trace-preserving LOCC with respect to
Alice and Bob (differently from before, this LOCC now\ depends on the cut $C$,
but we prefer not to complicate the notation). Similarly, in the presence of
bosonic channels, we may consider the approximate decomposition
\begin{equation}
\rho_{\mathbf{ab}}^{n,\mu}(C)=\bar{\Lambda}_{\mathbf{ab}}^{\mu}\left[
{\textstyle\bigotimes\limits_{(\mathbf{x},\mathbf{y})\in\tilde{C}}}
\sigma_{\mathbf{xy}}^{\mu\otimes n_{\mathbf{xy}}}\right]  , \label{streC2}%
\end{equation}
which converges in trace distance to $\rho_{\mathbf{ab}}^{n}(C)$ for large
$\mu$.

\subsection{Data processing and subadditivity}

Let us combine the stretching in Eq.~(\ref{streC1}) with two basic
properties of the entanglement measure $E_{\text{\textrm{M}}}$.
The first property is the monotonicity of $E_{\text{\textrm{M}}}$
under trace-preserving LOCCs; the second property is the
subadditivity of $E_{\text{\textrm{M}}}$ over tensor-product
states. Using these properties, we can simplify the general upper
bound of Eq.~(\ref{mention}) into a simple and computable
single-letter quantity. In fact, for any cut $C$ of the network
$\mathcal{N}$, we write
\begin{align}
E_{\text{\textrm{M}}}[\rho_{\mathbf{ab}}^{n}(C)] &  \leq E_{\text{\textrm{M}}%
}\left[
{\textstyle\bigotimes\limits_{(\mathbf{x},\mathbf{y})\in\tilde{C}}}
\sigma_{\mathbf{xy}}^{\otimes n_{\mathbf{xy}}}\right]  \label{forREG}\\
&  \leq\sum_{(\mathbf{x},\mathbf{y})\in\tilde{C}}n_{\mathbf{xy}}%
E_{\text{\textrm{M}}}(\sigma_{\mathbf{xy}}),
\end{align}
where $\bar{\Lambda}_{\mathbf{ab}}$ has disappeared. Let us introduce the
probability of using the generic edge $(\mathbf{x},\mathbf{y})$
\begin{equation}
p_{\mathbf{xy}}:=\lim_{n}\frac{n_{\mathbf{xy}}}{n},
\end{equation}
so that we may write the limit
\begin{equation}
\lim_{n}\frac{E_{\text{\textrm{M}}}[\rho_{\mathbf{ab}}^{n}(C)]}{n}\leq
\sum_{(\mathbf{x},\mathbf{y})\in\tilde{C}}p_{\mathbf{xy}}E_{\text{\textrm{M}}%
}(\sigma_{\mathbf{xy}}).
\end{equation}
Using the latter in Eq.~(\ref{mention}) allows us to write the following
bound, for any cut $C$%
\begin{equation}
E_{\text{\textrm{M}}}^{\star}(\mathcal{N})\leq E_{\text{\textrm{M}}}^{\star
}(\mathcal{N},C):=\sup_{\{p_{\mathbf{xy}}\}}\sum_{(\mathbf{x},\mathbf{y}%
)\in\tilde{C}}p_{\mathbf{xy}}E_{\text{\textrm{M}}}(\sigma_{\mathbf{xy}%
}).\label{toCc}%
\end{equation}

In the case of bosonic channels and asymptotic simulations, we may use the
triangle inequality
\begin{align}
||\rho_{\mathbf{ab}}^{n,\mu}-\phi^{n}||_{1}  &  \leq||\rho_{\mathbf{ab}%
}^{n,\mu}-\rho_{\mathbf{ab}}^{n}||_{1}\nonumber\\
+||\rho_{\mathbf{ab}}^{n}-\phi^{n}||_{1}  &  \leq\varepsilon^{\mu}%
+\varepsilon:=\Sigma^{\mu}{\rightarrow}0.
\end{align}
Then, we may repeat the derivations around Eqs.~(\ref{main})-(\ref{mention})
for $\rho_{\mathbf{ab}}^{n,\mu}$ instead of $\rho_{\mathbf{ab}}^{n}$, where we
also include the use of a suitable truncation of the states via a
trace-preserving LOCC $\boldsymbol{T}$ (see also Sec.~VIII.D of
Ref.~\cite{TQC}\ for a similar approach in the point-to-point case). This
leads to the $\mu$-dependent upper-bound
\begin{equation}
E_{\text{\textrm{M}}}^{\star}(\mathcal{N},\mu):=\sup_{\mathcal{P}}\lim
_{n}\frac{E_{\text{\textrm{M}}}(\rho_{\mathbf{ab}}^{n,\mu})}{n}.
\end{equation}
Because this is valid for any $\mu$, we may conservatively take the inferior
limit in $\mu$ and consider the upper bound%
\begin{equation}
E_{\text{\textrm{M}}}^{\star}(\mathcal{N}):=\underset{\mu}{\lim\inf
}~E_{\text{\textrm{M}}}^{\star}(\mathcal{N},\mu).
\end{equation}

Finally, by introducing the stretching of Eq.~(\ref{streC2}) with respect to
an entanglement cut $C$, and using the monotonicity and subadditivity of
$E_{\text{\textrm{M}}}$ with respect to the decomposition of $\rho
_{\mathbf{ab}}^{n,\mu}(C)$, we may repeat the previous reasonings and write%
\begin{align}
E_{\text{\textrm{M}}}^{\star}(\mathcal{N})  &  \leq E_{\text{\textrm{M}}%
}^{\star}(\mathcal{N},C)\nonumber\\
&  :=\sup_{\{p_{\mathbf{xy}}\}}\sum_{(\mathbf{x},\mathbf{y})\in\tilde{C}%
}p_{\mathbf{xy}}\left[  \underset{\mu}{\lim\inf}~E_{\text{\textrm{M}}}%
(\sigma_{\mathbf{xy}}^{\mu})\right]  , \label{ToCc2}%
\end{align}
which is a direct extension of the bound in Eq.~(\ref{toCc}).

We may formulate both Eqs.~(\ref{toCc}) and~(\ref{ToCc2}) in a compact way if
we define the entanglement measure $E_{\text{\textrm{M}}}$ over an asymptotic
state $\sigma:=\lim_{\mu}\sigma^{\mu}$ as
\begin{equation}
E_{\text{\textrm{M}}}(\sigma):=\underset{\mu}{\lim\inf}~E_{\text{\textrm{M}}%
}(\sigma^{\mu}). \label{eASS}%
\end{equation}
It is clear that, for a physical (non-asymptotic) state, we have the trivial
sequence $\sigma^{\mu}=\sigma$ for any $\mu$, so that Eq.~(\ref{eASS})
provides the standard definition. In the specific case of REE, we may write%
\begin{align}
E_{\text{\textrm{R}}}(\sigma)  &  =\underset{\mu}{\lim\inf}%
~E_{\text{\textrm{R}}}(\sigma^{\mu})\nonumber\\
&  =\inf_{\gamma^{\mu}}~\underset{\mu}{\lim\inf}~S(\sigma^{\mu}||\gamma^{\mu
}), \label{asyREE}%
\end{align}
where $\gamma^{\mu}$ is a sequence of separable states that
converges in trace norm; this means that there exists a separable
state $\gamma$ such that
$||\gamma^{\mu}-\gamma||_{1}\overset{\mu}{\rightarrow}0$.
Employing the extended definition of Eq.~(\ref{eASS}), we may
write Eq.~(\ref{toCc}) for both non-asymptotic
$\sigma_{\mathbf{xy}}$ and asymptotic states
$\sigma_{\mathbf{xy}}:=\lim_{\mu}\sigma_{\mathbf{xy}}^{\mu}$.

\subsection{Minimum entanglement cut and upper bounds}

By minimizing Eq.~(\ref{toCc}) over all possible cuts of the network, we find
the tightest upper bound, i.e.,
\begin{equation}
E_{\text{\textrm{M}}}^{\star}(\mathcal{N})\leq\min_{C}E_{\text{\textrm{M}}%
}^{\star}(\mathcal{N},C).
\end{equation}
Let us now specify this formula for different types of routing. For
single-path routing, we have $p_{\mathbf{xy}}\leq1$, so that we may use
\begin{equation}
\sup_{\{p_{\mathbf{xy}}\}}\sum_{(\mathbf{x},\mathbf{y})\in\tilde{C}%
}p_{\mathbf{xy}}(\cdots)\leq\max_{(\mathbf{x},\mathbf{y})\in\tilde{C}}%
(\cdots),
\end{equation}
in Eq.~(\ref{toCc}). Therefore, we derive the following upper
bound for the single-path SKC
\begin{equation}
\mathcal{K}(\mathcal{N})\leq\min_{C}E_{\text{\textrm{M}}}(C), \label{UB1}%
\end{equation}
where we introduce the single-edge flow of entanglement through the cut%
\begin{equation}
E_{\text{\textrm{M}}}(C):=\max_{(\mathbf{x},\mathbf{y})\in\tilde{C}%
}E_{\text{\textrm{M}}}(\sigma_{\mathbf{xy}}). \label{UBB1}%
\end{equation}
In particular, we may specify this result to a single chain of $N$
points and $N+1$ channels $\{\mathcal{E}_{i}\}$ with resource
states $\{\sigma_{i}\}$. This is a quantum network with a single
route, so that the cuts can be labelled by $i$ and the cut-sets
are just composed of a single edge. Therefore, Eqs.~(\ref{UB1})
and~(\ref{UBB1}) become
\begin{equation}
\mathcal{K}(\{\mathcal{E}_{i}\})\leq\min_{i}E_{\text{\textrm{M}}}(\sigma_{i}).
\label{chainAAA}%
\end{equation}

For multi-path routing, we have $p_{\mathbf{xy}}=1$ (flooding), so that we may
simplify
\begin{equation}
\sup_{\{p_{\mathbf{xy}}\}}\sum_{(\mathbf{x},\mathbf{y})\in\tilde{C}%
}p_{\mathbf{xy}}(\cdots)=\sum_{(\mathbf{x},\mathbf{y})\in\tilde{C}}(\cdots),
\end{equation}
in Eq.~(\ref{toCc}). Therefore, we can write the following upper
bound for the multi-path SKC
\begin{equation}
\mathcal{K}^{\text{\textrm{m}}}(\mathcal{N})\leq\min_{C}E_{\text{\textrm{M}}%
}^{\text{\textrm{m}}}(C), \label{UB2}%
\end{equation}
where we introduce the multi-edge flow of entanglement through the cut%
\begin{equation}
E_{\text{\textrm{M}}}^{\text{\textrm{m}}}(C):=\sum_{(\mathbf{x},\mathbf{y}%
)\in\tilde{C}}E_{\text{\textrm{M}}}(\sigma_{\mathbf{xy}}). \label{UBB2}%
\end{equation}
In these results, the definition of $E_{\text{\textrm{M}}}%
(\sigma_{\mathbf{xy}})$ is implicitly meant to be extended to asymptotic
states, according to Eq.~(\ref{eASS}). Then, note that the tightest values of
the upper bounds are achieved by extending the minimization to all network
simulations $\sigma(\mathcal{N})$, i.e., by enforcing $\min_{C}\rightarrow
\min_{\sigma(\mathcal{N})}\min_{C}$ in Eqs.~(\ref{UB1}) and~(\ref{UB2}).

Specifying Eqs.~(\ref{UB1}), (\ref{chainAAA}), and~(\ref{UB2}) to
the REE, we get the single-letter upper bounds
\begin{align}
\mathcal{C}(\{\mathcal{E}_{i}\})  &  \leq\mathcal{K}(\{\mathcal{E}_{i}%
\})\leq\min_{i}E_{\text{\textrm{R}}}(\sigma_{i}),\label{eqq0}\\
\mathcal{C}(\mathcal{N})  &  \leq\mathcal{K}(\mathcal{N})\leq\min
_{C}E_{\text{\textrm{R}}}(C),\label{eqq1}\\
\mathcal{C}^{\mathrm{m}}(\mathcal{N})  &  \leq\mathcal{K}^{\mathrm{m}%
}(\mathcal{N})\leq\min_{C}E_{\text{\textrm{R}}}^{\text{\textrm{m}}}(C),
\label{eqq2}%
\end{align}
which are Eqs.~(\ref{chain1}), (\ref{ubseqNN}) and~(\ref{mpBOUND}) of the main
text. The proofs of these upper bounds in terms of the REE\ can equivalently
be done following the \textquotedblleft converse part\textquotedblright%
\ derivations in Supplementary Note~1 (for chains), Supplementary Note~4 (for
networks under single-path routing), and Supplementary Note~5 (for networks
under multi-path routing). Differently from what presented in this Methods
section, such proofs exploit the lower semi-continuity of the quantum relative
entropy~\cite{HolevoBOOK} in order to deal with asymptotic simulations (e.g.
for bosonic channels).

\subsection{Lower bounds}

To derive lower bounds we combine the known results on two-way assisted
capacities~\cite{QKDpaper} with classical results in network information
theory. Consider the generic two-way assisted capacity $\mathcal{C}%
_{\mathbf{xy}}$ of the channel $\mathcal{E}_{\mathbf{xy}}$ (in particular,
this can be either $D_{2}=Q_{2}$ or $K$). Then, using the cut property of the
widest path (Supplementary Note~4), we derive the following achievable rate
for the generic single-path capacity of the network $\mathcal{N}$%
\begin{equation}
\mathcal{C}(\mathcal{N})\geq\min_{C}\max_{(\mathbf{x},\mathbf{y})\in\tilde{C}%
}\mathcal{C}_{\mathbf{xy}}~.
\end{equation}
For a chain $\{\mathcal{E}_{i}\}$, this simply specifies to%
\begin{equation}
\mathcal{C}(\{\mathcal{E}_{i}\})\geq\min_{i}\mathcal{C}(\mathcal{E}_{i}).
\end{equation}
Using the classical max-flow min-cut theorem (Supplementary
Note~5), we derive the following achievable rate for the generic
multi-path capacity of $\mathcal{N}$
\begin{equation}
\mathcal{C}^{\text{\textrm{m}}}(\mathcal{N})\geq\min_{C}\sum_{(\mathbf{x}%
,\mathbf{y})\in\tilde{C}}\mathcal{C}_{\mathbf{xy}}~.
\end{equation}

\subsection{Simplifications for teleportation-covariant and distillable
networks}

Recall that a quantum channel $\mathcal{E}$ is said to be
teleportation-covariant~\cite{QKDpaper} when, for any
teleportation unitary $U$ (Weyl-Pauli operator in finite dimension
or phase-space displacement in infinite dimension), we have
\begin{equation}
\mathcal{E}(U\rho U^{\dagger})=V\mathcal{E}(\rho)V^{\dagger},
\end{equation}
for some (generally-different) unitary transformation $V$. In this
case the quantum channel can be simulated by applying
teleportation over
its Choi matrix $\sigma_{\mathcal{E}}:=\mathcal{I}\otimes\mathcal{E}%
(\Phi)$, where $\Phi$ is a maximally-entangled state. Similarly,
if the teleportation-covariant channel is bosonic, we can write an
approximate
simulation by teleporting over the quasi-Choi matrix $\sigma_{\mathcal{E}%
}^{\mu}:=\mathcal{I}\otimes\mathcal{E}(\Phi^{\mu})$, where
$\Phi^{\mu}$ is a TMSV\ state. For a network of
teleportation-covariant channels, we therefore use teleportation
to simulate the network, so that the resource states in the upper
bounds of Eqs.~(\ref{eqq0})-(\ref{eqq2}) are Choi matrices
(physical or
asymptotic). In other words, we write the sandwich relations%
\begin{align}
\min_{i}\mathcal{C}(\mathcal{E}_{i})  &  \leq\mathcal{C}(\{\mathcal{E}%
_{i}\})\leq\min_{i}E_{\text{\textrm{R}}}(\sigma_{\mathcal{E}_{i}%
}),\label{sand1}\\
\min_{C}\max_{(\mathbf{x},\mathbf{y})\in\tilde{C}}\mathcal{C}_{\mathbf{xy}}
&  \leq\mathcal{C}(\mathcal{N})\leq\min_{C}\max_{(\mathbf{x},\mathbf{y}%
)\in\tilde{C}}E_{\text{\textrm{R}}}(\sigma_{\mathcal{E}_{\mathbf{xy}}}),\\
\min_{C}\sum_{(\mathbf{x},\mathbf{y})\in\tilde{C}}\mathcal{C}_{\mathbf{xy}}
&  \leq\mathcal{C}^{\mathrm{m}}(\mathcal{N})\leq\min_{C}\sum_{(\mathbf{x}%
,\mathbf{y})\in\tilde{C}}E_{\text{\textrm{R}}}(\sigma_{\mathcal{E}%
_{\mathbf{xy}}}), \label{sand3}%
\end{align}
with the REE taking the form of Eq.~(\ref{asyREE}) on an
asymptotic Choi
matrix $\sigma_{\mathcal{E}_{\mathbf{xy}}}:=\lim_{\mu}\sigma_{\mathcal{E}%
_{\mathbf{xy}}}^{\mu}$.

As a specific case, consider a quantum channel which is not only
teleportation-covariant but also distillable, so that it
satisfies~\cite{QKDpaper}
\begin{equation}
\mathcal{C}(\mathcal{E})=E_{\text{\textrm{R}}}(\sigma_{\mathcal{E}}%
)=D_{1}(\sigma_{\mathcal{E}}),
\end{equation}
where $D_{1}(\sigma_{\mathcal{E}})$ is the one-way distillability
of the Choi matrix $\sigma_{\mathcal{E}}$ (with a suitable
asymptotic expression for bosonic Choi matrices~\cite{QKDpaper}).
If a network (or a chain) is composed of these channels, then the
relations in Eqs.~(\ref{sand1})--(\ref{sand3}) collapse and we
fully determine the capacities
\begin{align}
\mathcal{C}(\{\mathcal{E}_{i}\})  &  =\min_{i}E_{\text{\textrm{R}}}(\sigma_{\mathcal{E}_{i}}),\\
\mathcal{C}(\mathcal{N})  &  =\min_{C}\max_{(\mathbf{x},\mathbf{y})\in
\tilde{C}}E_{\text{\textrm{R}}}(\sigma_{\mathcal{E}_{\mathbf{xy}}}),\\
\mathcal{C}^{\text{\textrm{m}}}(\mathcal{N})  &  =\min_{C}\sum_{(\mathbf{x}%
,\mathbf{y})\in\tilde{C}}E_{\text{\textrm{R}}}(\sigma_{\mathcal{E}%
_{\mathbf{xy}}}).
\end{align}
These capacities correspond to Eqs.~(\ref{DistillableCAPP}),
(\ref{distillableCCCC}), and (\ref{multiCC}) of the main text. They are
explicitly computed for chains and networks composed of lossy channels,
quantum-limited amplifiers, dephasing and erasure channels in Table~I of the
main text.

\subsection{Regularizations and other measures}

It is worth noticing that some of the previous formulas can be
re-formulated by using the regularization of the entanglement
measure, i.e.,
\begin{equation}
E_{\text{\textrm{M}}}^{\infty}(\sigma):=\lim_{n}\frac{E_{\text{\textrm{M}}%
}(\sigma^{\otimes n})}{n}.
\end{equation}
In fact, let us go back to the first upper bound in Eq.~(\ref{forREG}), which
implies%
\begin{equation}
E_{\text{\textrm{M}}}[\rho_{\mathbf{ab}}^{n}(C)]\leq\sum_{(\mathbf{x}%
,\mathbf{y})\in\tilde{C}}E_{\text{\textrm{M}}}(\sigma_{\mathbf{xy}}^{\otimes
n_{\mathbf{xy}}}). \label{condfff}%
\end{equation}
For a network under multi-path routing we have $n_{\mathbf{xy}}=n$, so that we
may write
\begin{equation}
\lim_{n}\frac{E_{\text{\textrm{M}}}[\rho_{\mathbf{ab}}^{n}(C)]}{n}\leq
\sum_{(\mathbf{x},\mathbf{y})\in\tilde{C}}E_{\text{\textrm{M}}}^{\infty
}(\sigma_{\mathbf{xy}}).
\end{equation}
By repeating previous steps, the latter equation implies the upper bound%
\begin{equation}
\mathcal{K}^{\text{\textrm{m}}}(\mathcal{N})\leq\min_{C}\sum_{(\mathbf{x}%
,\mathbf{y})\in\tilde{C}}E_{\text{\textrm{M}}}^{\infty}(\sigma_{\mathbf{xy}}),
\end{equation}
which is generally tighter than the result in Eqs.~(\ref{UB2}) and~(\ref{UBB2}%
). The same regularization can be written for a chain $\{\mathcal{E}_{i}\}$,
which can also be seen as a single-route network satisfying the flooding
condition $n_{\mathbf{xy}}=n$. Therefore, starting from the condition of
Eq.~(\ref{condfff}) with $n_{\mathbf{xy}}=n$, we may write
\begin{equation}
\mathcal{K}(\{\mathcal{E}_{i}\})\leq\min_{i}E_{\text{\textrm{M}}}^{\infty
}(\sigma_{i}),
\end{equation}
which is generally tighter than the result in Eq.~(\ref{chainAAA}). These
regularizations are important for the REE, but not for the squashed
entanglement which is known to be additive over tensor-products, so that
$E_{\text{\textrm{SQ}}}^{\infty}(\sigma)=E_{\text{\textrm{SQ}}}(\sigma)$.

Another extension is related to the use of the relative entropy
distance with respect to partial-positive-transpose (PPT) states.
This quantity can be denoted by RPPT and is defined by~\cite{KD2}
\begin{equation}
E_{\mathrm{P}}\left(  \sigma\right)  :=\inf_{\gamma\in\mathrm{PPT}}%
S(\sigma||\gamma),
\end{equation}
with an asymptotic extension similar to Eq.~(\ref{asyREE}) but in
terms of converging sequences of PPT states $\gamma^{\mu}$. The
RPPT is tighter than the REE but does not provide an upper bound
to the distillable key of a state, but rather to its distillable
entanglement. This means that it has normalization
$E_{\mathrm{P}}\left(  \phi^{n}\right)  \geq nR_{n}$ on a target
maximally-entangled state $\phi^{n}$\ with $nR_{n}$ ebits.

The RPPT is known to be monotonic under the action of PPT\
operations (and therefore LOCCs); it is continuous and subadditive
over tensor-product states. Therefore, we may repeat the
derivation that leads to Eq.~(\ref{mention}) but with respect to
protocols $\mathcal{P}$
of entanglement distribution. This means that we can write%
\begin{equation}
Q_{2}(\mathcal{N})=D_{2}(\mathcal{N})\leq E_{\text{\textrm{P}}}^{\star
}(\mathcal{N}):=\sup_{\mathcal{P}}\lim_{n}\frac{E_{\text{\textrm{P}}}%
(\rho_{\mathbf{ab}}^{n})}{n}.
\end{equation}
Using the decomposition of the output state $\rho_{\mathbf{ab}}^{n}$ as in
Eqs.~(\ref{streC1}) and~(\ref{streC2}), and repeating previous steps, we may
finally write%
\[
D_{2}(\{\mathcal{E}_{i}\})\leq\min_{i}E_{\text{\textrm{P}}}^{\infty}%
(\sigma_{i})\leq\min_{i}E_{\text{\textrm{P}}}(\sigma_{i}),
\]
for a chain $\{\mathcal{E}_{i}\}$ with resource states $\{\sigma_{i}\}$, and
\begin{align}
D_{2}(\mathcal{N})  &  \leq\min_{C}\max_{(\mathbf{x},\mathbf{y})\in\tilde{C}%
}E_{\text{\textrm{P}}}(\sigma_{\mathbf{xy}}),\\
D_{2}^{\mathrm{m}}(\mathcal{N})  &  \leq\min_{C}\sum_{(\mathbf{x}%
,\mathbf{y})\in\tilde{C}}E_{\text{\textrm{P}}}^{\infty}(\sigma_{\mathbf{xy}%
})\nonumber\\
&  \leq\min_{C}\sum_{(\mathbf{x},\mathbf{y})\in\tilde{C}}E_{\text{\textrm{P}}%
}(\sigma_{\mathbf{xy}}),
\end{align}
for the single- and multi-path entanglement distribution capacities of a
quantum network $\mathcal{N}$ with resource states $\sigma(\mathcal{N}%
)=\{\sigma_{\mathbf{xy}}\}_{(\mathbf{x},\mathbf{y})\in E}$.

\subsection*{Acknowledgements}

This work has been supported by the EPSRC\ via the `UK Quantum
Communications HUB' (EP/M013472/1) and `qDATA' (EP/L011298/1), and
by the European Union via Continuous Variable Quantum
Communications (CiViQ, Project ID: 820466). The author would like
to thank Seth Lloyd, Koji Azuma, Bill Munro, Richard Wilson, Edwin
Hancock, Rod Van Meter, Marco Lucamarini, Riccardo Laurenza,
Thomas Cope, Carlo Ottaviani, Gaetana Spedalieri, Cosmo Lupo,
Samuel Braunstein, Saikat Guha and Dirk Englund for feedback and
discussions.


\clearpage

\begin{center}
\textbf{{\large Supplementary Information}}
\end{center}

\setcounter{section}{0} 
\setcounter{figure}{0}
\renewcommand{\thefigure}{S\arabic{figure}}
\renewcommand{\bibnumfmt}[1]{[S#1]}
\renewcommand{\citenumfont}[1]{S#1}


\section{Supplementary Note 1:\ Chains of quantum
repeaters\label{SECrepeaters}}

Consider Alice and Bob to be end-points of a chain of $N+2$ points
with $N$ repeaters in the middle. For $i=0,\ldots,N$ we assume
that point $i$ is connected with point $i+1$ by a quantum channel
$\mathcal{E}_{i}$ which can be
forward or backward, for a total of $N+1$ channels $\{\mathcal{E}_{0}%
,\ldots\mathcal{E}_{i},\ldots\mathcal{E}_{N}\}$. Each point has a
local register which is a countable ensemble of quantum systems,
denoted by
$\mathbf{r}_{i}$ for the $i$-th point. In particular, we set $\mathbf{a}%
=\mathbf{r}_{0}$ for Alice and $\mathbf{b}=\mathbf{r}_{N+1}$ for
Bob. Registers are updated. For instance, if Alice sends a system
$a$, then we update $\mathbf{a}\rightarrow\mathbf{a}a$; if Bob
receives a system $b$, then we update
$b\mathbf{b}\rightarrow\mathbf{b}$. For this formalism see also
Ref.~\cite{QKDpapers}. The channels are completely arbitrary even
though our following formulas will simplify for
teleportation-covariant channels, and the sub-class of distillable
channels (see Ref.~\cite{QKDpapers} or the main paper\ for the
exact definitions of these channels).

The most general distribution protocol over the chain is based on
adaptive LOs and unlimited two-way CC involving all the points in
the chain. In other words, each point broadcasts classical
information and receives classical feedback from all the other
points, which is used to perform conditional LOs on the local
registers. In the following we always assume these
\textquotedblleft network\textquotedblright\ adaptive LOCCs,
unless we specify otherwise. The first step is the preparation of
the registers by an LOCC $\Lambda_{0}$ whose application to some
fundamental state provides an initial separable state
$\sigma_{\mathbf{ar}_{1}\cdots\mathbf{r}_{N}\mathbf{b}}$. Then,
Alice and the first repeater exchange a quantum system through
channel $\mathcal{E}_{0}$ (via forward or backward transmission).
This is followed by
an LOCC $\Lambda_{1}$ on the updated registers $\mathbf{ar}_{1}\mathbf{r}%
_{2}\ldots\mathbf{r}_{N}\mathbf{b}$. Next, the first and the
second repeaters exchange another quantum system through channel
$\mathcal{E}_{1}$ followed by another LOCC $\Lambda_{2}$, and so
on. Finally, Bob exchanges a system with the $N$th repeater
through channel $\mathcal{E}_{N}$ and the final LOCC
$\Lambda_{N+1}$ provides the output state $\rho_{\mathbf{ar}_{1}%
\cdots\mathbf{r}_{N}\mathbf{b}}$.

\begin{figure}[ptbh]
\vspace{-1.0cm}
\par
\begin{center}
\includegraphics[width=0.5\textwidth]{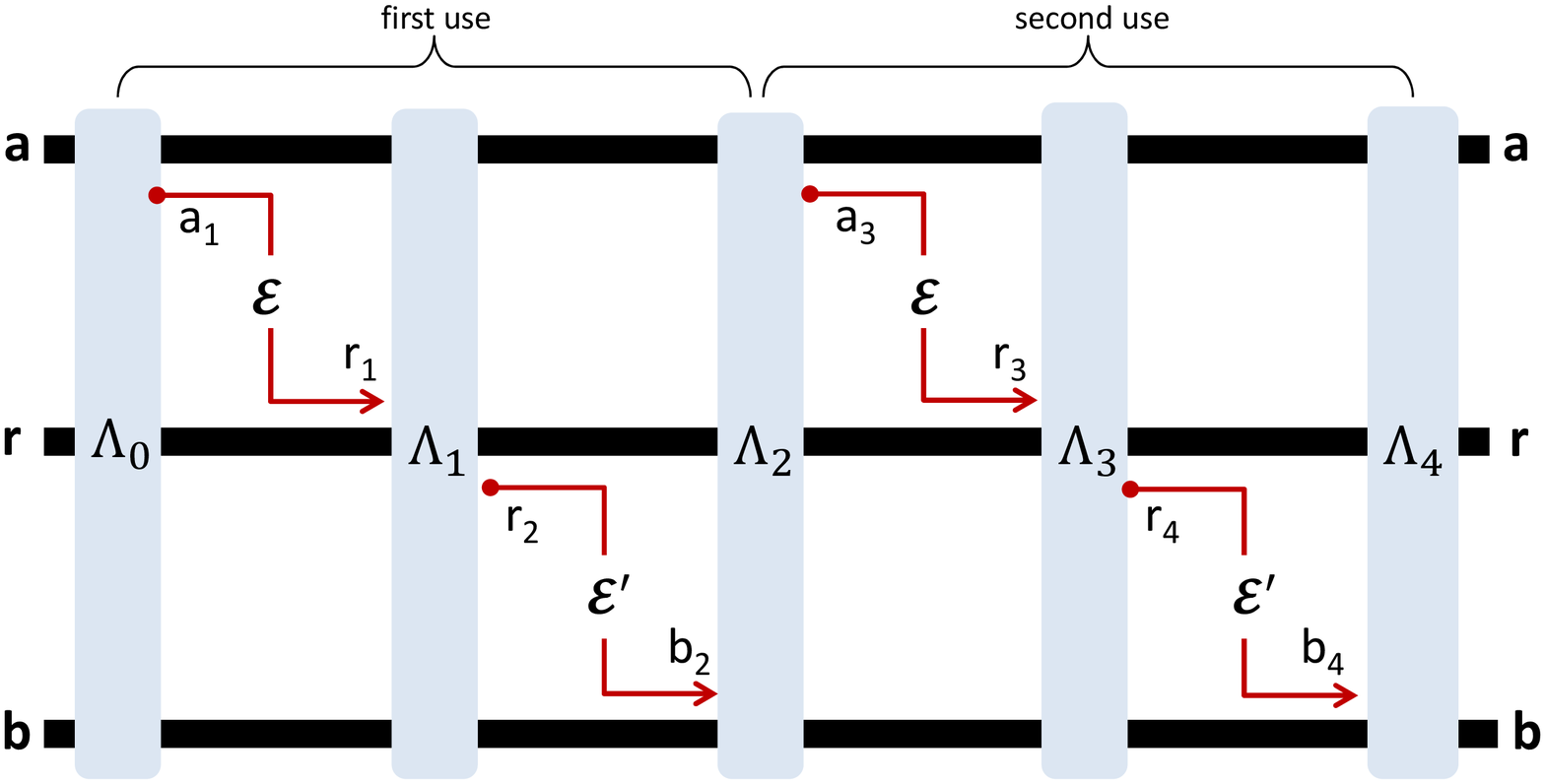} \vspace{-1.3cm}
\end{center}
\caption{Chain with a single repeater $\mathbf{r}$ and connected
by two forward channels $\mathcal{E}$ and $\mathcal{E}^{\prime}$.
Each transmission $k$ through one of the two channels occurs
between two instances of adaptive local operations and classical
communication (LOCCs) $\Lambda_{k-1}$ and $\Lambda_{k}$. In
particular, here we show two uses of the chain, with total output
state $\rho_{\mathbf{arb}}^{2}$. Note that, if the parties want to
distribute ebits or private bits, they may also use a different
order of transmissions in each use. For instance, in the first
use, the first transmission could be between the repeater
$\mathbf{r}$ and Bob $\mathbf{b}$, followed by that between Alice
$\mathbf{a}$ and the repeater $\mathbf{r}$. The order of the
transmissions and the sequence of LOCCs defines
the adaptive protocol $\mathcal{P}$ over the chain.}%
\label{repeaterSCHEME}%
\end{figure}

This procedure completes the first use of the chain. In the second
use, the initial state is the (non-separable) output state of the
first round
$\sigma_{\mathbf{ar}_{1}\cdots\mathbf{r}_{N}\mathbf{b}}^{2}=\rho
_{\mathbf{ar}_{1}\cdots\mathbf{r}_{N}\mathbf{b}}^{1}$. The
protocol goes as before with each pair of points $i$ and $i+1$
exchanging one system between
two LOCCs. The second use ends with the output state $\rho_{\mathbf{ar}%
_{1}\cdots\mathbf{r}_{N}\mathbf{b}}^{2}$ which is the input for
the third use and so on. After $n$ uses, the points share an
output state $\rho
_{\mathbf{ar}_{1}\cdots\mathbf{r}_{N}\mathbf{b}}^{n}$. By tracing
out the repeaters, we get Alice and Bob's final state
$\rho_{\mathbf{ab}}^{n}$, which depends on the sequence of LOCCs
$\mathcal{L}=\{\Lambda_{0},\cdots ,\Lambda_{n(N+1)}\}$. In
general, in each use of the chain, the order of the transmissions
can also be permuted. Both the order of these transmissions and
the sequence of LOCCs $\mathcal{L}$ defines the adaptive protocol
$\mathcal{P}_{\text{\textrm{chain}}}$ generating the output $\rho
_{\mathbf{ab}}^{n}$. See Fig.~\ref{repeaterSCHEME} for an example.

We say that an adaptive protocol
$\mathcal{P}_{\text{\textrm{chain}}}$ has rate
$R_{n}^{\varepsilon}$ if $\left\Vert \rho_{\mathbf{ab}}^{n}-\phi
_{n}\right\Vert _{1}\leq\varepsilon$, where $\phi_{n}$ is a target
state with $nR_{n}^{\varepsilon}$ bits. By taking the limit of
$n\rightarrow+\infty$,
$\varepsilon\rightarrow0$ (weak converse), and optimizing over $\mathcal{P}%
_{\text{\textrm{chain}}}$, we define the generic two-way capacity
of the chain, i.e.,
\begin{equation}
\mathcal{C}(\{\mathcal{E}_{i}\}):=\sup_{\mathcal{P}_{\text{\textrm{chain}}}%
}\lim_{\varepsilon,n}R_{n}^{\varepsilon}~.
\end{equation}
This capacity has different nature depending on the task of the
distribution protocol. For QKD, the target state is a private
state~\cite{KD2s} with secret key rate
$R_{n}^{\varepsilon,\mathrm{key}}$ (bits per chain use). In this
case $\mathcal{C}(\{\mathcal{E}_{i}\})$ is the secret key capacity
of the chain $K(\{\mathcal{E}_{i}\})$. Under two-way CCs, this is
also equal to the maximum rate at which Alice can
deterministically send a secret message to Bob through the chain,
i.e., its two-way private capacity $P_{2}(\{\mathcal{E}_{i}\})$.
For entanglement distribution (ED), the target state is a
maximally-entangled state with rate
$R_{n}^{\varepsilon,\mathrm{ED}}\leq R_{n}^{\varepsilon
,\mathrm{key}}$ (ebits per chain use). In this case, $\mathcal{C}%
(\{\mathcal{E}_{i}\})$ is an entanglement-distribution capacity $D_{2}%
(\{\mathcal{E}_{i}\})\leq K(\{\mathcal{E}_{i}\})$. Under two-way
CCs, $D_{2}$ is equal to the maximum rate at which Alice can
reliably send a qubits to Bob
through the chain, i.e., its two-way quantum capacity $Q_{2}(\{\mathcal{E}%
_{i}\})$.

We can build an upper bound for all the previous capacities, i.e.,
for the generic $\mathcal{C}(\{\mathcal{E}_{i}\})$. In fact, as
shown in the Methods section of our manuscript, we may write the
following weak converse bound in
terms of the relative entropy of entanglement (REE)%
\begin{equation}
\mathcal{C}(\{\mathcal{E}_{i}\})\leq E_{\mathrm{R}}^{\star}(\{\mathcal{E}%
_{i}\}):=\sup_{\mathcal{P}_{\text{\textrm{chain}}}}\underset{n}{\lim}%
~n^{-1}E_{\mathrm{R}}(\rho_{\mathbf{ab}}^{n}). \label{hard}%
\end{equation}
Recall that the REE is defined as~\cite{RMPrelents,VedFORMms,Plenioms}%
\begin{equation}
E_{\text{\textrm{R}}}(\sigma):=\inf_{\gamma\in\text{\textrm{SEP}}}%
~S(\sigma||\gamma), \label{REEdef}%
\end{equation}
where \textrm{SEP} is the set of separable bipartite states and
$S(\sigma ||\gamma):=\mathrm{Tr}\left[
\sigma(\log_{2}\sigma-\log_{2}\gamma)\right]  $ is the relative
entropy. In general, for an asymptotic state $\sigma
:=\lim_{\mu}\sigma^{\mu}$, we may extend the previous definition and consider%
\begin{equation}
E_{\text{\textrm{R}}}(\sigma):=\underset{\mu}{\lim\inf}~E_{\text{\textrm{R}}%
}(\sigma^{\mu})=\inf_{\gamma^{\mu}}~\underset{\mu}{\lim\inf}~S(\sigma^{\mu
}||\gamma^{\mu}), \label{REE_weaker}%
\end{equation}
where $\gamma^{\mu}$ is a converging sequence of separable
states~\cite{QKDpapers}, so that there is a separable $\gamma$
such that
$||\gamma^{\mu}-\gamma||_{1}\overset{\mu}{\rightarrow}0$. Both the
definitions in Eqs.~(\ref{REEdef}) and~(\ref{REE_weaker}) can be
regularized, so that we
have $E_{\text{\textrm{R}}}^{\infty}(\sigma)=\lim_{n}n^{-1}E_{\text{\textrm{R}%
}}(\sigma^{\otimes n})$.

In order to reduce the latter bound to a single-letter quantity we
simulate the chain, by replacing each channel $\mathcal{E}_{i}$
with a simulation $S_{i}=(\mathcal{T}_{i},\sigma_{i})$ for some
LOCC $\mathcal{T}_{i}$ and resource state $\sigma_{i}$. The next
step is to use teleportation stretching~\cite{QKDpapers} to
re-organize the adaptive protocol into a block version, where the
output state is expressed in terms of a tensor product of resource
states. A direct application of this procedure will allow us to
write
\begin{equation}
\rho_{\mathbf{ab}}^{n}=\bar{\Lambda}_{\mathbf{ab}}\left(  \otimes_{i=0}%
^{N}~\sigma_{i}^{\otimes n}\right)  ~, \label{UBnot}%
\end{equation}
for a trace-preserving LOCC $\bar{\Lambda}_{\mathbf{ab}}$ (this
reduction is proven afterwards). By using Eq.~(\ref{UBnot}), we
may then write
$E_{\mathrm{R}}(\rho_{\mathbf{ab}}^{n})\leq n\Pi_{i=0}^{N}E_{\mathrm{R}%
}(\sigma_{i})$, leading to the upper bound%
\begin{equation}
E_{\mathrm{R}}^{\star}(\{\mathcal{E}_{i}\})\leq\Pi_{i=0}^{N}E_{\mathrm{R}%
}(\sigma_{i})~.
\end{equation}

Unfortunately, this bound is too large. To improve it, we need to
perform cuts of the chain, such that Alice and Bob end up to be
disconnected. In a linear chain, the situation is particularly
simple, because any cut disconnects the two end-points. The
refined procedure consists of cutting channel $\mathcal{E}_{i}$,
stretching the protocol with respect to that channel and finally
minimizing over all cuts. Let us start with the formal definition
of cut of a chain.

\begin{definition}
[Cut of a chain]\label{defCUT}Consider a chain of $N$ repeaters $\{\mathbf{r}%
_{1},\ldots,\mathbf{r}_{N}\}$ connecting Alice $\mathbf{a}=\mathbf{r}_{0}%
$\ and Bob $\mathbf{b}=\mathbf{r}_{N+1}$\ by means of $N+1$
quantum channels
$\{\mathcal{E}_{0},\ldots,\mathcal{E}_{i},\ldots,\mathcal{E}_{N}\}$.
An entanglement cut \textquotedblleft$i$\textquotedblright\
disconnects channel $\mathcal{E}_{i}$ and induces a bipartition
$(\mathbf{A},\mathbf{B})$, where the set of points
$\mathbf{A}=\{\mathbf{r}_{0},\ldots,\mathbf{r}_{i}\}$ is
\textquotedblleft super-Alice\textquotedblright\ and $\mathbf{B}%
=\{\mathbf{r}_{i+1},\ldots,\mathbf{r}_{N}\}$ is \textquotedblleft
super-Bob\textquotedblright.
\end{definition}

By performing entanglement cuts in the chain, we may state the
following result which correctly extends teleportation stretching
to chains of quantum repeaters.

\begin{lemma}
[Chain stretching]\label{TheoCHAIN}Consider a chain of $N$
repeaters as in Definition~\ref{defCUT}. Given an arbitrary
entanglement cut $i$, consider the disconnected channel
$\mathcal{E}_{i}$\ and its simulation via a resource state
$\sigma_{i}$. For any such cut $i=0,\ldots,N$ the output of the
most general adaptive protocol
$\mathcal{P}_{\text{\textrm{chain}}}$ over $n$ uses of the chain
can be decomposed as
\begin{equation}
\rho_{\mathbf{ab}}^{n}=\bar{\Lambda}_{i}\left(
\sigma_{i}^{\otimes n}\right)
, \label{REPde}%
\end{equation}
where $\bar{\Lambda}_{i}$ is a trace-preserving LOCC. In
particular, for a chain of teleportation-covariant channels, we
may write Eq.~(\ref{REPde}) using the Choi-matrices
$\sigma_{\mathcal{E}_{i}}$ (with asymptotic formulations for
bosonic channels).
\end{lemma}

\begin{figure}[ptbh]
\vspace{0.0cm}
\par
\begin{center}
\includegraphics[width=0.5\textwidth] {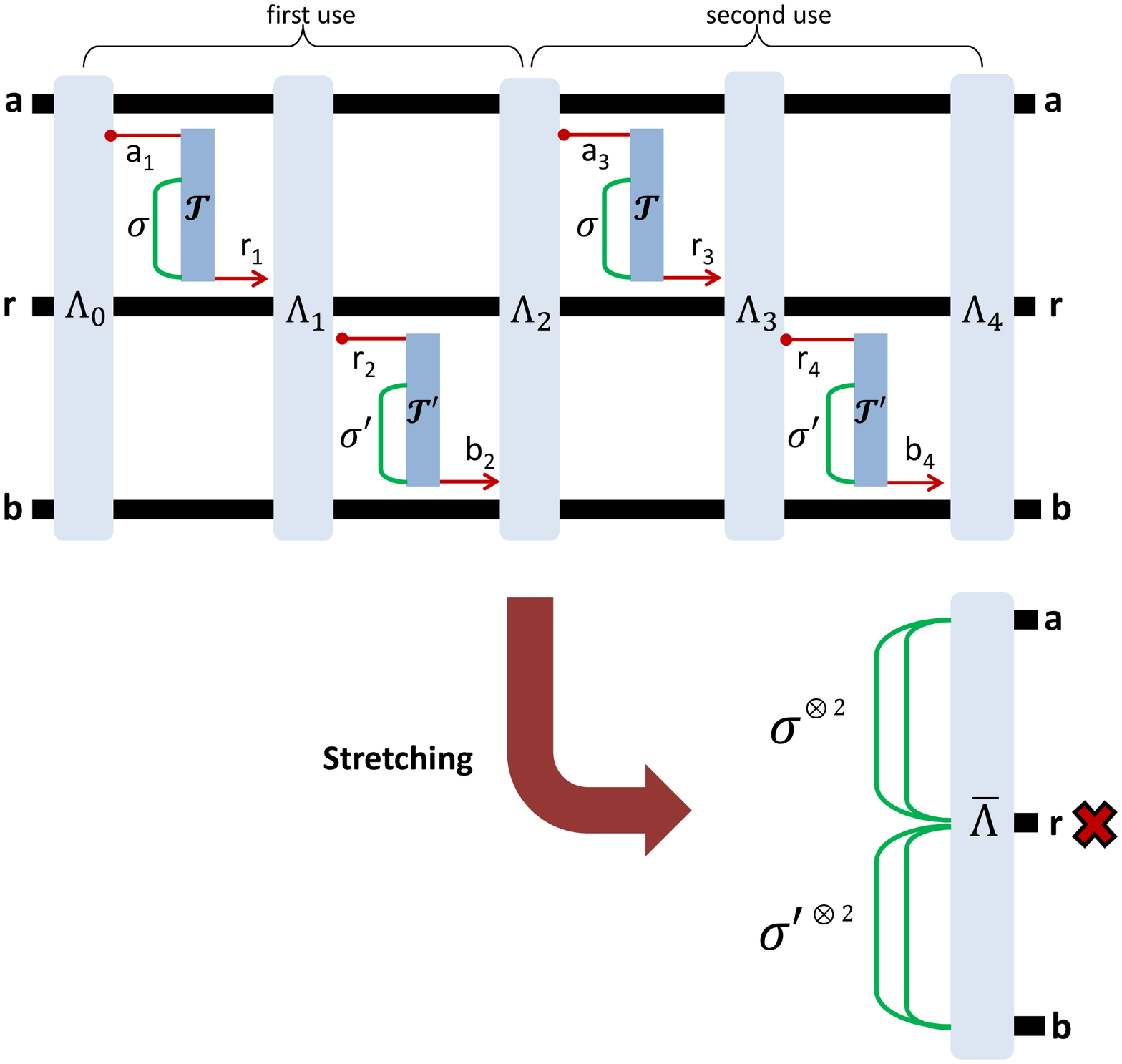}
\end{center}
\par
\vspace{-0.2cm}\caption{Teleportation stretching of a repeater.}%
\label{RePIC}
\end{figure}

\textbf{Proof.}~For simplicity let us start with the simple case
of a $3$-point chain ($N=1$), where Alice $\mathbf{a}$ and Bob
$\mathbf{b}$ are connected with a middle repeater $\mathbf{r}$ by
means of two channels $\mathcal{E}$ and $\mathcal{E}^{\prime}$ as
in Fig.~\ref{repeaterSCHEME} (the direction of the channels may be
different as well as the order in which they are used). Assume two
adaptive uses of the
chain ($n=2$) starting from a fundamental state $\rho_{\mathbf{a}}^{0}%
\otimes\rho_{\mathbf{r}}^{0}\otimes\rho_{\mathbf{b}}^{0}$. As
depicted in Fig.~\ref{RePIC}, we replace each channel with a
corresponding
simulation: $\mathcal{E}\rightarrow(\mathcal{T},\sigma)$ and $\mathcal{E}%
^{\prime}\rightarrow(\mathcal{T}^{\prime},\sigma^{\prime})$. Then,
the resource states are stretched back in time before the LOCCs
which are all collapsed into a single LOCC $\bar{\Lambda}$
(trace-preserving after averaging over all measurements). After
two uses of the repeater we have the output
state $\rho_{\mathbf{arb}}^{2}=\bar{\Lambda}\left(  \sigma^{\otimes2}%
\otimes\sigma^{\prime\otimes2}\right)  $. By tracing the repeater $\mathbf{r}%
$, we derive
$\rho_{\mathbf{ab}}^{2}=\bar{\Lambda}_{\mathbf{ab}}\left(
\sigma^{\otimes2}\otimes\sigma^{\prime\otimes2}\right)  $ up to
re-defining the LOCC. By extending the procedure to an arbitrary
number of repeaters $N$
and uses $n$, we get%
\begin{equation}
\rho_{\mathbf{ar}_{1}\ldots\mathbf{r}_{N}\mathbf{b}}^{n}=\bar{\Lambda}\left(
\otimes_{i=0}^{N}~\sigma_{i}^{\otimes n}\right)  , \label{decoPROOF}%
\end{equation}
and tracing out all the repeaters, we derive Eq.~(\ref{UBnot}).

Therefore, thanks to teleportation stretching, the quantum
transmissions between each pair of near-neighbor points have been
replaced with tensor-products of resource states, followed by a
single but complicated trace-preserving LOCC. In this reduction,
the resource states are responsible for distributing entanglement
between the points of the chain. In order to get tight upper
bounds we need to perform entanglement cuts.

\begin{figure}[ptbh]
\vspace{-1.2cm}
\par
\begin{center}
\includegraphics[width=0.5\textwidth] {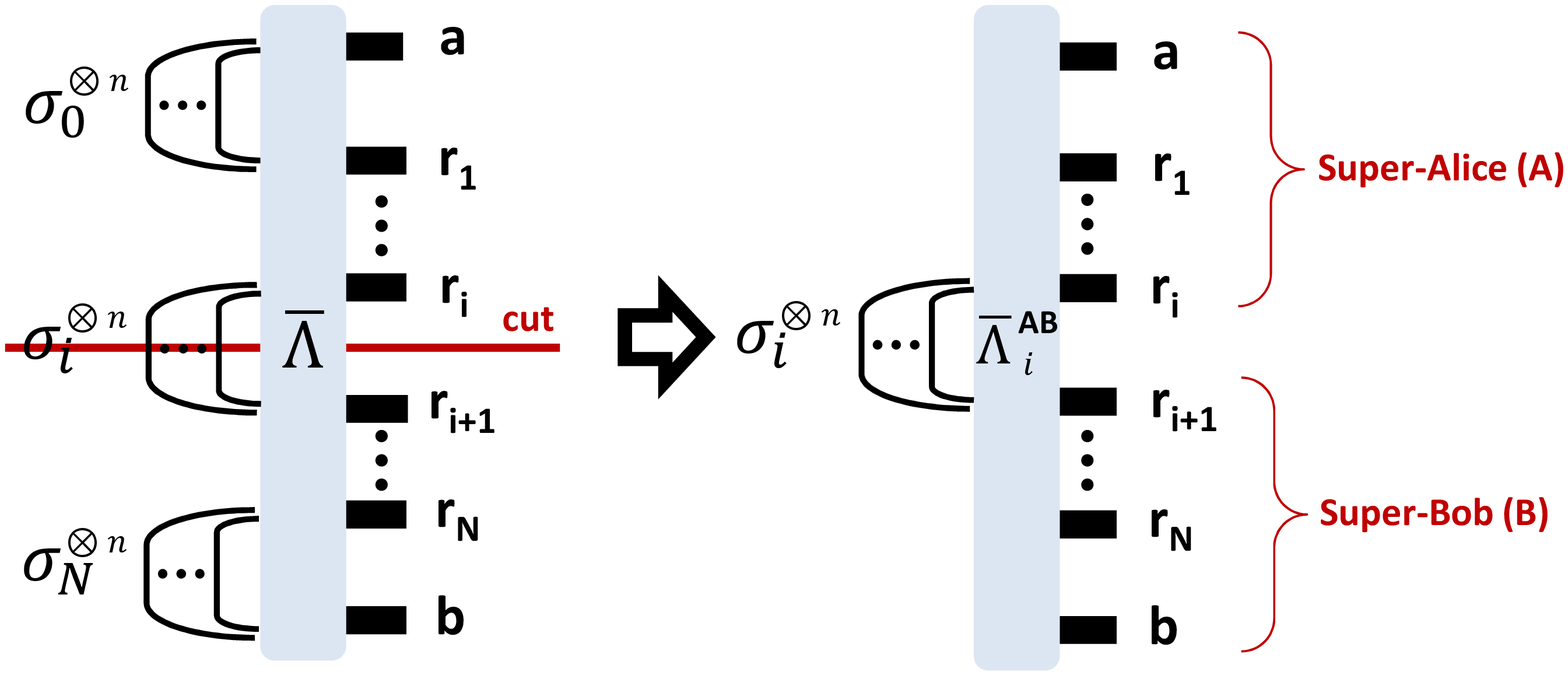}
\end{center}
\par
\vspace{-1.5cm}\caption{Reduction of the stretched scenario}%
\label{Reduction}%
\end{figure}

Let us perform a cut \textquotedblleft$i$\textquotedblright\ of
the chain, so that channel $\mathcal{E}_{i}$ is disconnected
between $\mathbf{r}_{i}$ and $\mathbf{r}_{i+1}$. This cut can be
done directly on the stretched chain as in Fig.~\ref{Reduction}.
This cut defines super-Alice $\mathbf{A}$ and super-Bob
$\mathbf{B}$. Now, let us include all the resource states
$\sigma_{k}^{\otimes n}$ with $k<i$ in the LOs of super-Alice, and
all the resource states with $k>i+1$ in the LOs of super-Bob. This
operation has two outcomes: (i) it defines a novel
trace-preserving LOCC $\bar{\Lambda}_{i}$ which is local with
respect to the super-parties; and (ii) it leaves with a reduced
number of resource states $\sigma_{i}^{\otimes n}$, i.e., only
those associated with the cut. For the super-parties, we may write
$\rho
_{\mathbf{AB}}^{n}=\bar{\Lambda}_{\mathbf{AB}}^{i}(\sigma_{i}^{\otimes
n})$.
By tracing out all the middle repeaters $\mathbf{r}_{1}\mathbf{r}_{2}%
\ldots\mathbf{r}_{N}$, the resulting LOCC $\bar{\Lambda}_{i}$
remains local with respect to $\mathbf{a}$ and $\mathbf{b}$, and
we get the end-to-end output $\rho_{\mathbf{ab}}^{n}$ as in
Eq.~(\ref{REPde}), for any cut $i$.

The extension of the proof to bosonic channels exploits asymptotic
simulations. For each channel $\mathcal{E}_{i}$ in the chain we
may consider
its approximation $\mathcal{E}_{i}^{\mu}$ with simulation $(\mathcal{T}%
_{i}^{\mu},\sigma_{i}^{\mu})$. This leads to the output state
$\rho
_{\mathbf{ab}}^{n,\mu}=\bar{\Lambda}_{i}^{\mu}(\sigma_{i}^{\mu\otimes
n})$ for a trace-preserving LOCC $\bar{\Lambda}_{i}^{\mu}$. Since
$\mathcal{E}_{i}$ is the point-wise limit of
$\mathcal{E}_{i}^{\mu}$ for large $\mu$, if we
consider the energy-constrained diamond distance $\varepsilon_{\bar{N}}%
^{i}:=\left\Vert \mathcal{E}_{i}-\mathcal{E}_{i}^{\mu}\right\Vert
_{\diamond\bar{N}}$, we have
$\varepsilon_{\bar{N}}^{i}\rightarrow0$ for any energy (mean
number of photons) $\bar{N}$ and cut $i$ (see Ref.~\cite[Eq.
(98)]{QKDpapers}\ or the Methods section of the main manuscript
for the definition of this distance). By directly extending a
\textquotedblleft peeling\textquotedblright\ argument given in
Ref.~\cite[Eq. (103)]{QKDpapers}, we easily show that the
trace-distance between the actual output $\rho_{\mathbf{ab}}^{n}$
and the simulated one $\rho_{\mathbf{ab}}^{n,\mu}$ is
controlled as follows%
\begin{equation}
\left\Vert
\rho_{\mathbf{ab}}^{n}-\rho_{\mathbf{ab}}^{n,\mu}\right\Vert
_{1}\leq n%
{\textstyle\sum\limits_{i=0}^{N}}
\left\Vert \mathcal{E}_{i}-\mathcal{E}_{i}^{\mu}\right\Vert _{\diamond\bar{N}%
}.
\end{equation}
Clearly, this distance goes to zero in $\mu$, for any number of
uses $n$, number of repeaters $N$ and energy $\bar{N}$. In other
words, given an
arbitrary cut $i$ we have%
\begin{equation}
\left\Vert \rho_{\mathbf{ab}}^{n}-\bar{\Lambda}_{i}^{\mu}(\sigma_{i}%
^{\mu\otimes n})\right\Vert _{1}\overset{\mu}{\rightarrow}0,\label{stretcvv}%
\end{equation}
or, more compactly,\
\begin{equation}
\rho_{\mathbf{ab}}^{n}=\lim_{\mu}\bar{\Lambda}_{i}^{\mu}(\sigma_{i}%
^{\mu\otimes n}),\label{stretcvvv}%
\end{equation}
for any number of uses $n$, repeaters $N$, and energy
$\bar{N}$.$~\blacksquare $

By using the previous lemma, we can now prove the following result
which establishes a single-letter REE\ upper bound for the generic
two-way capacity $\mathcal{C}(\{\mathcal{E}_{i}\})$ of a chain of
quantum repeaters. This is a bound for the maximal rates for
entanglement distribution ($D_{2}$), quantum communication
($Q_{2}$), secret key generation ($K$) and private communication
($P_{2}$) through the repeater chain. The formula simplifies for a
teleportation-covariant chain and even more for a distillable
chain, for which the repeater-assisted capacity is found to be the
minimum among the two-way capacities of the individual distillable
channels.

\begin{theorem}
[Single-letter REE bound]\label{singleLETTtheorem}Consider a chain
of $N$ repeaters as in Definition~\ref{defCUT}. The generic
two-way capacity of the
chain must satisfy the following minimization over the entanglement cuts%
\begin{equation}
\mathcal{C}(\{\mathcal{E}_{i}\})\leq\min_{i}E_{\mathrm{R}}(\sigma_{i}),
\label{TheoRE}%
\end{equation}
where $\sigma_{i}$\ is the resource state of an arbitrary LOCC
simulation of $\mathcal{E}_{i}$. For a chain of
teleportation-covariant channels (e.g. Pauli, Gaussian channels),
we may write the bound in terms of their Choi
matrices, i.e.,%
\begin{equation}
\mathcal{C}(\{\mathcal{E}_{i}\})\leq\min_{i}E_{\mathrm{R}}(\sigma
_{\mathcal{E}_{i}}), \label{TheoREtelecov}%
\end{equation}
where the REE\ is intended to be asymptotic for bosonic channels.
In particular, for a chain of distillable channels (e.g., lossy
channels, quantum-limited amplifiers, dephasing and erasure
channels), we establish the
capacity as%
\begin{equation}
\mathcal{C}(\{\mathcal{E}_{i}\})=\min_{i}E_{\mathrm{R}}(\sigma_{\mathcal{E}%
_{i}})=\min_{i}\mathcal{C}(\mathcal{E}_{i}), \label{reppAssCC}%
\end{equation}
where $\mathcal{C}(\mathcal{E}_{i})$ are the individual two-way
capacities associated with each distillable channel
$\mathcal{E}_{i}$ in the chain. In
this case, we also have $\mathcal{C}(\{\mathcal{E}_{i}\})=\min_{i}D_{1}%
(\sigma_{\mathcal{E}_{i}})$, so that the capacity may be achieved
by using one-way entanglement distillation followed by
entanglement swapping.
\end{theorem}

\textbf{Proof.}~For an arbitrary chain, perform the stretching of
the protocol for any entanglement cut $i$, so that we may write
Eq.~(\ref{REPde}). Because
the REE\ is non-decreasing under trace-preserving LOCCs, we get $E_{\mathrm{R}%
}(\rho_{\mathbf{ab}}^{n})\leq E_{\mathrm{R}}(\sigma_{i}^{\otimes
n})$. By replacing the latter inequality in the general weak
converse bound of
Eq.~(\ref{hard}), we may drop the supremum over the protocols $\mathcal{P}%
_{\text{\textrm{chain}}}$\ and derive the following bound in terms
of the
regularized REE\ of the resource state%
\begin{equation}
\mathcal{C}(\{\mathcal{E}_{i}\})\leq E_{\mathrm{R}}^{\infty}(\sigma_{i}%
):=\lim_{n}~n^{-1}E_{\mathrm{R}}(\sigma_{i}^{\otimes n}). \label{regull}%
\end{equation}
By minimizing over all the entanglement cuts, we get%
\begin{equation}
\mathcal{C}(\{\mathcal{E}_{i}\})\leq\min_{i}E_{\mathrm{R}}^{\infty}(\sigma
_{i})\leq\min_{i}E_{\mathrm{R}}(\sigma_{i}), \label{regull2}%
\end{equation}
where the last inequality is due to the subadditivity of the REE
over tensor-product states.

For teleportation-covariant channels, we may set
$\sigma_{i}=\sigma _{\mathcal{E}_{i}}$~\cite{QKDpapers}, so that
Eq.~(\ref{TheoREtelecov}) holds.
Then, for distillable channels, we have~\cite{QKDpapers} $\mathcal{C}%
(\mathcal{E}_{i})=E_{\mathrm{R}}(\sigma_{\mathcal{E}_{i}})=D_{1}%
(\sigma_{\mathcal{E}_{i}})$, so that
$\mathcal{C}(\{\mathcal{E}_{i}\})\leq
\min_{i}D_{1}(\sigma_{\mathcal{E}_{i}})$. It is clear that $\min_{i}%
D_{1}(\sigma_{\mathcal{E}_{i}})$ is also an achievable lower bound
so that it provides the capacity and we may also write
Eq.~(\ref{reppAssCC}). In fact, in the $i$th point-to-point
connection, points $\mathbf{r}_{i}$ and $\mathbf{r}_{i+1}$ may
distill $D_{1}(\sigma_{\mathcal{E}_{i}})$ ebits via one-way CCs.
After this is done in all the connections, sessions of
entanglement swapping will transfer at least $\min_{i}D_{1}(\sigma
_{\mathcal{E}_{i}})$ ebits to the end points.

To extend the result to bosonic channels with asymptotic
simulations, we adopt a weaker definition of REE as given in
Eq.~(\ref{REE_weaker}). Consider the asymptotic stretching of the
output state $\rho_{\mathbf{ab}}^{n}$\ as in Eq.~(\ref{stretcvv})
which holds for any number of uses $n$, repeaters $N$, and energy
$\bar{N}$. Then, for any cut $i$, the simplification of the REE
bound $E_{\mathrm{R}}(\rho_{\mathbf{ab}}^{n})$ goes as follows
\begin{align}
E_{\mathrm{R}}(\rho_{\mathbf{ab}}^{n})  &  =\inf_{\gamma\in\mathrm{SEP}}%
S(\rho_{\mathbf{ab}}^{n}||\gamma)\nonumber\\
&  \overset{(1)}{\leq}\inf_{\gamma^{\mu}}S\left[  \lim_{\mu}\bar{\Lambda}%
_{i}^{\mu}(\sigma_{i}^{\mu\otimes
n})~||~\lim_{\mu}\gamma^{\mu}\right]
\nonumber\\
&  \overset{(2)}{\leq}\inf_{\gamma^{\mu}}\underset{\mu\rightarrow+\infty}%
{\lim\inf}~S\left[  \bar{\Lambda}_{i}^{\mu}(\sigma_{i}^{\mu\otimes
n})~||~\gamma^{\mu}\right] \nonumber\\
&  \overset{(3)}{\leq}\inf_{\gamma^{\mu}}\underset{\mu\rightarrow+\infty}%
{\lim\inf}~S\left[  \bar{\Lambda}_{i}^{\mu}(\sigma_{i}^{\mu\otimes n}%
)~||~\bar{\Lambda}_{i}^{\mu}(\gamma^{\mu})\right] \nonumber\\
&  \overset{(4)}{\leq}\inf_{\gamma^{\mu}}\underset{\mu\rightarrow+\infty}%
{\lim\inf}~S\left(  \sigma_{i}^{\mu\otimes
n}~||~\gamma^{\mu}\right)
\nonumber\\
&  \overset{(5)}{=}E_{\mathrm{R}}(\sigma_{i}^{\otimes n})
\end{align}
where: (1)$~\gamma^{\mu}$ is a generic sequence of separable
states converging in trace norm, i.e., such that there is a
separable state $\gamma:=\lim_{\mu
}\gamma^{\mu}$ so that $\Vert\gamma-\gamma^{\mu}\Vert\overset{\mu}%
{\rightarrow}0$; (2)~we use the lower semi-continuity of the
relative
entropy~\cite{HolevoBOOKs}; (3)~we use that $\bar{\Lambda}_{i}^{\mu}%
(\gamma^{\mu})$ are specific types of converging separable
sequences within the set of all such sequences; (4)~we use the
monotonicity of the relative entropy under trace-preserving LOCCs;
and (5)~we use the regularized definition of REE for asymptotic
states.

For any energy $\bar{N}$, we may apply the general weak converse
bound of Eq.~(\ref{hard}), so that we may again write
Eq.~(\ref{regull}) in terms of the regularized REE
$E_{\mathrm{R}}^{\infty}(\sigma_{i})$. Since this upper
bound does no longer depend on the protocols $\mathcal{P}%
_{\text{\textrm{chain}}}$, it applies to both energy-constrained
and energy-unconstrained registers (i.e., we may relax the
constraint $\bar{N}$). The proof of the further condition
$E_{\mathrm{R}}^{\infty}(\sigma_{i})\leq
E_{\mathrm{R}}(\sigma_{i})$ is based on the subadditivity of the
REE over tensor product states, which holds for asymptotic states
too~\cite{QKDpapers}. Thus, the minimization over the cuts
provides again Eq.~(\ref{regull2}). The remaining steps of the
proof for teleportation and distillable channels are trivially
extended to asymptotic simulations. In particular, one can define
an asymptotic notion of one-way distillable entanglement $D_{1}$
for an unbounded Choi matrix as explained in
Ref.~\cite{QKDpapers}.~$\blacksquare$

\subsection{Capacities for distillable chains}

Let us specify our results for various types of distillable
chains. Let us start by considering a lossy chain, where Alice and
Bob are connected by $N$ repeaters and each connection
$\mathcal{E}_{i}$ is a lossy (pure-loss) channel with
transmissivity $\eta_{i}$. By combining Eq.~(\ref{reppAssCC}) of
Theorem~\ref{singleLETTtheorem} with the PLOB bound
$\mathcal{C}(\eta _{i})=-\log_{2}(1-\eta_{i})$~\cite{QKDpapers},
we find that the capacity of the
lossy chain is given by%
\begin{equation}
\mathcal{C}_{\text{loss}}(\{\eta_{i}\})=\min_{i}\mathcal{C}(\eta_{i}%
)=-\log_{2}(1-\eta_{\text{min}}), \label{LossyCHAIN}%
\end{equation}
where $\eta_{\text{min}}:=\min_{i}\eta_{i}$. Therefore, no matter
how many repeaters we use, the minimum transmissivity in the chain
fully determines the ultimate rate of quantum or private
communication between the two end-points.

Suppose that we require a minimum performance of $1$ bit per use
of the chain (this could be $1$ secret bit or $1$ ebit or $1$
qubit). From Eq.~(\ref{LossyCHAIN}), we see that we need to ensure
at least $\eta _{\text{min}}=1/2$, which means at most $3$dB of
loss in each link. This \textquotedblleft$3$dB
rule\textquotedblright\ implies that $1$ bit rate communication
can occur in chains whose maximum point-to-point distance is 15km
(assuming fiber connections at the loss rate of 0.2dB/km).

Consider now an amplifying chain, i.e., a chain connected by
quantum-limited amplifiers with gains $\{g_{i}\}$. Using
Eq.~(\ref{reppAssCC}) and
$\mathcal{C}(g_{i})=-\log_{2}(1-g_{i}^{-1})$~\cite{QKDpapers}, we
find that the repeater-assisted capacity is fully determined by
the highest gain $g_{\max }:=\max_{i}g_{i}$, so that
\begin{equation}
\mathcal{C}_{\text{amp}}(\{g_{i}\})=-\log_{2}\left(
1-g_{\max}^{-1}\right)  .
\end{equation}

In the DV setting, start with a spin chain where the state
transfer between the $i$th spin and the next one is modeled by a
dephasing channel with
probability $p_{i}\leq1/2$. Using Eq.~(\ref{reppAssCC}) and $\mathcal{C}%
(p_{i})=1-H_{2}(p_{i})$~\cite{QKDpapers}, we find the
repeater-assisted
capacity%
\begin{equation}
\mathcal{C}_{\text{deph}}(\{p_{i}\})=1-H_{2}(p_{\max}),
\end{equation}
where $p_{\max}:=\max_{i}p_{i}$ is the maximum probability of
phase flipping in the chain, and $H_{2}$ is the binary Shannon
entropy. When the spins are connected by erasure channels with
probabilities $\{p_{i}\}$, we combine Eq.~(\ref{reppAssCC}) and
$\mathcal{C}(p_{i})=1-p_{i}$~\cite{QKDpapers}. Therefore we derive
\begin{equation}
\mathcal{C}_{\text{erase}}(\{p_{i}\})=1-p_{\max},
\end{equation}
where $p_{\max}$ is the maximum probability of an erasure.

Note that the latter results for the spin chains can be readily
extended from qubits to qudits of arbitrary dimension $d$, by
using the corresponding two-way capacities proven in
Ref.~\cite{QKDpapers}. See Table~I of the main paper for a
schematic representations of these formulas. Finally, note that
Eq.~(\ref{reppAssCC}) of Theorem~\ref{singleLETTtheorem} may be
applied to hybrid distillable chains, where channels are
distillable but of different kind between each pair of repeaters,
e.g., we might have erasure channels alternated with dephasing
channels or lossy channels, etc.

\subsection{Quantum repeaters in optical communications\label{OptimalUSE}}

Let us discuss in more detail the use of quantum repeaters in the
bosonic setting. Suppose that we are given a long communication
line with transmissivity $\eta$, such as an optical/telecom fiber.
A cut of this line generates two lossy channels with
transmissivities $\eta^{\prime}$ and $\eta^{\prime\prime}$ such
that $\eta=\eta^{\prime}\eta^{\prime\prime}$. Suppose that we are
also given a number $N$ of repeaters that we could potentially
insert along the line. The question is: \textit{What is the
optimal way to cut the line and insert the repeaters?}

From the formula in Eq.~(\ref{LossyCHAIN}), we can immediately see
that the optimal solution is to insert $N$\ equidistant repeaters,
so that the resulting $N+1$ lossy channels have identical
transmissivities
\begin{equation}
\eta_{i}=\eta_{\text{min}}=\sqrt[N+1]{\eta}~.
\end{equation}
This leads to the maximum repeater-assisted capacity%
\begin{equation}
\mathcal{C}_{\text{loss}}(\eta,N)=-\log_{2}\left(
1-\sqrt[N+1]{\eta}\right)
~. \label{optLOSScap1}%
\end{equation}
This capacity has been plotted in Fig.~2 of the main text for
increasing number of repeaters $N$ as a function of the total loss
of the line, which is expressed in decibel (dB) by
$\eta_{\text{dB}}:=-10\log_{10}\eta$. In particular, we compare
the repeater-assisted capacity with the point-to-point benchmark,
i.e., the maximum performance achievable in the absence of
repeaters (PLOB bound~\cite{QKDpapers}).

Let us study two opposite regimes that we may call
repeater-dominant and loss-dominant. In the former, we fix the
total transmissivity $\eta$ of the
line and use many equidistant repeaters $N\gg1$. We then have%
\begin{equation}
\mathcal{C}_{\text{loss}}(\eta,N\gg1)\simeq\log_{2}N-\log_{2}\ln\frac{1}{\eta
}~,
\end{equation}
which means that the capacity scales logarithmically in the number
of repeaters, independently from the loss. In the second regime
(loss-dominant), we fix the number of repeaters $N$ and we
consider high loss $\eta\simeq0$, in such a way that each link of
the chain is very lossy, i.e., we may set
$\sqrt[N+1]{\eta}\simeq0$. We then find%
\begin{equation}
\mathcal{C}_{\text{loss}}(\eta\simeq0,N)\simeq\frac{\sqrt[N+1]{\eta}}{\ln
2}\simeq1.44~\sqrt[N+1]{\eta},
\end{equation}
which is also equal to $\sqrt[N+1]{\eta}$ nats per use. This is
the fundamental rate-loss scaling which affects long-distance
repeater-assisted quantum optical communications.

In the bosonic setting, it is interesting to compare the use of
quantum repeaters with the performance of a multi-band
communication, where Alice and Bob can exploit a communication
line which is composed of $M$\ parallel and independent lossy
channels with identical transmissivity $\eta$. For instance, $M$
can be interpreted as the frequency bandwidth of a multimode
optical fiber. The capacity of a multiband lossy channel is given
by~\cite{QKDpapers}
\begin{equation}
\mathcal{C}_{\text{loss}}(\eta,M)=-M\log_{2}(1-\eta). \label{Mband}%
\end{equation}

\begin{figure}[ptbh]
\vspace{0.11cm}
\par
\begin{center}
\includegraphics[width=0.45\textwidth] {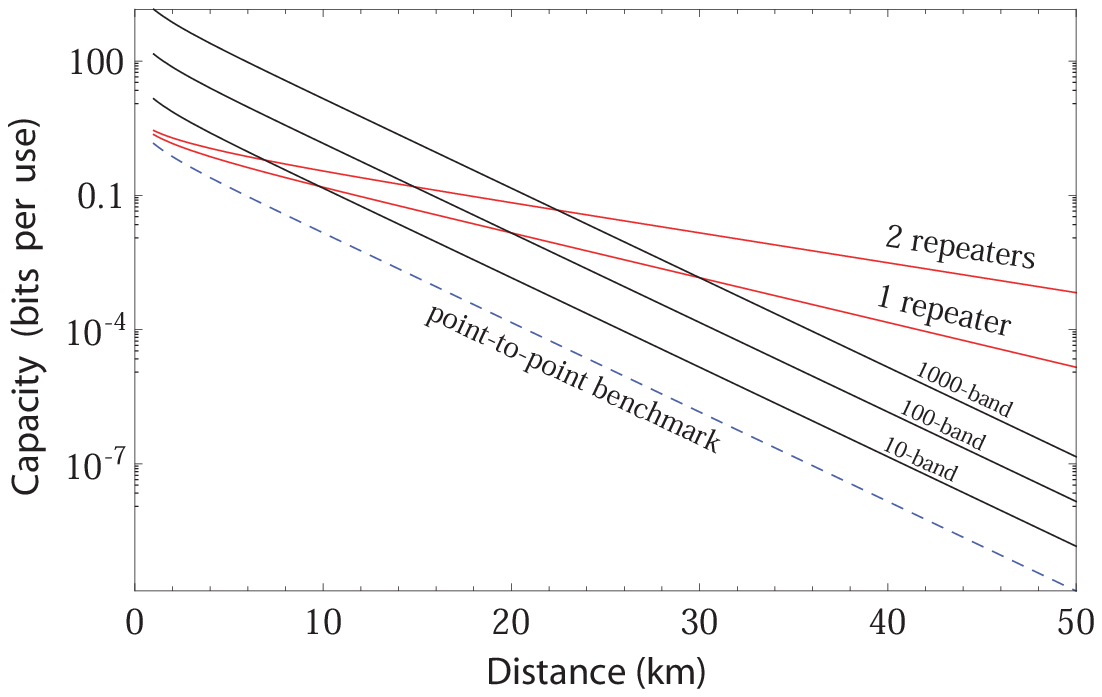} \vspace{2cm}
\vspace{-1.4cm} \vspace{1cm}
\end{center}
\par
\vspace{-2.0cm}\caption{Capacity (bits per use) versus distance
(km) assuming the standard loss rate of $0.2$ dB/km. We compare
the use of repeaters ($N=1,2$) with that of a point-to-point
multiband communication (for $M=10$, $100$, and $1000$ bands or
parallel channels). Dashed line is the point-to-point benchmark
(single-band, no repeaters). We see how the multiband strategy
increases the capacity in an additive way but it clearly suffers
from a poor long-distance rate-loss scaling with respect to the
use of quantum repeaters.} \label{compara}
\end{figure}

Using Eqs.~(\ref{optLOSScap1}) and~(\ref{Mband}) we may compare
the use of $N$ equidistant repeaters with the use of $M$ bands. In
Fig.~\ref{compara}, we clearly see that multiband quantum
communication provides an additive effect on the capacity which is
very useful at short-intermediate distances. However, at long
distances, this solution is clearly limited by the same rate-loss
scaling which affects the single-band quantum channel
(point-to-point benchmark) and, therefore, it cannot compete with
the long-distance performance of repeater-assisted quantum
communication.

\subsection{Multiband repeater chains}

In general, the most powerful approach consists of relaying
multiband quantum communication, i.e., combining multiband
channels with quantum repeaters. In this regard, let us first
discuss how Theorem~\ref{singleLETTtheorem} can be easily extended
to repeater chains which are connected by multiband quantum
channels. Then, we describe the performances in the bosonic
setting.

Consider a multiband channel $\mathcal{E}^{\text{band}}$ which is
composed of $M$ independent channels (or bands) $\mathcal{E}_{k}$,
i.e.,
\begin{equation}
\mathcal{E}^{\text{band}}=%
{\textstyle\bigotimes\nolimits_{k=1}^{M}}
\mathcal{E}_{k}~. \label{multibandGEN}%
\end{equation}
Assume that each band $\mathcal{E}_{k}$ can be LOCC-simulated with
some resource state $\sigma_{k}$. From Ref.~\cite{QKDpapers} and
the subadditivity of the REE, we may write the following bound for
its two-way capacity
\begin{align}
\mathcal{C}(\mathcal{E}^{\text{band}})  &  \leq
E_{\mathrm{R}}\left(
{\textstyle\bigotimes\nolimits_{k=1}^{M}}
\sigma_{k}\right) \nonumber\\
&  \leq%
{\textstyle\sum_{k=1}^{M}}
E_{\mathrm{R}}(\sigma_{k}):=\Psi(\mathcal{E}^{\text{band}}).
\end{align}
A multiband channel $\mathcal{E}^{\text{band}}$ is said\ to be
teleportation-covariant (distillable) if all its components
$\mathcal{E}_{k}$
are teleportation-covariant (distillable). In a distillable $\mathcal{E}%
^{\text{band}}$, for each band $\mathcal{E}_{k}$ we may write $\mathcal{C}%
(\mathcal{E}_{k})=D_{1}(\sigma_{\mathcal{E}_{k}})=E_{\mathrm{R}}%
(\sigma_{\mathcal{E}_{k}})$ where $\sigma_{\mathcal{E}_{k}}$ is
its Choi matrix (with suitable asymptotic description in the
bosonic case). Then, it is
straightforward to prove that~\cite{QKDpapers}%

\begin{equation}
\mathcal{C}(\mathcal{E}^{\text{band}})=%
{\textstyle\sum_{k=1}^{M}}
\mathcal{C}(\mathcal{E}_{k}).
\end{equation}

Similarly, we can extend Theorem~\ref{singleLETTtheorem}. Consider
an adaptive
protocol over a repeater chain connected by multiband channels $\{\mathcal{E}%
_{i}^{\text{band}}\}$. We can define a corresponding two-way
capacity for the multiband chain
$\mathcal{C}(\{\mathcal{E}_{i}^{\text{band}}\})$ and derive
the upper bound%
\begin{equation}
\mathcal{C}(\{\mathcal{E}_{i}^{\text{band}}\})\leq\min_{i}\Psi(\mathcal{E}%
_{i}^{\text{band}})~. \label{band1}%
\end{equation}
For a distillable multiband chain, we then have%
\begin{equation}
\mathcal{C}(\{\mathcal{E}_{i}^{\text{band}}\})=\min_{i}\mathcal{C}%
(\mathcal{E}_{i}^{\text{band}}). \label{bandDISTILL11}%
\end{equation}

In the bosonic setting, consider a chain of $N$ quantum repeaters
with $N+1$ channels $\{\mathcal{E}_{i}\}$, where $\mathcal{E}_{i}$
is a multiband lossy channel with $M_{i}$ bands and constant
transmissivity $\eta_{i}$ (over the bands). The two-way capacity
of the $i$th link is therefore given by
$\mathcal{C}_{\text{loss}}(\eta_{i},M_{i})$ as specified by
Eq.~(\ref{Mband}). Because multiband lossy channels are
distillable, we can apply Eq.~(\ref{bandDISTILL11}) and derive the
following repeater-assisted capacity
of the multiband lossy chain%
\begin{align}
\mathcal{C}_{\text{loss}}(\{\eta_{i},M_{i}\})  &  =\min_{i}\mathcal{C}%
_{\text{loss}}(\eta_{i},M_{i})\nonumber\\
&  =\min_{i}\left[  -M_{i}\log_{2}(1-\eta_{i})\right] \nonumber\\
&  =-\log_{2}\left[  \max_{i}(1-\eta_{i})^{M_{i}}\right] \nonumber\\
&  :=-\log_{2}\theta_{\max}~. \label{preBANDR}%
\end{align}

As before, it is interesting to discuss the symmetric scenario
where the $N$ repeaters are equidistant, so that entire
communication line is split into
$N+1$ links of the same optical length. Each link \textquotedblleft%
$i$\textquotedblright\ is therefore associated with a multiband
lossy channel, with bandwidth $M_{i}$ and constant transmissivity
$\eta_{i}=\sqrt[N+1]{\eta} $ (equal for all its bands). In this
case, we have $\theta_{\max
}=(1-\sqrt[N+1]{\eta})^{\min_{i}M_{i}}$ in previous
Eq.~(\ref{preBANDR}). In
other words, the repeater-assisted capacity of the chain becomes%
\[
\mathcal{C}_{\text{loss}}(\eta,N,\{M_{i}\})=-M_{\min}\log_{2}(1-\sqrt[N+1]%
{\eta}),
\]
where $M_{\min}:=\min_{i}M_{i}$ is the minimum bandwidth along the
line, as intuitively expected.

In general, the capacity is determined by an interplay between
transmissivity and bandwidth of each link. This is particularly
evident in the regime of high
loss. By setting $\eta_{i}\simeq0$ in Eq.~(\ref{preBANDR}), we in fact derive%
\begin{equation}
\mathcal{C}_{\text{loss}}(\{\eta_{i}\simeq0,M_{i}\})\simeq
c~\min_{i}\left( M_{i}\eta_{i}\right)  ,
\end{equation}
where the constant $c$ is equal to $1.44$ bits or $1$ nat.

\section{Supplementary Note 2: Quantum networks\label{SECnetworks}}

We now consider the general case of a quantum network, where two
end-users are connected by an arbitrary ensemble of routes through
intermediate points or repeaters. Our analysis combines tools from
quantum information theory (in particular, the generalization of
the tools developed in Ref.~\cite{QKDpapers}, needed for the
converse part) and elements from classical network information
theory (necessary for the achievability part). In this section, we
start by introducing the main adaptive protocols based on
sequential (single-path) or parallel (multi-path) routing of
quantum systems. We also give the corresponding definitions of
network capacities. Then, in Supplementary Note~3, we will show
how to simulate and \textquotedblleft stretch\textquotedblright\
quantum networks, so that the output of an adaptive protocol is
completely simplified into a decomposition of tensor-product
states. This tool will be exploited to derive single-letter REE
upper bounds in the subsequent sections. In particular, in
Supplementary Note~4 we\ will present the results for single-path
routing, while in Supplementary Note~5 we will present results for
multi-path routing. The upper bounds will be combined with
suitable lower bounds, and exact formulas will be established for
quantum networks connected by distillable channels.

\subsection{Notation and general definitions}

Consider a quantum communication network $\mathcal{N}$ whose
points are connected by memoryless quantum channels. The quantum
network can be represented as an undirected finite
graph~\cite{Slepians} $\mathcal{N}=(P,E)$ where $P$ is the finite
set of points of the network, also known as vertices, and $E$ is
the set of all connections, also known as edges (without loss of
generality, the graph may be considered to be acyclic). Every
point $x\in P$ has a local register of quantum systems
$\mathbf{x}$ to be used for the quantum communication. To simplify
notation, we identify a point with its local register
$x=\mathbf{x}$. Two points $\mathbf{x},\mathbf{y}\in P$ are
connected by an undirected edge $(\mathbf{x},\mathbf{y})\in E$ if
there is a
memoryless quantum channel $\mathcal{E}_{\mathbf{xy}}$ between $\mathbf{x}%
$\ and $\mathbf{y}$, which may be forward
$\mathcal{E}_{\mathbf{x\rightarrow y}}$\ or backward
$\mathcal{E}_{\mathbf{y}\rightarrow\mathbf{x}}$.

In general, there may be multiple edges between two points, with
each edge representing an independent quantum channel. For
instance, two undirected edges between $\mathbf{x}$\ and
$\mathbf{y}$ represent two channels
$\mathcal{E}_{\mathbf{xy}}\otimes\mathcal{E}_{\mathbf{xy}}^{\prime}$\
and these may be associated with a double-band quantum
communication (in one of the two directions) or a two-way quantum
communication (forward and backward channels). While we allow for
the possibility of multiple edges in the graph (so that it is more
generally a multi-graph) we may also collapse multiple edges into
a single edges to simplify the complexity of the network and
therefore notation.

In the following, we also use the labeled notation
$\mathbf{p}_{i}$ for the generic point of the graphical network,
so that two points $\mathbf{p}_{i}$ and $\mathbf{p}_{j}$ are
connected by an edge if there is a quantum channel
$\mathcal{E}_{ij}:=\mathcal{E}_{\mathbf{p}_{i}\mathbf{p}_{j}}$. We
also adopt the specific notation $\mathbf{a}$ and $\mathbf{b}$ for
the two end-points, Alice and Bob. An end-to-end route is an
undirected path between Alice and
Bob, which is specified by a sequence of edges $\{(\mathbf{a},\mathbf{p}%
_{i}),\cdots,(\mathbf{p}_{j},\mathbf{b})\}$, simply denoted as $\mathbf{a}%
-\mathbf{p}_{i}-\cdots-\mathbf{p}_{j}-\mathbf{b}$. This may be
interpreted as a linear chain of $N$ repeaters between Alice and
Bob, connected by a sequence
of $N+1$ channels $\{\mathcal{E}_{k}\}$, i.e.,%
\begin{equation}
\mathbf{a}\overset{\mathcal{E}_{0}}{-}(\mathbf{p}_{i}:=\mathbf{r}_{1}%
)-\cdots\overset{\mathcal{E}_{k}}{-}\cdots-(\mathbf{p}_{j}:=\mathbf{r}%
_{N})\overset{\mathcal{E}_{N}}{-}\mathbf{b},
\end{equation}
where the same repeater may appear at different positions (in
particular, this occurs when the route is not a simple path, so
that there are cycles).

In general, the two end-points may transmit quantum systems
through an ensemble of routes $\Omega=\{1,\ldots,\omega,\ldots\}$.
Note that this ensemble is generally large but can always be made
finite in a finite network, by just reducing the routes to be
simple paths, void of cycles (without losing generality).
Different routes $\omega$ and $\omega^{\prime}$ may have
collisions, i.e., repeaters and channels in common. Generic route
$\omega$
involves the transmission through $N_{\omega}+1$ channels $\{\mathcal{E}%
_{0}^{\omega},\ldots,\mathcal{E}_{k}^{\omega},\ldots,\mathcal{E}_{N_{\omega}%
}^{\omega}\}$. In general, we assume that each quantum
transmission through each channel is alternated with network
LOCCs: These are defined as adaptive LOs performed by all points
of the network on their local registers, which are assisted by
unlimited two-way CC involving the entire network.

Finally, we consider two possible fundamental strategies for
routing the systems through the network: Sequential or parallel.
In a sequential or single-path routing, quantum systems are
transmitted from Alice to Bob through a single route for each use
of the network. This process is generally stochastic, so that
route $\omega$ is chosen with some probability $p_{\omega }$. By
contrast, in a parallel or multi-path routing, systems are
simultaneously transmitted through multiple routes for each use of
the network. This may be seen as a \textquotedblleft broadband
use\textquotedblright\ of the quantum network. We now explain
these two strategies in detail.

\subsection{Sequential (single-path) routing}

The most general network protocol for sequential quantum
communication involves the use of generally-different routes,
accessed one after the other. The network is initialized by means
of a first LOCC $\Lambda_{0}$ which prepares an initial separable
state. With probability $\pi_{0}^{1}$, Alice $\mathbf{a}$
exchanges one system with repeater $\mathbf{p}_{i}$. This is
followed by another LOCC $\Lambda_{1}$. Next, with probability
$\pi_{1}^{1}$, repeater $\mathbf{p}_{i}$ exchanges one system with
repeater $\mathbf{p}_{j}$ and so on. Finally, with probability
$\pi_{N_{1}}^{1}$, repeater $\mathbf{p}_{k}$ exchanges one system
with Bob $\mathbf{b}$, followed by a final LOCC
$\Lambda_{N_{1}+1}$. Thus, with probability $p_{1}=\Pi_{i}\pi
_{i}^{1}$, the end-points exchange one system which has undergone
$N_{1}+1$ transmissions $\{\mathcal{E}_{i}^{1}\}$ along the first
route. Let us remark that the various probabilities $\pi_{i}^{1}$
are more precisely conditional probabilities, so that each
repeater generally updates its probability distribution on the
basis of the previous steps and the CCs received from all the
other repeaters.

The next uses may involve different routes. After many uses $n$,
the random
process defines a sequential routing table $\mathcal{R}=\{\omega,p_{\omega}%
\}$, where route $\omega$ is picked with probability $p_{\omega}$
and involves $N_{\omega}+1$ transmissions
$\{\mathcal{E}_{i}^{\omega}\}$. Thus, we have a total of
$N_{\text{tot}}=\Sigma_{\omega}np_{\omega}(N_{\omega}+1)$
transmissions and a sequence of LOCCs $\mathcal{L}=\{\Lambda_{0}%
,\ldots,\Lambda_{N_{\text{tot}}}\}$, whose output provides Alice
and Bob's final state $\rho_{\mathbf{ab}}^{n}$. Note that we may
weaken the previous description: While maintaining the sequential
use of the routes, in each route we may permute the order of the
transmissions (as before for the case of a linear chain of
repeaters).

The sequential network protocol
$\mathcal{P}_{\text{\textrm{seq}}}$ is characterized by
$\mathcal{R}$ and $\mathcal{L}$, and its average rate is
$R_{n}^{\varepsilon}$ if $\left\Vert \rho_{\mathbf{ab}}^{n}-\phi
_{n}\right\Vert _{1}\leq\varepsilon$, where $\phi_{n}$ is a target
state of $nR_{n}^{\varepsilon}$ bits. By taking the asymptotic
rate for large $n$, small $\varepsilon$ (weak converse), and
optimizing over all the sequential
protocols, we define the sequential or single-path capacity of the network%
\begin{equation}
\mathcal{C}(\mathcal{N}):=\sup_{\mathcal{P}_{\text{\textrm{seq}}}}%
\lim_{\varepsilon,n}R_{n}^{\varepsilon}. \label{necCAPdef}%
\end{equation}
The capacity $\mathcal{C}(\mathcal{N})$ provides the maximum
number of (quantum, entanglement, or secret) bits which are
distributed per sequential use of the network or single-path
transmission. In particular, by specifying the target state, we
define the corresponding network capacities for quantum
communication, entanglement distillation, key generation and
private communication, which satisfy
\begin{equation}
Q_{2}(\mathcal{N})=D_{2}(\mathcal{N})\leq
K(\mathcal{N})=P_{2}(\mathcal{N}).
\end{equation}

\begin{figure}[ptbh]
\vspace{-1.5cm}
\par
\begin{center}
\includegraphics[width=0.5\textwidth] {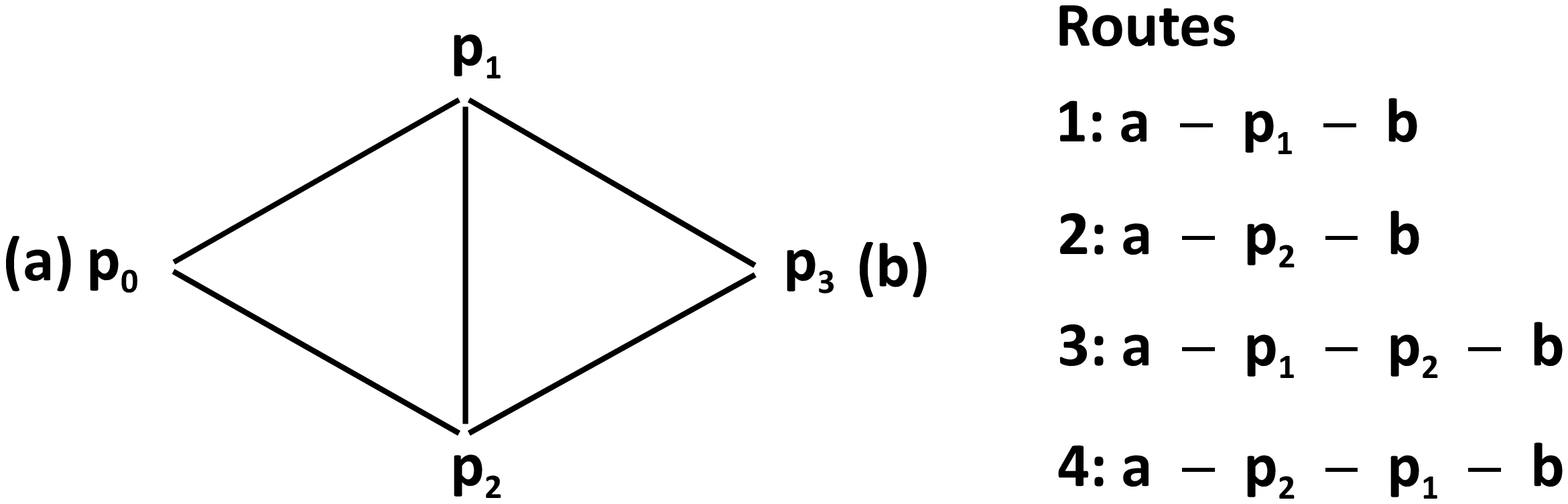}
\end{center}
\par
\vspace{-2.0cm}\caption{Sequential use of a diamond quantum
network. Each use of the network corresponds to routing a quantum
system between the two end-points Alice $\mathbf{a}$ and Bob
$\mathbf{b}$. In a diamond network with four points
$\mathbf{p}_{0}=\mathbf{a}$, $\mathbf{p}_{1}$, $\mathbf{p}_{2}$,
and $\mathbf{p}_{3}=\mathbf{b}$, we may identify four basic routes
$\omega=1,2,3,4$ (see list on the right). These are simple paths
between Alice and Bob with the middle points $\mathbf{p}_{1}$ and
$\mathbf{p}_{2}$ acting as quantum repeaters in different
succession. For instance, $\mathbf{p}_{1}$ is the first repeater
in route $3$ and the second repeater in route $4$. Note
that we may consider further routes by including loops between $\mathbf{p}%
_{1}$ and $\mathbf{p}_{2}$. These other solutions are non-simple
paths that we may discard without losing generality.} \label{NET1}
\end{figure}

It is important to note that the sequential use is the best
practical strategy when Alice and the other points of the network
aim to optimize the use of their quantum resources.\ In fact,
$\mathcal{C}(\mathcal{N})$ can also be expressed as maximum number
of target bits per quantum system routed. Assuming that the points
have deterministic control on the routing, they can adaptively
select the best routes based on the CCs received by the other
repeaters. Under such hypothesis, they can optimize the protocol
on the fly and adapt the routing table so that it converges to the
use of an optimal route $\omega_{\ast}$. See Fig.~\ref{NET1} for
an example of sequential use of a simple network.

\subsection{Parallel (multi-path) routing\label{BroadNET}}

Here we consider a different situation where Alice, Bob and the
other points of the network do not have restrictions or costs
associated with the use of their quantum resources, so that they
can optimize the use of the quantum network without worrying if
some of their quantum systems are inefficiently transmitted or
even lost (this may be the practical scenario of many optical
implementations, e.g., based on cheap resources like coherent
states). In such a case, the optimal use of the quantum network is
parallel or broadband, meaning that the quantum systems are
simultaneously routed through multiple paths each time the quantum
network is accessed.

In a parallel network protocol, Alice sends quantum systems to all
repeaters she has a connection with. Such a simultaneous
transmission to her \textquotedblleft neighbor\textquotedblright\
repeaters can be denoted by
$\mathbf{a}\rightarrow\{\mathbf{p}_{k}\}$ and may be called
``multipoint (quantum) communication''. In turn, each of the
receiving repeaters sends quantum systems to another set of
neighbor repeaters $\mathbf{p}_{k}\rightarrow\{\mathbf{p}_{j}\}$
and so on, until Bob $\mathbf{b}$ is reached as an end-point. This
is done in such a way that each multipoint communication occurs
between two network LOCCs, and different multipoint communications
do not overlap, so that all edges of the network are used exactly
once at the end of each end-to-end transmission. This condition is
assured by imposing that new multpoint communications may only
involve unused edges, a strategy commonly known as
\textquotedblleft flooding\textquotedblright ~\cite{floodings}.

In general, each multipoint communication must be intended in a
weaker sense as a point-to-multipoint \textit{connection} where
quantum systems may be exchanged through forward or backward
transmissions, following different physical directions of the
available quantum channels. Independently from these physical
directions, we may always assign a common sender-receiver
direction to all the edges involved in the process, so that there
will be a \textit{logical} sender-receiver orientation associated
with the multipoint communication. For this reason, the notation
$\mathbf{a}\rightarrow\{\mathbf{p}_{k}\}$ must be generally
interpreted as a process where Alice \textquotedblleft connects
to\textquotedblright\ repeaters $\{\mathbf{p}_{k}\}$. As a result
of these multiple connection, Alice may share ebits or secret bits
with each of the receivers, or she may teleport qubits to each of
them (independently from the actual physical direction of the
quantum channels).

\begin{figure}[ptbh]
\vspace{-1.2cm}
\par
\begin{center}
\includegraphics[width=0.5\textwidth] {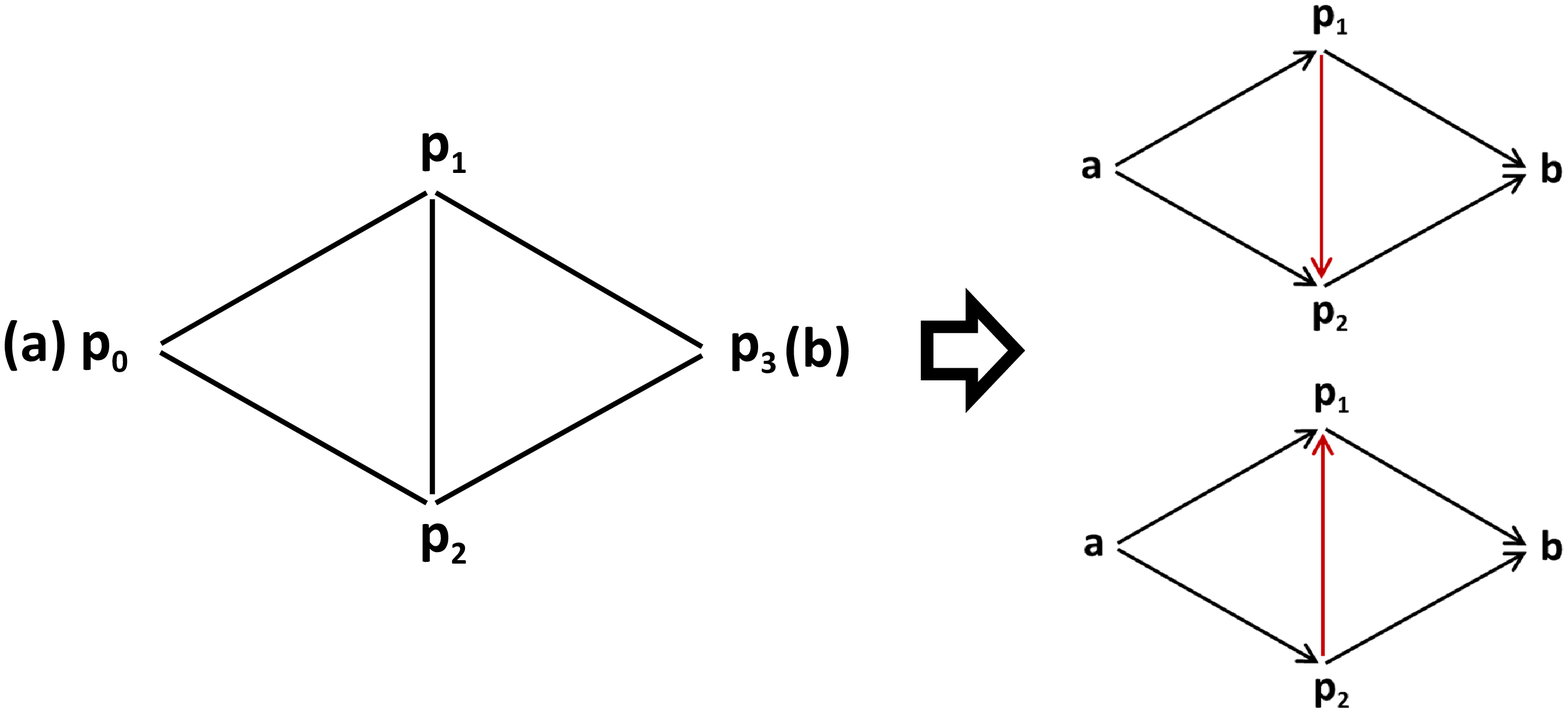}
\end{center}
\par
\vspace{-1.2cm}\caption{Orientations of a diamond quantum
network.\ There are only two possible orientations that transform
the original undirected network (left) into a flow network
(right). Within an orientation, there is a well-defined logical
multipoint communication from each point to all its
out-neighborhood (empty for Bob). A multi-path routing strategy
(flooding) is defined as a sequence of such multipoint
communications. Therefore, in the upper orientation, we may
identify the basic multi-path routing $\mathbf{a}\rightarrow
\{\mathbf{p}_{1},\mathbf{p}_{2}\}$, $\mathbf{p}_{1}\rightarrow\{\mathbf{p}%
_{2},\mathbf{b}\}$, and $\mathbf{p}_{2}\rightarrow\mathbf{b}$.
Other routings are given by permutation in the sequence. For
instance, we may have the different sequence
$\mathbf{p}_{1}\rightarrow\{\mathbf{p}_{2},\mathbf{b}\}$,
$\mathbf{p}_{2}\rightarrow\mathbf{b}$ and $\mathbf{a}\rightarrow
\{\mathbf{p}_{1},\mathbf{p}_{2}\}$ for the upper orientation. In
the lower orientation, we have the basic multi-path routing
$\mathbf{a}\rightarrow
\{\mathbf{p}_{1},\mathbf{p}_{2}\}$, $\mathbf{p}_{2}\rightarrow\{\mathbf{p}%
_{1},\mathbf{b}\}$ and $\mathbf{p}_{1}\rightarrow\mathbf{b}$, plus
all the possible permutations.} \label{NET2b} \end{figure}

To better explain this broadband use, let us formalize the notion
of orientation. Recall that a directed edge is an ordered pair
$(\mathbf{x},\mathbf{y})$, where the initial vertex $\mathbf{x}$
is called \textquotedblleft tail\textquotedblright\ and the
terminal vertex $\mathbf{y}$ is called \textquotedblleft
head\textquotedblright. Let us transform the undirected graph of
the network $\mathcal{N}=(P,E)$ into a directed graph by randomly
choosing a direction for all the edges, while keeping Alice as
tail and Bob as head. The goal is to represent the quantum network
as a flow network where Alice is the\textit{\ source} and Bob is
the\textit{\ sink}~\cite{Dinics,Karps}. In general, there are many
solutions for this random orientation. In fact, consider the
sub-network where Alice and Bob have been disconnected, i.e.,
$\mathcal{N}^{\prime}=(P^{\prime},E^{\prime})$ with
$P^{\prime}=P\setminus \{\mathbf{a,b}\}$. There are
$2^{|E^{\prime}|}$ possible directed graphs that can be generated,
where $|E^{\prime}|$ is the number of undirected edges in
$\mathcal{N}^{\prime}$. Thus, we have $2^{|E^{\prime}|}$\
orientations of the original network $\mathcal{N}$. Each of these
orientations defines a flow network and provides possible
strategies for multi-path routing. See Fig.~\ref{NET2b} for an
example.

Then, let us us introduce the notions of in- and
out-neighborhoods. Given an orientation of $\mathcal{N}$, we have
a corresponding flow network, denoted by
$\mathcal{N}_{D}=(P,E_{D})$, where $E_{D}$ is the set of directed
edges. For arbitrary point $\mathbf{p}$, we
define its out-neighborhood as the set of heads going from $\mathbf{p}$%
\begin{equation}
N^{\text{out}}(\mathbf{p})=\{\mathbf{x}\in
P:(\mathbf{p},\mathbf{x})\in E_{D}\},
\end{equation}
and its in-neighborhood as the set of tails going into $\mathbf{p}$%
\begin{equation}
N^{\text{in}}(\mathbf{p})=\{\mathbf{x}\in
P:(\mathbf{x},\mathbf{p})\in E_{D}\}.
\end{equation}
A multipoint communication from point $\mathbf{p}$ is
\textit{logically} defined as a point-to-multipoint connection
from $\mathbf{p}$ to all its out-neighborhood
$N^{\text{out}}(\mathbf{p})$, i.e., $\mathbf{p}\rightarrow N^{\text{out}%
}(\mathbf{p})$, with quantum systems exchanged along the available
quantum channels. A multi-path routing strategy can therefore be
defined as an ordered sequence of such multipoint communications.
See Fig.~\ref{NET2b}.

\begin{figure}[ptbh]
\begin{center}
\vspace{-0.6cm} \includegraphics[width=0.45\textwidth]{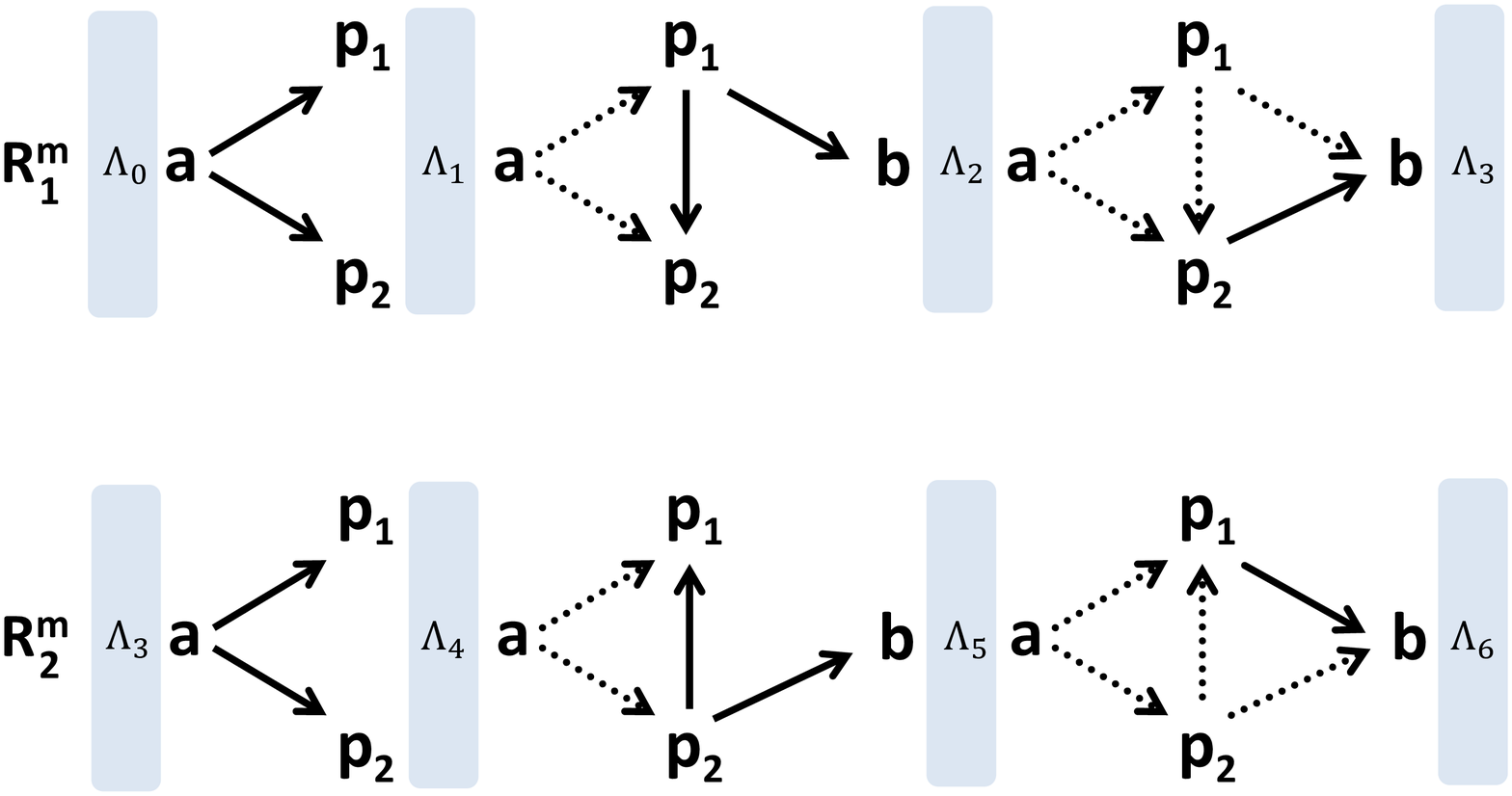}
\end{center}
\par
\vspace{-1.1cm}\caption{Two possible parallel uses of a diamond
quantum network. In the upper multi-path routing
$R_{1}^{\text{m}}$, after the initial LOCC $\Lambda_{0}$, there is
the first multipoint communication $\mathbf{a}\rightarrow
\{\mathbf{p}_{1},\mathbf{p}_{2}\}$, followed by the LOCC
$\Lambda_{1}$. Then,
we have the second multipoint communication $\mathbf{p}_{1}\rightarrow\{\mathbf{b}%
,\mathbf{p}_{2}\}$ followed by $\Lambda_{2}$. Finally, we have $\mathbf{p}%
_{2}\rightarrow\mathbf{b}$ followed by the final LOCC
$\Lambda_{3}$. This completes a single end-to-end transmission. In
the lower multi-path routing $R_{2}^{\text{m}}$, $\mathbf{p}_{1}$ and $\mathbf{p}_{2}$ are inverted.}%
\label{duo}
\end{figure}

Using these definitions we may easily formalize the multi-path
network protocol that we may simply call \textquotedblleft
flooding protocol\textquotedblright. Suppose that we have
$|P|=Z+2$ points in the network ($Z$ repeaters plus the two
end-points). The first step of the protocol is the agreement of a
multi-path routing strategy $R_{1}^{\text{m}}$ by means of
preliminary CCs among all the points. This is part of an
initialization LOCC $\Lambda_{0}$ which prepares an initial
separable state for the entire network. Then, Alice $\mathbf{a}$\
exchanges quantum systems with all her out-neighborhood
$N^{+}(\mathbf{a})$. This multipoint communication is followed by
a network LOCC $\Lambda _{1}$. Next, repeater $\mathbf{p}_{1}\in
N^{+}(\mathbf{a})$ exchanges quantum systems with all its
out-neighborhood $N^{+}(\mathbf{p}_{1})$, which is followed by
another LOCC $\Lambda_{2}$ and so on. At some step $Z+1$, Bob
$\mathbf{b}$ will have exchanged quantum systems with all his
in-neighborhood $N^{-}(\mathbf{b})$, after which there is a final
LOCC $\Lambda_{Z+1}$. This completes the first multi-path
transmission between the end-points by means of
the routing $R_{1}^{\text{m}}$ and the sequence of LOCCs $\{\Lambda_{0}%
,\ldots,\Lambda_{Z+1}\}$. Then, there is the second use of the
network with a generally different routing strategy
$R_{2}^{\text{m}}$ etc. See Fig.~\ref{duo}.

Let us note that the points of the network may generally update
their routing strategy \textquotedblleft on the
fly\textquotedblright, i.e., while the protocol is running; then,
the various multipoint communications may be suitably permuted in
their order. In any case, for large number of uses $n$, we will
have a
sequence of multi-path routings $\mathcal{R}^{\text{m}}=\{R_{1}^{\text{m}%
},\ldots,R_{n}^{\text{m}}\}$ and network LOCCs
$\mathcal{L}=\{\Lambda _{0},\ldots,\Lambda_{n(Z+1)}\}$ whose
output provides Alice and Bob's final
state $\rho_{\mathbf{ab}}^{n}$. The flooding protocol $\mathcal{P}%
_{\mathrm{flood}}$ will be fully described by
$\mathcal{R}^{\text{m}}$ and $\mathcal{L}$. By definition, its
average rate is $R_{n}^{\varepsilon}$ if $\left\Vert
\rho_{\mathbf{ab}}^{n}-\phi_{n}\right\Vert _{1}\leq\varepsilon$,
where $\phi_{n}$ is a target state of $nR_{n}^{\varepsilon}$ bits.
The multi-path capacity of the network is defined by optimizing
the weak-converse asymptotic rate over all flooding protocols,
i.e.,
\begin{equation}
\mathcal{C}^{\text{m}}(\mathcal{N}):=\sup_{\mathcal{P}_{\mathrm{flood}}}%
\lim_{\varepsilon,n}R_{n}^{\varepsilon}.\label{bbnetCAPdef}%
\end{equation}
By specifying the target state, we define corresponding capacities
for quantum communication, entanglement distillation, key
generation and private communication, satisfying
\begin{equation}
Q_{2}^{\text{m}}(\mathcal{N})=D_{2}^{\text{m}}(\mathcal{N})\leq K^{\text{m}%
}(\mathcal{N})=P_{2}^{\text{m}}(\mathcal{N}).
\end{equation}

Before proceeding, some other considerations are in order. Note
that the parallel uses of the network may also be re-arranged in
such a way that each point performs all its multipoint
communications before another point. For instance, in the example
of Fig.~\ref{duo}, we may consider Alice performing all her $n$
multipoint communications
$\mathbf{a}\rightarrow\{\mathbf{p}_{1},\mathbf{p}_{2}\}$ as
a first step. Suppose that routes $R_{1}^{\text{m}}$ and $R_{2}^{\text{m}}%
$\ are chosen with probability $p$ and $1-p$. Then, after Alice
has finished,\ point $\mathbf{p}_{1}$ performs its $np$ multipoint
communications and $\mathbf{p}_{2}$\ performs its $n(1-p)$ ones,
and so on. We may always re-arrange the protocol and adapt the
LOCC sequence $\mathcal{L}$ to include this variant.

Then, there is a simplified formulation to keep in mind. In fact,
we may consider a special case where the various multipoint
communications, within the same routing strategy, are not
alternated with network LOCCs but they are all performed
simultaneously, with only the initial and final LOCCs to be
applied. For instance, for the routing $R_{1}^{\text{m}}$ of
Fig.~\ref{duo}, this means to set $\Lambda_{1}=\Lambda_{2}=I$ and
assume that the multipoint communications
$\mathbf{a}\rightarrow\{\mathbf{p}_{1},\mathbf{p}_{2}\}$, $\mathbf{p}%
_{1}\rightarrow\{\mathbf{b},\mathbf{p}_{2}\}$ and $\mathbf{p}_{2}%
\rightarrow\mathbf{b}$ occur simultaneously, after the
initialization $\Lambda_{0}$ and before $\Lambda_{3}$. In general,
any variant of the protocol may be considered as long as each
quantum channel (edge) is used exactly $n$ times at the end of the
communication, i.e., after $n$ uses of the quantum network.

In the following Supplementary Note~3, we show how to simulate a
quantum network and then exploit teleportation stretching to
reduce adaptive protocols (based on single- or multi-path
routings) into much simpler block versions. By combining this
technique with entanglement cuts of the quantum network, we will
derive very useful decompositions for Alice and Bob's output
state. These decompositions will be later exploited in
Supplementary Notes~4 and~5 to
derive single-letter upper bounds for the network capacities $\mathcal{C}%
(\mathcal{N})$ and $\mathcal{C}^{\text{m}}(\mathcal{N})$.
Corresponding lower bounds will also be derived by combining
point-to-point quantum protocols with classical routing
strategies, with exact results for distillable networks.

\section{Supplementary Note 3: Simulation and stretching of a quantum
network\label{SecSTRETCHINGNET}}

\subsection{General approach}

Consider a quantum network $\mathcal{N}$\ which is connected by
arbitrary quantum channels. Given two points $\mathbf{x}$\ and
$\mathbf{y}$ connected by channel $\mathcal{E}_{\mathbf{xy}}$, we
consider its simulation
$S_{\mathbf{xy}}=(\mathcal{T}_{\mathbf{xy}},\sigma_{\mathbf{xy}})$
for some LOCC $\mathcal{T}_{\mathbf{xy}}$ and resource state
$\sigma_{\mathbf{xy}}$. Repeating this for all connected points
$(\mathbf{x},\mathbf{y})\in E$, we
define an LOCC simulation of the entire network $S(\mathcal{N}%
)=\{S_{\mathbf{xy}}\}_{(\mathbf{x},\mathbf{y})\in E}$ and a
corresponding resource representation of the network
$\sigma(\mathcal{N})=\{\sigma
_{\mathbf{xy}}\}_{(\mathbf{x},\mathbf{y})\in E}$. For a network of
teleportation-covariant channels, its simulation $S(\mathcal{N})$
is based on teleportation over Choi matrices, so that we may
consider $\sigma
(\mathcal{N})=\{\sigma_{\mathcal{E}_{\mathbf{xy}}}\}_{(\mathbf{x}%
,\mathbf{y})\in E}$, i.e., we have a \textquotedblleft
Choi-representation\textquotedblright\ of the network. Note that
the simulation may be asymptotic for a network of bosonic
channels, following the same treatment previously explained for a
linear chain of repeaters.

\begin{figure}[ptbh]
\vspace{-1.4cm}
\par
\begin{center}
\includegraphics[width=0.48\textwidth] {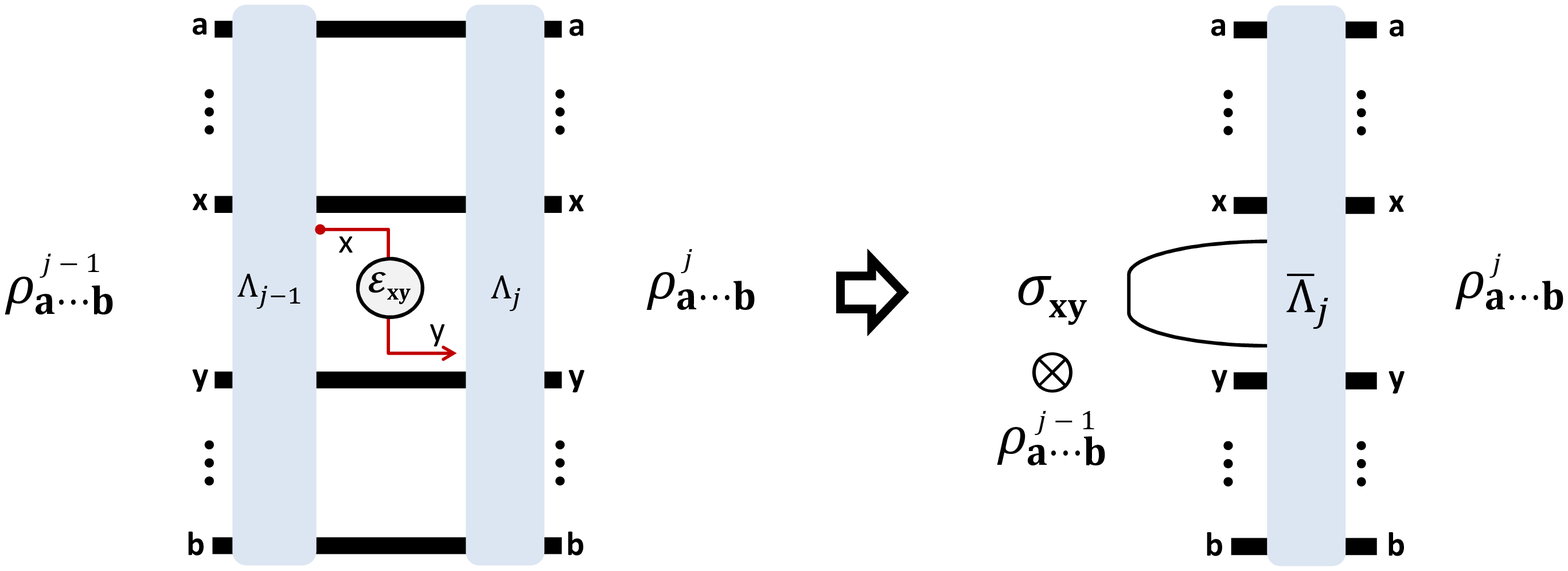}
\end{center}
\par
\vspace{-1.7cm}\caption{Stretching of a network. Consider the
$j$th transmission between points $\mathbf{x}$ and $\mathbf{y}$,
so that the network state $\rho_{\mathbf{a\ldots b}}^{j-1}$ is
transformed into $\rho
_{\mathbf{a\ldots b}}^{j}$. By introducing the simulation $(\mathcal{T}%
_{\mathbf{xy}},\sigma_{\mathbf{xy}})$ of channel
$\mathcal{E}_{\mathbf{xy}}$, we may stretch the resource state
$\sigma_{\mathbf{xy}}$ out of the LOCCs and collapse
$\Lambda_{j-1}$, $\mathcal{T}_{\mathbf{xy}}$ and $\Lambda_{j}$
into a
single LOCC $\bar{\Lambda}_{j}$ applied to $\rho_{\mathbf{a\ldots b}}%
^{j-1}\otimes\sigma_{\mathbf{xy}}$, as in Eq.~(\ref{Net_IT}).}
\label{NET3}
\end{figure}

By adopting a network simulation $S(\mathcal{N})$, we may simplify
adaptive protocols via teleportation stretching, by extending the
procedure employed for a linear chain of quantum repeaters, with
the important difference that we now have many possible chains
(the network routes) and these may also have collisions, i.e.,
repeaters and channels in common. The stretching of a quantum
network is performed iteratively, i.e., transmission after
transmission. Suppose that the $j$th transmission in the network
occurs
between points $\mathbf{x}$ and $\mathbf{y}$ via channel $\mathcal{E}%
_{\mathbf{xy}}$ with associated resource state
$\sigma_{\mathbf{xy}}$. Call $\rho_{\mathbf{a\ldots b}}^{j}$ the
global state of the network after this
transmission. Then, we may write%
\begin{equation}
\rho_{\mathbf{a\ldots b}}^{j}=\bar{\Lambda}_{j}\left(
\rho_{\mathbf{a\ldots
b}}^{j-1}\otimes\sigma_{\mathbf{xy}}\right)  ,\label{Net_IT}%
\end{equation}
where $\bar{\Lambda}_{j}$\ is a trace-preserving LOCC (see
Fig.~\ref{NET3} for a schematic visualization).

By iterating Eq.~(\ref{Net_IT}) and considering that the initial
state of network $\rho_{\mathbf{a\ldots b}}^{0}$ is separable, we
may then write the
network output state after $n$ transmissions as%
\begin{equation}
\rho_{\mathbf{a\ldots b}}^{n}=\bar{\Lambda}\left[  \underset{(\mathbf{x}%
,\mathbf{y})\in E}{%
{\textstyle\bigotimes}
}~\sigma_{\mathbf{xy}}^{\otimes n_{\mathbf{xy}}}\right]  ,
\label{NetREDUCTION}%
\end{equation}
where $n_{\mathbf{xy}}$ is the number of uses of channel $\mathcal{E}%
_{\mathbf{xy}}$ or, equivalently, edge $(\mathbf{x},\mathbf{y})$.
Then, by tracing out all the points but Alice and Bob, we get
their final shared state
\begin{equation}
\rho_{\mathbf{ab}}^{n}=\bar{\Lambda}_{\mathbf{ab}}\left[
\underset
{(\mathbf{x},\mathbf{y})\in E}{%
{\textstyle\bigotimes}
}~\sigma_{\mathbf{xy}}^{\otimes n_{\mathbf{xy}}}\right]  , \label{NET2RED}%
\end{equation}
for another trace-preserving LOCC $\bar{\Lambda}_{\mathbf{ab}}$.

\begin{figure}[ptbh]
\vspace{-1.3cm}
\par
\begin{center}
\includegraphics[width=0.48\textwidth] {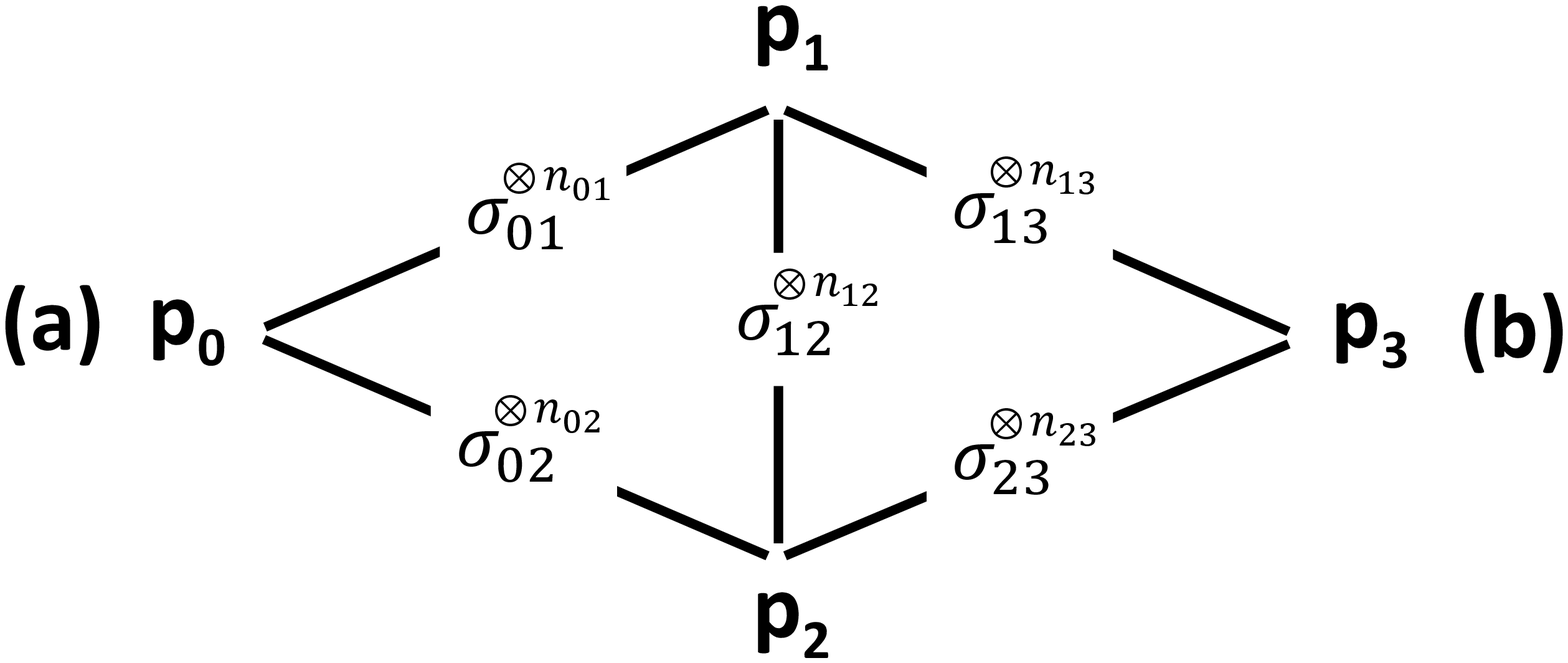} \vspace{-2.0cm}
\end{center}
\caption{Network stretching. Consider a diamond quantum network $\mathcal{N}%
^{\diamond}=(\{\mathbf{p}_{0},\mathbf{p}_{1},\mathbf{p}_{2},\mathbf{p}%
_{3}\},E)$ with resource representation
$\sigma(\mathcal{N}^{\diamond
})=\{\sigma_{01},\sigma_{02},\sigma_{12},\sigma_{13},\sigma_{23}\}$.
Before stretching, an arbitrary edge $(\mathbf{x},\mathbf{y})$
with channel $\mathcal{E}_{\mathbf{xy}}$ is used $n_{\mathbf{xy}}$
times. After stretching, the same edge $(\mathbf{x},\mathbf{y})$
is associated with $n_{\mathbf{xy}}$ copies of the resource state
$\sigma_{\mathbf{xy}}$. The latter is the Choi matrix
$\sigma_{\mathcal{E}_{\mathbf{xy}}}$ if
$\mathcal{E}_{\mathbf{xy}}$ is teleportation-covariant. The global
state of the network is expressed as in
Eq.~(\ref{LemmaNETstretching}), which may take an asymptotic form
for a network of bosonic channels.} \label{NET4}
\end{figure}

Note that the decompositions of Eqs.~(\ref{NetREDUCTION}) and
(\ref{NET2RED}) can be written for any adaptive network protocol
(sequential or flooding). For a sequential protocol
$n_{\mathbf{xy}}=np_{\mathbf{xy}}\leq n$, where $p_{\mathbf{xy}}$
is the probability of using edge $(\mathbf{x},\mathbf{y})$. For a
flooding protocol, we instead have $n_{\mathbf{xy}}=n$, because
each edge is used exactly once in each end-to-end transmission. In
particular, in a
flooding protocol, we have the parallel use of several channels $\mathcal{E}%
_{\mathbf{x}_{1}\mathbf{y}_{1}}$, $\mathcal{E}_{\mathbf{x}_{2}\mathbf{y}_{2}}%
$, \ldots\ within each multipoint communication, which means that
trivial LOCCs (identities) are applied between every two
transmissions within the same multipoint communication. We have
therefore proven the following result (see also Fig.~\ref{NET4}
for a simple example).

\begin{lemma}
[Network stretching]\label{LemmaNET}Consider a quantum network $\mathcal{N}%
=(P,E)$ which is simulable with some resource representation
$\sigma
(\mathcal{N})=\{\sigma_{\mathbf{xy}}\}_{(\mathbf{x},\mathbf{y})\in
E}$. Then,
consider $n$ uses of an adaptive protocol so that edge $(\mathbf{x}%
,\mathbf{y})\in E$ is used $n_{\mathbf{xy}}$ times. We may write
the global
output state of the network as%
\begin{equation}
\rho_{\mathbf{a\ldots b}}^{n}=\bar{\Lambda}\left[  \underset{(\mathbf{x}%
,\mathbf{y})\in E}{%
{\textstyle\bigotimes}
}~\sigma_{\mathbf{xy}}^{\otimes n_{\mathbf{xy}}}\right]  ,
\label{LemmaNETstretching}%
\end{equation}
for a trace-preserving LOCC $\bar{\Lambda}$. Similarly, Alice and
Bob's output state $\rho_{\mathbf{ab}}^{n}$ is given by
Eq.~(\ref{LemmaNETstretching}) up to a different trace-preserving
LOCC $\bar{\Lambda}_{\mathbf{ab}}$. In particular, we have
$n_{\mathbf{xy}}\leq n$ ($n_{\mathbf{xy}}=n$) for a sequential
(flooding) protocol. Formulations may be asymptotic for bosonic
channels.
\end{lemma}

As we state in the lemma, the stretching procedure also applies to
networks of bosonic channels with asymptotic simulations. This can
be understood by extending the argument already given for linear
chains. For the sake of clarity, we make this argument explicit
here. Consider again the $j$th transmission in the network
occurring via channel $\mathcal{E}_{\mathbf{xy}}$ as in
Fig.~\ref{NET3}. For the global state of the network, we may write
\begin{equation}
\rho_{\mathbf{a\ldots b}}^{j}=\Lambda_{j}\circ\mathcal{E}_{\mathbf{xy}}%
\circ\Lambda_{j-1}(\rho_{\mathbf{a\ldots b}}^{j-1})~.
\end{equation}
Suppose that we replace each channel $\mathcal{E}_{\mathbf{xy}}$
in the network with an approximation
$\mathcal{E}_{\mathbf{xy}}^{\mu}$, with
point-wise limit $\mathcal{E}_{\mathbf{xy}}=\lim_{\mu}\mathcal{E}%
_{\mathbf{xy}}^{\mu}$, meaning that $\left\Vert \mathcal{E}_{\mathbf{xy}}%
(\rho)-\mathcal{E}_{\mathbf{xy}}^{\mu}(\rho)\right\Vert
_{1}\overset{\mu }{\rightarrow}0$ for any state $\rho$. We may
build the approximate network state
\begin{equation}
\rho_{\mathbf{a\ldots b}}^{j,\mu}=\Lambda_{j}\circ\mathcal{E}_{\mathbf{xy}%
}^{\mu}\circ\Lambda_{j-1}(\rho_{\mathbf{a\ldots b}}^{j-1,\mu})~.
\end{equation}

Now assume that all the registers in the network are bounded by a
large but finite mean number of photons $\bar{N}$, so that we may
write $\left\Vert
\mathcal{E}_{\mathbf{xy}}-\mathcal{E}_{\mathbf{xy}}^{\mu}\right\Vert
_{\diamond\bar{N}}\overset{\mu}{\rightarrow}0$ in
energy-constrained diamond distance. By using the monotonicity
under CPTP maps and the triangle
inequality, we then compute%
\begin{align}
&  \left\Vert \rho_{\mathbf{a\ldots b}}^{j}-\rho_{\mathbf{a\ldots
b}}^{j,\mu
}\right\Vert _{1}\nonumber\\
&  \leq\left\Vert \mathcal{E}_{\mathbf{xy}}\circ\Lambda_{j-1}(\rho
_{\mathbf{a\ldots
b}}^{j-1})-\mathcal{E}_{\mathbf{xy}}^{\mu}\circ\Lambda
_{j-1}(\rho_{\mathbf{a\ldots b}}^{j-1,\mu})\right\Vert _{1}\nonumber\\
&  \leq\left\Vert \mathcal{E}_{\mathbf{xy}}\circ\Lambda_{j-1}(\rho
_{\mathbf{a\ldots
b}}^{j-1})-\mathcal{E}_{\mathbf{xy}}^{\mu}\circ\Lambda
_{j-1}(\rho_{\mathbf{a\ldots b}}^{j-1})\right\Vert _{1}\nonumber\\
&  +\left\Vert \mathcal{E}_{\mathbf{xy}}^{\mu}\circ\Lambda_{j-1}%
(\rho_{\mathbf{a\ldots
b}}^{j-1})-\mathcal{E}_{\mathbf{xy}}^{\mu}\circ
\Lambda_{j-1}(\rho_{\mathbf{a\ldots b}}^{j-1,\mu})\right\Vert _{1}\nonumber\\
&  \leq\left\Vert
\mathcal{E}_{\mathbf{xy}}-\mathcal{E}_{\mathbf{xy}}^{\mu
}\right\Vert _{\diamond\bar{N}}+\left\Vert \rho_{\mathbf{a\ldots b}}%
^{j-1}-\rho_{\mathbf{a\ldots b}}^{j-1,\mu}\right\Vert _{1}.
\end{align}
By iterating the previous formula for all the transmissions in the
network, we
derive%
\begin{equation}
\left\Vert \rho_{\mathbf{a\ldots b}}^{n}-\rho_{\mathbf{a\ldots
b}}^{n,\mu
}\right\Vert _{1}\leq%
{\textstyle\sum\limits_{(\mathbf{x},\mathbf{y})\in E}}
n_{\mathbf{xy}}\left\Vert \mathcal{E}_{\mathbf{xy}}-\mathcal{E}_{\mathbf{xy}%
}^{\mu}\right\Vert _{\diamond\bar{N}}~.\label{bbb1}%
\end{equation}
This distance goes to zero in $\mu$ for any number of uses $n$,
any finite number of edges $|E|$, and any energy $\bar{N}$.

Now suppose that the generic approximate channel $\mathcal{E}_{\mathbf{xy}%
}^{\mu}$ has LOCC simulation with some resource state $\sigma_{\mathbf{xy}%
}^{\mu}$. Then, we may write the approximate network stretching
\begin{equation}
\rho_{\mathbf{a\ldots b}}^{n,\mu}=\bar{\Lambda}^{\mu}\left[
\underset
{(\mathbf{x},\mathbf{y})\in E}{%
{\textstyle\bigotimes}
}~\sigma_{\mathbf{xy}}^{\mu\otimes n_{\mathbf{xy}}}\right]  , \label{bbb2}%
\end{equation}
for a trace-preserving LOCC\ $\bar{\Lambda}^{\mu}$. Combining Eqs.~(\ref{bbb1}%
) and~(\ref{bbb2}), we may therefore write the asymptotic version
of network
stretching%
\begin{equation}
\rho_{\mathbf{a\ldots b}}^{n}=\lim_{\mu}\bar{\Lambda}^{\mu}\left[
\underset{(\mathbf{x},\mathbf{y})\in E}{%
{\textstyle\bigotimes}
}~\sigma_{\mathbf{xy}}^{\mu\otimes n_{\mathbf{xy}}}\right]  ,
\label{totalASYMP}%
\end{equation}
where the limit in $\mu$\ is intended in trace norm and holds for
any finite $n$, $|E|$ and $\bar{N}$.

Similarly, let us consider Alice and Bob's reduced state $\rho_{\mathbf{ab}%
}^{n}$ and its approximation $\rho_{\mathbf{ab}}^{n,\mu}$. As a
result of the partial trace, we may write
\begin{equation}
\left\Vert
\rho_{\mathbf{ab}}^{n}-\rho_{\mathbf{ab}}^{n,\mu}\right\Vert
_{1}\leq\left\Vert \rho_{\mathbf{a\ldots b}}^{n}-\rho_{\mathbf{a\ldots b}%
}^{n,\mu}\right\Vert _{1},
\end{equation}
so that we may apply the bound in Eq.~(\ref{bbb1}) and write%
\begin{equation}
\left\Vert
\rho_{\mathbf{ab}}^{n}-\rho_{\mathbf{ab}}^{n,\mu}\right\Vert
_{1}\leq%
{\textstyle\sum\limits_{(\mathbf{x},\mathbf{y})\in E}}
n_{\mathbf{xy}}\left\Vert \mathcal{E}_{\mathbf{xy}}-\mathcal{E}_{\mathbf{xy}%
}^{\mu}\right\Vert _{\diamond\bar{N}}~.
\end{equation}
If the generic channel $\mathcal{E}_{\mathbf{xy}}^{\mu}$ has LOCC
simulation
with some resource state $\sigma_{\mathbf{xy}}^{\mu}$, then we may write%
\begin{equation}
\rho_{\mathbf{ab}}^{n}=\lim_{\mu}\bar{\Lambda}_{\mathbf{ab}}^{\mu}\left[
\underset{(\mathbf{x},\mathbf{y})\in E}{%
{\textstyle\bigotimes}
}~\sigma_{\mathbf{xy}}^{\mu\otimes n_{\mathbf{xy}}}\right]  ,
\end{equation}
where the limit in $\mu$\ is intended in trace norm and holds for
any finite $n$, $|E|$ and $\bar{N}$.

\subsection{Network stretching with entanglement cuts}

We may achieve a non-trivial simplification of previous
Lemma~\ref{LemmaNET} in such a way that we greatly reduce the
number of resource states in the decomposition of Alice and Bob's
output state $\rho_{\mathbf{ab}}^{n}$. This is possible using
Alice-Bob entanglement cuts of the quantum network. These types of
cuts will enable us to include many resource states in Alice's and
Bob's LOs, while preserving the locality between the two
end-points.

By definition, an Alice-Bob entanglement cut $C$ of the quantum
network is a bipartition $(\mathbf{A},\mathbf{B})$ of all the
points $P$\ of the network such that $\mathbf{a}\in\mathbf{A}$ and
$\mathbf{b}\in\mathbf{B}$. Then, the cut-set $\tilde{C}$ of $C$ is
the set of edges with one end-point in each subset of the
bipartition, so that the removal of these edges disconnects the
network. Explicitly,
\begin{equation}
\tilde{C}=\{(\mathbf{x},\mathbf{y})\in E:\mathbf{x}\in\mathbf{A},\mathbf{y}%
\in\mathbf{B}\}.
\end{equation}
Note that the cut-set $\tilde{C}$\ identifies an ensemble of
channels
$\{\mathcal{E}_{\mathbf{xy}}\}_{(\mathbf{x},\mathbf{y})\in\tilde{C}}$.
Similarly, we may define the following complementary sets
\begin{align}
\tilde{A}  &  =\{(\mathbf{x},\mathbf{y})\in E:\mathbf{x,y}\in\mathbf{A}\},\\
\tilde{B}  &  =\{(\mathbf{x},\mathbf{y})\in
E:\mathbf{x,y}\in\mathbf{B}\},
\end{align}
so that $\tilde{A}\cup\tilde{B}\cup\tilde{C}=E$.

To simplify the stretching of the network, we then adopt the
following procedure. Given an arbitrary cut
$C=(\mathbf{A},\mathbf{B})$, we extend Alice and Bob to their
corresponding partitions. This means that we consider super-Alice
with global register $\mathbf{A}$, and super-Bob with global
register $\mathbf{B}$. Then, all the resource states $\{\sigma_{\mathbf{xy}%
}\}_{(\mathbf{x},\mathbf{y})\in\tilde{A}}$ are included in the LOs
of
super-Alice, and all those $\{\sigma_{\mathbf{xy}}\}_{(\mathbf{x}%
,\mathbf{y})\in\tilde{B}}$ are included in the LOs of super-Bob.
Note that the only resource states not absorbed in LOs are those
in the cut-set
$\{\sigma_{\mathbf{xy}}\}_{(\mathbf{x},\mathbf{y})\in\tilde{C}}$.
These states are the only ones responsible for distributing
entanglement between the super-parties. The inclusion of all the
other resource states into the global LOCC $\bar{\Lambda}$ leads
to another trace-preserving quantum operation
$\bar{\Lambda}_{\mathbf{AB}}$ which remains local with respect to
$\mathbf{A}$ and $\mathbf{B}$. Thus, for any cut $C$, we may write
the following output state for super-Alice $\mathbf{A}$ and Bob
$\mathbf{B}$ after $n$ uses of an
adaptive protocol%
\begin{equation}
\rho_{\mathbf{AB}}^{n}(C)=\bar{\Lambda}_{\mathbf{AB}}\left[
\underset
{(\mathbf{x},\mathbf{y})\in\tilde{C}}{%
{\textstyle\bigotimes}
}~\sigma_{\mathbf{xy}}^{\otimes n_{\mathbf{xy}}}\right]  .
\end{equation}

\begin{figure}[tbh]
\begin{center}
\vspace{-0.8cm}
\includegraphics[width=0.45\textwidth]{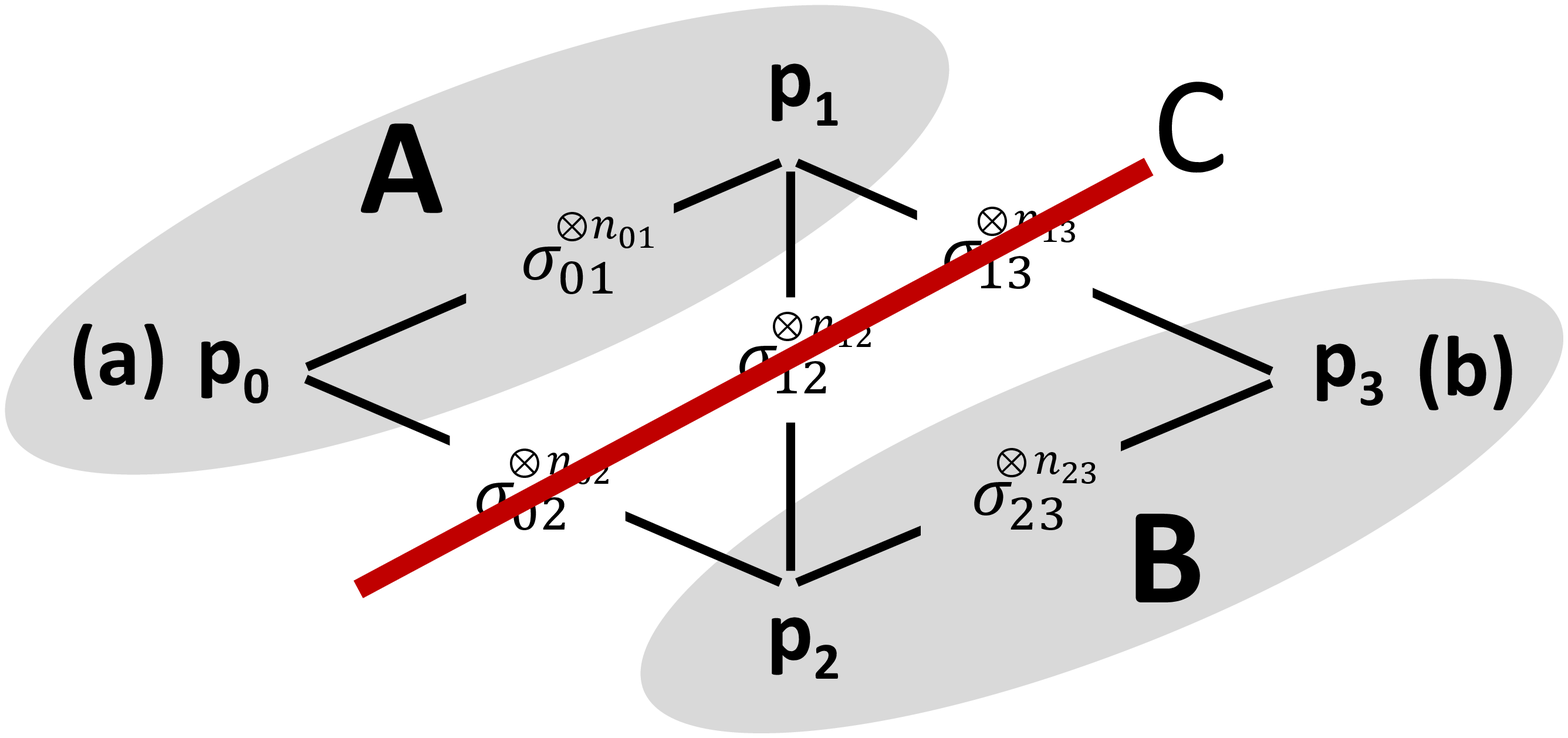}
\vspace{-1.0cm}
\end{center}
\caption{Network stretching with entanglement cuts. We show one of
the
possible entanglement cuts $C$ of the diamond quantum network $\mathcal{N}%
^{\diamond}$. This cut creates super-Alice $\mathbf{A}=\{\mathbf{a}%
,\mathbf{p}_{1}\}$ and super-Bob
$\mathbf{B}=\{\mathbf{b},\mathbf{p}_{2}\}$. The resource states
$\sigma_{01}^{\otimes n_{01}}$ are absorbed in the local
operations (LOs) of $\mathbf{A}$, while the resource states
$\sigma_{23}^{\otimes n_{23}}$ are absorbed in the LOs of
$\mathbf{B}$. The cut-set is composed by the set of
edges $\tilde{C}=\{(\mathbf{p}_{0},\mathbf{p}_{2}),(\mathbf{p}_{1}%
,\mathbf{p}_{2}),(\mathbf{p}_{1},\mathbf{p}_{3})\}$ with
corresponding resource states $\sigma_{02}^{\otimes n_{02}}$,
$\sigma_{12}^{\otimes n_{12}}$ and $\sigma_{13}^{\otimes n_{13}}$.
This subset of states can be used to decompose the output state of
Alice and Bob $\rho_{\mathbf{ab}}^{n}(C)$ according to
Eq.~(\ref{cutEQ}).} \label{cut}
\end{figure}

The next step is tracing out all registers but the original
Alice's $\mathbf{a}$ and Bob's $\mathbf{b}$. This operation
preserves the locality between $\mathbf{a}$ and $\mathbf{b}$. In
other words, we may write the
following reduced output state for the two end-points%
\begin{align}
\rho_{\mathbf{ab}}^{n}(C)  &
=\mathrm{Tr}_{P\setminus\{\mathbf{a,b}\}}\left[
\rho_{\mathbf{AB}}^{n}(C)\right] \nonumber\\
&  =\bar{\Lambda}_{\mathbf{ab}}\left[  \underset{(\mathbf{x},\mathbf{y}%
)\in\tilde{C}}{%
{\textstyle\bigotimes}
}~\sigma_{\mathbf{xy}}^{\otimes n_{\mathbf{xy}}}\right]  ,
\end{align}
where $\bar{\Lambda}_{\mathbf{ab}}$ is a trace-preserving LOCC.
All these reasonings automatically transform Lemma~\ref{LemmaNET}
into the following improved Lemma. See also Fig.~\ref{cut} for an
example.

\begin{lemma}
[Network stretching with cuts]\label{reduceCHOI}Consider a quantum
network $\mathcal{N}=(P,E)$ simulable with a resource
representation $\sigma
(\mathcal{N})=\{\sigma_{\mathbf{xy}}\}_{(\mathbf{x},\mathbf{y})\in
E}$. For a teleportation-covariant network, $\sigma(\mathcal{N})$
is a
Choi-representation, i.e., $\sigma_{\mathbf{xy}}=\sigma_{\mathcal{E}%
_{\mathbf{xy}}}$. Then, consider $n$ uses of an adaptive protocol
so that edge $(\mathbf{x},\mathbf{y})\in E$ is used
$n_{\mathbf{xy}}$ times. For any entanglement cut $C$ and
corresponding cut-set $\tilde{C}$, we may write Alice
and Bob's output state as%
\begin{equation}
\rho_{\mathbf{ab}}^{n}(C)=\bar{\Lambda}_{\mathbf{ab}}\left[
\underset
{(\mathbf{x},\mathbf{y})\in\tilde{C}}{%
{\textstyle\bigotimes}
}~\sigma_{\mathbf{xy}}^{\otimes n_{\mathbf{xy}}}\right]  , \label{cutEQ}%
\end{equation}
for a trace-preserving LOCC $\bar{\Lambda}_{\mathbf{ab}}$. In
particular, we have $n_{\mathbf{xy}}\leq n$ ($n_{\mathbf{xy}}=n$)
for a sequential (flooding) protocol. Formulations may be
asymptotic for bosonic channels.
\end{lemma}

As stated in this improved lemma,\ the decomposition in
Eq.~(\ref{cutEQ}) can be extended to networks of bosonic channels
with asymptotic simulations. We can adapt the previous reasoning
to find the cut-version of
Eq.~(\ref{totalASYMP}), i.e., the trace-norm limit%
\begin{equation}
\left\Vert
\rho_{\mathbf{ab}}^{n}(C)-\bar{\Lambda}_{\mathbf{ab}}^{\mu}\left[
\underset{(\mathbf{x},\mathbf{y})\in\tilde{C}}{%
{\textstyle\bigotimes}
}~\sigma_{\mathbf{xy}}^{\mu\otimes n_{\mathbf{xy}}}\right]
\right\Vert
_{1}\overset{\mu}{\rightarrow}0, \label{cutASYpproof}%
\end{equation}
for suitable sequences of trace-preserving LOCC $\bar{\Lambda}_{\mathbf{ab}%
}^{\mu}$ and resource states $\sigma_{\mathbf{xy}}^{\mu}$ (with
the result holding for any $n$, number of edges $|E|$ and mean
number of photons $\bar {N}$).

With Lemma~\ref{reduceCHOI} in our hands, we have the necessary
tool to derive our single-letter upper bounds for the single- and
multi-path capacities of an arbitrary quantum network. This tool
needs to be combined with a general weak converse upper bound
based on the REE. In the following Supplementary Note~4, we derive
our results for the case of single-path routing over the network.
The results for multi-path routing will be given in Supplementary
Note~5. In both these Supplementary Notes, the upper bounds will
be compared with suitable lower bounds that are derived by mixing
point-to-point quantum protocols with classical routing strategies
(widest path and maximum flow of a network).

\section{Supplementary Note 4: Results for single-path
routing\label{secSINGLE}}

\subsection{Converse part (upper bound)}

In order to write a single-letter upper bound for the single-path
capacity of the quantum network, we need to introduce the notion
of REE flowing through a cut under some simulation. Consider an
arbitrary quantum network
$\mathcal{N}=(P,E)$ with a resource representation $\sigma(\mathcal{N}%
)=\{\sigma_{\mathbf{xy}}\}_{(\mathbf{x},\mathbf{y})\in E}$. Then,
consider an arbitrary entanglement cut $C$ with corresponding
cut-set $\tilde{C}$. Under the simulation considered, we define
the single-edge flow of REE through the
cut as the following quantity%
\begin{equation}
E_{\mathrm{R}}(C):=\max_{(\mathbf{x},\mathbf{y})\in\tilde{C}}E_{\mathrm{R}%
}(\sigma_{\mathbf{xy}})~. \label{ERC}%
\end{equation}
By minimizing $E_{\mathrm{R}}(C)$ over all possible entanglement
cuts of the network, we build our upper bound for the single-path
capacity. In fact, we may prove the following.

\begin{theorem}
[Converse for single-path capacity]\label{theoUBnet}Consider an
arbitrary quantum network $\mathcal{N}=(P,E)$ with some resource
representation
$\sigma(\mathcal{N})=\{\sigma_{\mathbf{xy}}\}_{(\mathbf{x},\mathbf{y})\in
E}$. In particular, $\sigma(\mathcal{N})$ may be a
Choi-representation for a teleportation-covariant network. Then,
the single-path capacity of
$\mathcal{N}$ must satisfy the single-letter bound%
\begin{equation}
\mathcal{C}(\mathcal{N})\leq\min_{C}E_{\mathrm{R}}(C)~, \label{UBnetmain}%
\end{equation}
where the single-edge flow of REE in Eq.~(\ref{ERC}) is minimized
across all cuts of the network. Formulations may be asymptotic for
networks of bosonic channels.
\end{theorem}

\textbf{Proof.}~~We start from the general weak converse upper
bound proven in the Methods section of the paper. In terms of the
REE\ and for network sequential protocols
$\mathcal{P}_{\text{\textrm{seq}}}$, this bound takes the
form%
\begin{equation}
\mathcal{C}(\mathcal{N})\leq E_{\mathrm{R}}^{\bigstar}(\mathcal{N}%
):=\sup_{\mathcal{P}_{\text{\textrm{seq}}}}\underset{n}{\lim}\frac
{E_{\mathrm{R}}(\rho_{\mathbf{ab}}^{n})}{n}.\label{ccc2}%
\end{equation}
According to previous Lemma~\ref{reduceCHOI}, for any sequential
protocol $\mathcal{P}_{\text{\textrm{seq}}}$ and entanglement cut
$C$ of the network, we may write Eq.~(\ref{cutEQ}). Computing the
REE on this decomposition and exploiting basic properties
(monotonicity of REE under $\bar{\Lambda }_{\mathbf{ab}}$ and
subadditivity over tensor products), we derive the
following inequality%
\begin{equation}
E_{\mathrm{R}}\left[  \rho_{\mathbf{ab}}^{n}(C)\right]  \leq\sum
_{(\mathbf{x},\mathbf{y})\in\tilde{C}}n_{\mathbf{xy}}E_{\mathrm{R}}%
(\sigma_{\mathbf{xy}}),\label{ccc3}%
\end{equation}
where $n_{\mathbf{xy}}=np_{\mathbf{xy}}$\ and $p_{\mathbf{xy}}$\
being the probability of using edge $(\mathbf{x},\mathbf{y})$\
according to protocol $\mathcal{P}_{\text{\textrm{seq}}}$.\ By
maximizing over the convex
combination, we get rid of $p_{\mathbf{xy}}$ and write%
\begin{equation}
E_{\mathrm{R}}\left[  \rho_{\mathbf{ab}}^{n}(C)\right]  \leq n\max
_{(\mathbf{x},\mathbf{y})\in\tilde{C}}E_{\mathrm{R}}(\sigma_{\mathbf{xy}%
})=nE_{\mathrm{R}}(C)~.\label{mmllll}%
\end{equation}
By using Eq.~(\ref{mmllll}) in Eq.~(\ref{ccc2}), we see that both
the optimization over $\mathcal{P}_{\text{\textrm{seq}}}$ and the
limit over $n$
disappear, and we are left with the bound%
\begin{equation}
\mathcal{C}(\mathcal{N})\leq E_{\mathrm{R}}(C),~~\text{for any~}%
C\text{.}\label{preRESS}%
\end{equation}
By minimizing over all cuts, we therefore prove
Eq.~(\ref{UBnetmain}).

Note that, from Eq.~(\ref{ccc3}) we may also derive
\begin{equation}
\mathcal{C}(\mathcal{N})\leq\bar{E}_{\mathrm{R}}(C):=\sum_{(\mathbf{x}%
,\mathbf{y})\in\tilde{C}}\bar{p}_{\mathbf{xy}}E_{\mathrm{R}}(\sigma
_{\mathbf{xy}}), \label{boundAVE}%
\end{equation}
where $\bar{p}_{\mathbf{xy}}$ is the optimal use of edge $(\mathbf{x}%
,\mathbf{y})$ over all possible
$\mathcal{P}_{\text{\textrm{seq}}}$. Here
$\bar{E}_{\mathrm{R}}(C)$ represents the \textit{average flow} of
REE through
$C$ under the chosen simulation and optimized over $\mathcal{P}%
_{\text{\textrm{seq}}}$. By minimizing over all cuts, we get%
\begin{equation}
\mathcal{C}(\mathcal{N})\leq\min_{C}\bar{E}_{\mathrm{R}}(C).
\end{equation}
This may be tighter than Eq.~(\ref{UBnetmain}) but difficult to
compute due to residual optimization over the protocols.

Finally, note that Eq.~(\ref{UBnetmain}) can be extended to
considering asymptotic simulations, following the same ideas in
the proof of Theorem~\ref{singleLETTtheorem}. Let us compute the
REE on the asymptotic state $\rho_{\mathbf{ab}}^{n}(C)$ of
Eq.~(\ref{cutASYpproof}). We may write
\begin{align}
&  E_{\mathrm{R}}[\rho_{\mathbf{ab}}^{n}(C)]\nonumber\\
&
=\inf_{\gamma\in\mathrm{SEP}}S[\rho_{\mathbf{ab}}^{n}(C)||\gamma
]\nonumber\\
&  \overset{(1)}{\leq}\inf_{\gamma^{\mu}}S\left.  \left\{  \lim_{\mu}%
\bar{\Lambda}_{\mathbf{ab}}^{\mu}\left[  \underset{(\mathbf{x},\mathbf{y}%
)\in\tilde{C}}{%
{\textstyle\bigotimes}
}~\sigma_{\mathbf{xy}}^{\mu\otimes n_{\mathbf{xy}}}\right]
\right\Vert
~\lim_{\mu}\gamma^{\mu}\right\} \nonumber\\
&  \overset{(2)}{\leq}\inf_{\gamma^{\mu}}\underset{\mu\rightarrow+\infty}%
{\lim\inf}S\left.  \left\{
\bar{\Lambda}_{\mathbf{ab}}^{\mu}\left[
\underset{(\mathbf{x},\mathbf{y})\in\tilde{C}}{%
{\textstyle\bigotimes}
}~\sigma_{\mathbf{xy}}^{\mu\otimes n_{\mathbf{xy}}}\right]
\right\Vert
~\gamma^{\mu}\right\} \nonumber\\
&  \overset{(3)}{\leq}\inf_{\gamma^{\mu}}\underset{\mu\rightarrow+\infty}%
{\lim\inf}S\left.  \left\{
\bar{\Lambda}_{\mathbf{ab}}^{\mu}\left[
\underset{(\mathbf{x},\mathbf{y})\in\tilde{C}}{%
{\textstyle\bigotimes}
}~\sigma_{\mathbf{xy}}^{\mu\otimes n_{\mathbf{xy}}}\right]
\right\Vert
~\bar{\Lambda}_{\mathbf{ab}}^{\mu}(\gamma^{\mu})\right\} \nonumber\\
&  \overset{(4)}{\leq}\inf_{\gamma^{\mu}}\underset{\mu\rightarrow+\infty}%
{\lim\inf}S\left.  \left[  \underset{(\mathbf{x},\mathbf{y})\in\tilde{C}}{%
{\textstyle\bigotimes}
}~\sigma_{\mathbf{xy}}^{\mu\otimes n_{\mathbf{xy}}}\right\Vert
~\gamma^{\mu
}\right] \nonumber\\
&  \overset{(5)}{=}E_{\mathrm{R}}\left[  \underset{(\mathbf{x},\mathbf{y}%
)\in\tilde{C}}{%
{\textstyle\bigotimes}
}~\sigma_{\mathbf{xy}}^{\otimes n_{\mathbf{xy}}}\right] \nonumber\\
&  \overset{(6)}{\leq}\sum_{(\mathbf{x},\mathbf{y})\in\tilde{C}}%
n_{\mathbf{xy}}E_{\mathrm{R}}(\sigma_{\mathbf{xy}}),
\end{align}
where: (1)$~\gamma^{\mu}$ is a generic sequence of separable
states converging in trace norm, i.e., such that there is a
separable state $\gamma:=\lim_{\mu }\gamma^{\mu}$ so that
$\Vert\gamma-\gamma^{\mu}\Vert_{1}\overset{\mu }{\rightarrow}0$;
(2)~we use the lower semi-continuity of the relative
entropy~\cite{HolevoBOOKs}; (3)~we use that
$\bar{\Lambda}_{\mathbf{ab}}^{\mu }(\gamma^{\mu})$ are specific
types of converging separable sequences within the set of all such
sequences; (4)~we use the monotonicity of the relative entropy
under trace-preserving LOCCs; (5)~we use the definition of REE for
asymptotic states $\sigma_{\mathbf{xy}}:=\lim_{\mu}\sigma_{\mathbf{xy}}^{\mu}%
$; (6)~we use the subadditivity over tensor products.

Therefore, we have again Eq.~(\ref{ccc3}) but where the REE is
written as in the weaker formulation for asymptotic states given
in Eq.~(\ref{REE_weaker}). The next steps of the proof are exactly
as before, and they lead to Eq.~(\ref{preRESS}).~$\blacksquare$

\subsection{Direct part (achievable rate)}

In this section, we derive an achievable asymptotic rate for the
end-to-end quantum/private communication via single-path routing.
This rate will provide a lower bound to the single-path capacity
of an arbitrary quantum network, i.e., with arbitrary topology and
arbitrary quantum channels. The non-trivial result is that the
achievable rate can be written in terms of a capacity minimized
over the entanglement cuts in the network. This step will allow us
to exactly establish the single-path capacity of distillable
networks in the next subsection.

Consider an arbitrary quantum network $\mathcal{N}=(P,E)$ where
edge
$(\mathbf{x},\mathbf{y})\in E$ is connected by channel $\mathcal{E}%
_{\mathbf{xy}}$ with associated two-way capacity $\mathcal{C}_{\mathbf{xy}%
}=\mathcal{C}(\mathcal{E}_{\mathbf{xy}})$. Given an arbitrary
entanglement cut $C$ of the network, we define its single-edge
capacity as the maximum number
of target bits distributed by a single edge across the cut, i.e.,%
\begin{equation}
\mathcal{C}(C):=\max_{(\mathbf{x},\mathbf{y})\in\tilde{C}}\mathcal{C}%
_{\mathbf{xy}}~. \label{cutEQUIV}%
\end{equation}
A minimum cut $C_{\text{min}}$ is such that
\begin{equation}
\mathcal{C}(C_{\text{min}})=\min_{C}\mathcal{C}(C). \label{minCUUTT}%
\end{equation}
Then, given a route $\omega\in\Omega$ with an associated chain of
channels $\{\mathcal{E}_{i}^{\omega}\}$, we define its capacity as
the minimum capacity
among its channels, i.e.,%
\begin{equation}
\mathcal{C}(\omega):=\min_{i}\mathcal{C}(\mathcal{E}_{i}^{\omega})~.
\label{genRR}%
\end{equation}
An optimal route $\omega_{\ast}$\ is such that%
\begin{equation}
\mathcal{C}(\omega_{\ast})=\max_{\omega\in\Omega}\mathcal{C}(\omega)~.
\end{equation}

It is clear that $\mathcal{C}(\omega_{\ast})$ is an achievable
end-to-end rate. In fact, consider independent point-to-point
protocols between pairs of consecutive points along route
$\omega_{\ast}$. An optimal adaptive protocol
between points $\mathbf{r}_{i}^{\omega_{\ast}}$ and $\mathbf{r}_{i+1}%
^{\omega_{\ast}}$ (connected by $\mathcal{E}_{i}^{\omega_{\ast}}$)
achieves the capacity value
$\mathcal{C}(\mathcal{E}_{i}^{\omega_{\ast}})$. Then, by composing
all outputs via a network LOCCs (e.g., swapping the distilled
states or relaying the secret keys via one-time pad sessions),
Alice and Bob obtain
an achievable rate of $\min_{i}\mathcal{C}(\mathcal{E}_{i}^{\omega_{\ast}%
})=\mathcal{C}(\omega_{\ast})$.

Thus, we may write the lower bound $\mathcal{C}(\mathcal{N})\geq
\mathcal{C}(\omega_{\ast})=\max_{\omega}\mathcal{C}(\omega)$. The
crucial observation is that this bound is also equal to the
minimization in Eq.~(\ref{minCUUTT}) over all entanglement cuts.
In fact, we may prove the following.

\begin{theorem}
[Lower bound]\label{LBtheor}Consider an arbitrary quantum network
$\mathcal{N}=(P,E)$ where two end-points are connected by an
ensemble of routes $\Omega=\{\omega\}$ and may be disconnected by
an entanglement cut $C$.
The single-path capacity of the network satisfies%
\begin{equation}
\mathcal{C}(\mathcal{N})\geq\max_{\omega\in\Omega}\mathcal{C}(\omega)=\min
_{C}\mathcal{C}(C).
\end{equation}
Thus, the capacity $\mathcal{C}(\omega_{\ast})$ of an optimal
route $\omega_{\ast}$ not only is an achievable rate but it is
also equal to the single-edge capacity
$\mathcal{C}(C_{\text{min}})$ of a minimum cut $C_{\text{min}}$.
Furthermore, the optimal route $\omega_{\ast}$ is a simple path
within a maximum spanning tree of the network.
\end{theorem}

\textbf{Proof.}~It is easy to show the inequality
$\mathcal{C}(\omega_{\ast
})\geq\mathcal{C}(C_{\text{min}})$. In fact, an edge $(\mathbf{\tilde{x}%
},\mathbf{\tilde{y}})$ of the optimal route $\omega_{\ast}$ must
belong to the cut-set $\tilde{C}_{\text{min}}$. Thus, the capacity
of that edge must
simultaneously satisfy $\mathcal{C}_{\mathbf{\tilde{x}\tilde{y}}}%
\geq\mathcal{C}(\omega_{\ast})$ and $\mathcal{C}_{\mathbf{\tilde{x}\tilde{y}}%
}\leq\mathcal{C}(C_{\text{min}})$. In order to show the opposite
inequality
$\mathcal{C}(\omega_{\ast})\leq\mathcal{C}(C_{\text{min}})$, we
need to exploit some basic results from graph theory. Consider the
maximum spanning tree of the connected undirected graph $(P,E)$.
This is a subgraph $\mathcal{T}=(P,E_{\text{tree}})$ which
connects all the points in such a way
that the sum of the capacities associated with each edge $(\mathbf{x}%
,\mathbf{y})\in E_{\text{tree}}$ is the maximum. In other words,
it maximizes
the following quantity%
\begin{equation}
\mathcal{C}(\mathcal{T}):=%
{\textstyle\sum\nolimits_{(\mathbf{x},\mathbf{y})\in
E_{\text{tree}}}}
\mathcal{C}_{\mathbf{xy}}~.
\end{equation}

Note that the optimal route $\omega_{\ast}$ between Alice and Bob
is the unique path between Alice and Bob within this
tree~\cite{tree}. Let us call $e(\omega_{\ast})$ the critical edge
in $\omega_{\ast}$, i.e., that specific edge which realizes the
minimization
\begin{equation}
\mathcal{C}_{e(\omega_{\ast})}=\mathcal{C}(\omega_{\ast})=\min_{i}%
\mathcal{C}(\mathcal{E}_{i}^{\omega_{\ast}})~.
\end{equation}
Since this edge is part of a spanning tree, there is always an
Alice-Bob cut $C_{\ast}$ of the network which crosses
$e(\omega_{\ast})$ and no other edges of the spanning tree. In
fact, this condition would fail only if there was a cycle in the
tree, which is not possible by definition.

\begin{figure}[ptbh]
\vspace{-2.2cm}
\par
\begin{center}
\includegraphics[width=0.5\textwidth] {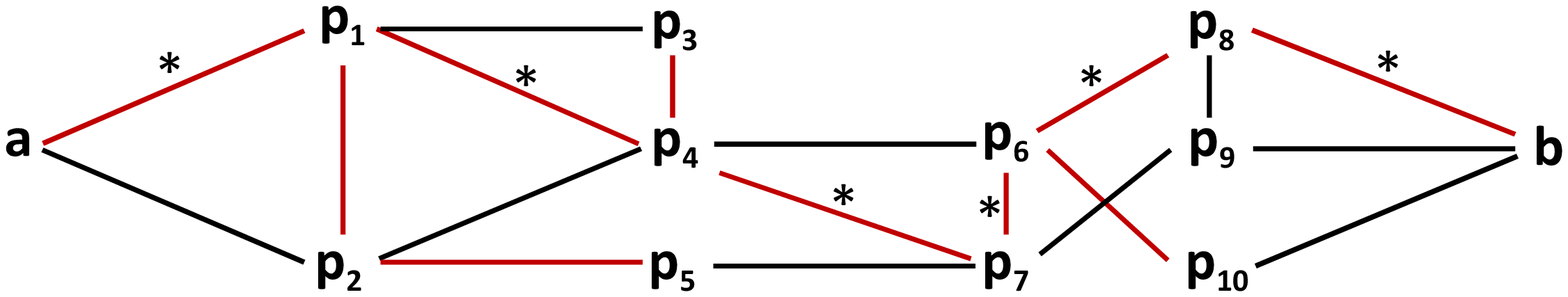}
\end{center}
\par
\vspace{-2.8cm}\caption{Example of a network and its maximum
spanning tree (red edges). The optimal route $\omega_{\ast}$
between Alice and Bob is a unique path within this tree
(highlighted by the asterisks). The critical edge
$e(\omega_{\ast})$ is the one maximizing the capacity, i.e.,
realizing the condition
$\mathcal{C}_{e(\omega_{\ast})}=\mathcal{C}(\omega_{\ast})$.
Wherever the critical edge might be along the optimal route, we
can always make an Alice-Bob entanglement cut $C_{\ast}$ which
crosses that specific edge
and no other edge of the spanning tree. This property leads to $\mathcal{C}%
(C_{\ast})=\mathcal{C}(\omega_{\ast})$.}%
\label{cutP}
\end{figure}

Then, we must also have that $e(\omega_{\ast})$ is the optimal
edge in the
cut-set $\tilde{C}_{\ast}$, i.e., $\mathcal{C}_{e(\omega_{\ast})}%
=\mathcal{C}(C_{\ast})$. By absurd, assume this is not the case.
This implies that there is another edge
$e^{\prime}\in\tilde{C}_{\ast}$, not belonging to $\mathcal{T}$,
such that $\mathcal{C}_{e^{\prime}}=\mathcal{C}(C_{\ast})$. For
the cut property of the maximum spanning trees~\cite{Dijkstra}, we
have that an edge in $C_{\ast}$ with maximum capacity must belong
to all the maximum spanning trees of the network. Therefore
$e^{\prime}$ must belong to $\mathcal{T}$ which leads to a
contradiction. In conclusion, we have found an Alice-Bob cut
$C_{\ast}$ which realizes the condition $\mathcal{C}(C_{\ast
})=\mathcal{C}(\omega_{\ast})$. For an example see
Fig.~\ref{cutP}.~$\blacksquare$

Note that the previous result applies not only to quantum networks
but to any graphical weighted network. It is sufficient to replace
the capacity of the edge with a generic weight. In fact,
Theorem~\ref{LBtheor} can be restated as
follows, which represents a \textquotedblleft single-flow\textquotedblright%
\ formulation of the max-flow min-cut
theorem~\cite{Harriss,Fords,ShannonFLOWs,netflows}.

\begin{proposition}
[Cut property of the widest path]\label{widePROPO}Consider a
network described by an undirected graph $\mathcal{N}=(P,E)$,
whose edge $e\in E$\ has weight $W(e)$. Denote by
$\Omega=\{\omega\}$\ the ensemble of undirected paths between the
end-points, Alice and Bob. Define the weight of a path
$\omega=\{e_{i}\}$ as $W(\omega)=\min_{i}W(e_{i})$, and the weight
of an Alice-Bob cut $C$ as $W(C)=\max_{e\in\tilde{C}}W(e)$. The
weight of the widest
path is equal to that of the minimum cut%
\begin{align}
W(\omega_{\text{wide}})  &  :=\max_{\omega}W(\omega)\nonumber\\
&  =\min_{C}W(C):=W(C_{\text{min}}).
\end{align}
\
\end{proposition}

Finding the optimal route $\omega_{\ast}$ in a quantum network
(Theorem~\ref{LBtheor}) is equivalent to finding the widest path
$\omega_{\text{wide}}$\ in a weighted network
(Proposition~\ref{widePROPO}), i.e.,\ solving the well-known
widest path problem~\cite{MITps}. Using a modified Dijkstra's
algorithm, the solution is found in time $O(\left\vert
E\right\vert \log_{2}\left\vert P\right\vert )$ (see Chapter 2.7.1
of Ref.~\cite{newBOOKs}, and below for a description of this
modified algorithm). In practical cases, this algorithm can be
optimized and its asymptotic performance becomes $O(\left\vert
E\right\vert +\left\vert P\right\vert \log_{2}\left\vert
P\right\vert )$~\cite{Fred}. Another possibility is using an
algorithm for finding a maximum spanning tree of the network, such
as the Kruskal's algorithm~\cite{Kruskal,MITps}. The latter has
the asymptotic complexity $O(\left\vert E\right\vert
\log_{2}\left\vert P\right\vert )$ for building the tree. This
step is then followed by the search of the route within the tree
which takes linear time $O(\left\vert P\right\vert )$~\cite{tree}.

For clarity here we briefly recall the modified Dijkstra's
algorithm for computing the widest path, which is not so known as
the most popular version for computing the shortest path. Consider
an undirected graph $\mathcal{N}=(P,E)$ where each edge $e\in E$
has an associated width $\mathrm{w}(e)$ and consider a start point
$s$. Given another point $p\in P$, let us call $\mathrm{w\_to}(p)$
the width of a path from $s$ to $p$ (as given by the minimum width
of the edges along the path). We impose
$\mathrm{w\_to}(s)=\infty$. Then, let us initialize a tree
$T=\{s\}$ with no edges. A point $p\neq s$ will be inserted in the
tree if it has maximum $\mathrm{w\_to}(p)$. This is done by
repeating the following steps:

\begin{enumerate}
\item For each neighbor-point $p$ of the tree $T$, compute:
\begin{equation}
\mathrm{w\_to}(p)=\max_{e=(q,p):q\in T}\left\{  \min\left[  \mathrm{w\_to}%
(q),\mathrm{w}(e)\right]  \right\}  .\nonumber
\end{equation}

\item Insert the neighbor-point $p$ with the maximum $\mathrm{w\_to}(p)$ into
the tree $T$.
\end{enumerate}

\noindent After iteration, this algorithm creates a tree $T$ which
specifies the widest path in the graph. The running time is the
same of the original Dijkstra's algorithm.

\subsection{Formulas for teleportation-covariant and distillable
networks\label{SecDISTILLABLENET}}

The results of Theorems~\ref{theoUBnet} and~\ref{LBtheor} can be
specified for quantum networks which are connected by
teleportation-covariant channels. Given a teleportation-covariant
network $\mathcal{N}=(P,E)$ whose teleportation simulation has an
associated Choi-representation $\sigma
(\mathcal{N})=\{\sigma_{\mathcal{E}_{\mathbf{xy}}}\}_{(\mathbf{x}%
,\mathbf{y})\in E}$, we may write the following for the single-path capacity%
\begin{equation}
\min_{C}\mathcal{C}(C)\leq\mathcal{C}(\mathcal{N})\leq\min_{C}E_{\mathrm{R}%
}(C)~, \label{sandmmm}%
\end{equation}
with $\mathcal{C}(C)$ being defined in Eq.~(\ref{cutEQUIV}), and%
\begin{equation}
E_{\mathrm{R}}(C)=\max_{(\mathbf{x},\mathbf{y})\in\tilde{C}}E_{\mathrm{R}%
}(\sigma_{\mathcal{E}_{\mathbf{xy}}})~.
\end{equation}
The latter may have an asymptotic formulation for networks of
bosonic channels, with the REE\ taking the form as in
Eq.~(\ref{REE_weaker}) over
$\sigma_{\mathcal{E}_{\mathbf{xy}}}:=\lim_{\mu}\sigma_{\mathcal{E}%
_{\mathbf{xy}}}^{\mu}$, where
$\sigma_{\mathcal{E}_{\mathbf{xy}}}^{\mu}$ is a sequence of Choi
approximating states with finite energy.

In particular, consider a network connected by distillable
channels. This means that for any edge $(\mathbf{x},\mathbf{y})\in
E$, we may write (exactly or asymptotically)
\begin{equation}
\mathcal{C}_{\mathbf{xy}}:=\mathcal{C}(\mathcal{E}_{\mathbf{xy}}%
)=E_{\mathrm{R}}(\sigma_{\mathcal{E}_{\mathbf{xy}}})=D_{1}(\sigma
_{\mathcal{E}_{\mathbf{xy}}}). \label{dueEELL}%
\end{equation}
By imposing this condition in Eq.~(\ref{sandmmm}), we find that
upper and lower bounds coincide. We have therefore the following
result which establishes the single-path capacity
$\mathcal{C}(\mathcal{N})$\ of a distillable network and fully
extends the widest path problem~\cite{Pollacks} to quantum
communications.

\begin{corollary}
[Single-path capacities]\label{coroNETseq}Consider a distillable
network $\mathcal{N}=(P,E)$, where two end-points are connected by
an ensemble\ of routes $\Omega=\{\omega\}$ and may be disconnected
by an entanglement cut $C$. An arbitrary edge
$(\mathbf{x},\mathbf{y})\in E$\ is connected by a distillable
channel $\mathcal{E}_{\mathbf{xy}}$\ with two-way capacity
$\mathcal{C}_{\mathbf{xy}}$\ and Choi matrix $\sigma_{\mathcal{E}%
_{\mathbf{xy}}}$. Then, the single-path capacity of the network is
equal to
\begin{equation}
\mathcal{C}(\mathcal{N})=\min_{C}E_{\mathrm{R}}(C)=\min_{C}\max_{(\mathbf{x}%
,\mathbf{y})\in\tilde{C}}E_{\mathrm{R}}(\sigma_{\mathcal{E}_{\mathbf{xy}}}),
\label{StretcCOR}%
\end{equation}
with an implicit asymptotic formulation for bosonic channels.
Equivalently, $\mathcal{C}(\mathcal{N})$ is also equal to the
minimum (single-edge) capacity
of the entanglement cuts and the maximum capacity of the routes, i.e.,%
\begin{equation}
\mathcal{C}(\mathcal{N})=\min_{C}\mathcal{C}(C)=\max_{\omega}\mathcal{C}%
(\omega)~.
\end{equation}
The optimal end-to-end route $\omega_{\ast}$ achieving the
capacity can be found in time $O(\left\vert E\right\vert
\log_{2}\left\vert P\right\vert )$, where $\left\vert E\right\vert
$ is the number of edges and $\left\vert P\right\vert $ is the
number of points. Over this route, a capacity-achieving protocol
is based on one-way entanglement distillation sessions between
consecutive points, followed by entanglement swapping.
\end{corollary}

The proof of this corollary is a direct application of the
previous reasonings. We see that it first reduces the routing
problem to a classical optimization problem, i.e., finding the
widest path. Then, over this optimal route, the single-path
capacity is achieved by a non-adaptive protocol based on one-way
CCs. In fact, we have that any two consecutive points
$\mathbf{r}_{i}$ and $\mathbf{r}_{i+1}$ along $\omega_{\ast}$ may
distill ebits at the rate of
$D_{1}(\sigma_{\mathcal{E}_{i}^{\omega_{\ast}}})$, where
$\mathcal{E}_{i}^{\omega_{\ast}}$ is the connecting channel. Then,
sessions of entanglement swapping (also based on one-way CCs),
distribute ebits at the
end-points with a rate of at least $\min_{i}D_{1}(\sigma_{\mathcal{E}%
_{i}^{\omega_{\ast}}})$. Due to Eq.~(\ref{dueEELL}), this rate is
equal to
$\min_{i}\mathcal{C}(\mathcal{E}_{i}^{\omega_{\ast}})=\mathcal{C}(\omega
_{\ast})$, which corresponds to the capacity
$\mathcal{C}(\mathcal{N})$.

\subsection{Single-path capacities of fundamental networks}

Let us specify the result of Corollary~\ref{coroNETseq}\ to
fundamental scenarios such as bosonic networks subject to
pure-loss or quantum-limited amplification, or spin networks
affected by dephasing or erasure. These are in fact all
distillable networks. We find extremely simple formulas for their
single-path capacities, setting their ultimate limit for quantum
communication, entanglement distribution, key generation and
private communication under single-path routing.

Start with a network connected by lossy channels
$\mathcal{N}_{\text{loss}}$, which well describes both free-space
or fiber-based optical communications. According to
Corollary~\ref{coroNETseq}, we may compute its capacity
$\mathcal{C}(\mathcal{N}_{\text{loss}})$ by minimizing over the
cuts or maximizing over the routes. Generic edge
$(\mathbf{x},\mathbf{y})\in E$ has an associated lossy channel
with transmissivity $\eta_{\mathbf{xy}}$ and capacity
$\mathcal{C}_{\mathbf{xy}}=-\log_{2}(1-\eta_{\mathbf{xy}})$.
Therefore, an entanglement cut has single-edge capacity
\begin{align}
\mathcal{C}(C)  &
=\max_{(\mathbf{x},\mathbf{y})\in\tilde{C}}\left[
-\log_{2}(1-\eta_{\mathbf{xy}})\right]  =-\log_{2}(1-\eta_{C}),\nonumber\\
\eta_{C}  &
:=\max_{(\mathbf{x},\mathbf{y})\in\tilde{C}}\eta_{\mathbf{xy}},
\end{align}
where $\eta_{C}$ may be identified as the (single-edge)
transmissivity of the cut. By minimizing over the cuts, we may
write the single-path capacity of the lossy network as
\begin{equation}
\mathcal{C}(\mathcal{N}_{\text{loss}})=-\log_{2}(1-\tilde{\eta}_{C}%
),~\tilde{\eta}_{C}:=\min_{C}\eta_{C},
\end{equation}
where $\tilde{\eta}_{C}$ is the minimum transmissivity of the
cuts.

Consider now a generic end-to-end route $\omega$ along the lossy
network. This route is associated with a sequence of lossy
channels with transmissivities
$\{\eta_{i}^{\omega}\}$. We then compute the route capacity as%
\begin{align}
\mathcal{C}_{\omega}  &  =\min_{i}\left[
-\log_{2}(1-\eta_{i}^{\omega
})\right]  =-\log_{2}(1-\eta_{\omega}),\nonumber\\
\eta_{\omega}  &  :=\min_{i}\eta_{i}^{\omega},
\end{align}
where $\eta_{\omega}$ is the route transmissivity. By maximizing
over the routes, we may equivalently write the single-path
capacity of the lossy
network as%
\begin{equation}
\mathcal{C}(\mathcal{N}_{\text{loss}})=-\log_{2}(1-\tilde{\eta}),~\tilde{\eta
}:=\max_{\omega}\eta_{\omega},
\end{equation}
where $\tilde{\eta}$ is the maximum transmissivity of the routes.

Similar conclusions can be derived for bosonic networks which are
composed of other distillable Gaussian channels, such as multiband
lossy channels, quantum-limited amplifiers or even hybrid
combinations. In particular, consider a network of quantum-limited
amplifiers $\mathcal{N}_{\text{amp}}$,
where the generic edge $(\mathbf{x},\mathbf{y})\in E$ has gain $g_{\mathbf{xy}%
}$ with capacity $\mathcal{C}_{\mathbf{xy}}=-\log_{2}(1-g_{\mathbf{xy}}^{-1}%
)$, and the generic end-to-end route $\omega$ is associated with a
sequence of gains $\{g_{i}^{\omega}\}$. We can repeat the previous
steps of the lossy network but setting $g^{-1}=\eta$, so that
$\max\eta=\min g$. Thus, for an entanglement cut $C$, we may write
\begin{align}
\mathcal{C}(C)  &
=\max_{(\mathbf{x},\mathbf{y})\in\tilde{C}}\left[
-\log_{2}(1-g_{\mathbf{xy}}^{-1})\right]  =-\log_{2}(1-g_{C}^{-1}),\nonumber\\
g_{C}  &
:=\min_{(\mathbf{x},\mathbf{y})\in\tilde{C}}g_{\mathbf{xy}}~.
\end{align}
For a route $\omega$, we have the capacity%
\begin{align}
\mathcal{C}_{\omega}  &  =\min_{i}\{-\log_{2}[1-(g_{i}^{\omega})^{-1}%
]\}=-\log_{2}(1-g_{\omega}^{-1}),\nonumber\\
g_{\omega}  &  :=\max_{i}g_{i}^{\omega}~.
\end{align}
By minimizing over the cuts or maximizing over the routes, we
derive the two
equivalent formulas%
\begin{equation}
\mathcal{C}(\mathcal{N}_{\text{amp}})=-\log_{2}(1-\tilde{g}_{C}^{-1}%
)=-\log_{2}(1-\tilde{g}^{-1}),
\end{equation}
where $\tilde{g}_{C}:=\max_{C}g_{C}$ and
$\tilde{g}:=\min_{\omega}g_{\omega}$.

We can also compute the single-path capacities of DV networks
where links between qudits are affected by dephasing or erasure or
a mix of the two errors. For simplicity, consider the case of
qubits, such as spin $1/2$ or
polarized photons. In a qubit network with dephasing channels $\mathcal{N}%
_{\text{deph}}$, the generic edge $(\mathbf{x},\mathbf{y})\in E$
has a
dephasing probability $p_{\mathbf{xy}}\leq1/2$ and capacity $\mathcal{C}%
_{\mathbf{xy}}=1-H_{2}(p_{\mathbf{xy}})$. The generic end-to-end
route $\omega$ is associated with a sequence of such dephasing
probabilities $\{p_{i}^{\omega}\}$. For an entanglement cut $C$,
we have
\begin{align}
\mathcal{C}(C)  &
=\max_{(\mathbf{x},\mathbf{y})\in\tilde{C}}\left[
1-H_{2}(p_{\mathbf{xy}})\right]  =1-H_{2}(p_{C}),\nonumber\\
p_{C}  &
:=\min_{(\mathbf{x},\mathbf{y})\in\tilde{C}}p_{\mathbf{xy}}.
\label{pcDEP}%
\end{align}
For a generic route $\omega$, we may write%
\begin{align}
\mathcal{C}_{\omega}  &  =\min_{i}\left[
1-H_{2}(p_{i}^{\omega})\right]
=1-H_{2}(p_{\omega}),\nonumber\\
p_{\omega}  &  :=\max_{i}p_{i}^{\omega}~. \label{pomegaDEP}%
\end{align}
By minimizing over the cuts or maximizing over the routes, we then
derive the single-path capacity
\begin{equation}
\mathcal{C}(\mathcal{N}_{\text{deph}})=1-H_{2}(\tilde{p}_{C})=1-H_{2}%
(\tilde{p}),
\end{equation}
where we have set
\begin{equation}
\tilde{p}_{C}:=\max_{C}p_{C},~~\tilde{p}:=\min_{\omega}p_{\omega}.
\label{pcset}%
\end{equation}

Finally, for a qubit network affected by erasures
$\mathcal{N}_{\text{erase}}$ we have that edge
$(\mathbf{x},\mathbf{y})\in E$ is associated with an erasure
channel with probability $p_{\mathbf{xy}}$ and corresponding
capacity $\mathcal{C}_{\mathbf{xy}}=1-p_{\mathbf{xy}}$. As a
result, we may repeat all the previous derivation for the
dephasing network $\mathcal{N}_{\text{deph}}$
up to replacing $H_{2}(p)$ with $p$. For a cut and a route, we have%
\begin{equation}
\mathcal{C}(C)=1-p_{C},~~\mathcal{C}_{\omega}=1-p_{\omega},
\end{equation}
where $p_{C}$ and $p_{\omega}$ are defined as in
Eqs.~(\ref{pcDEP}) and~(\ref{pomegaDEP}). Thus, the single-path
capacity of the erasure network
simply reads%
\begin{equation}
\mathcal{C}(\mathcal{N}_{\text{erase}})=1-\tilde{p}_{C}=1-\tilde{p},~.
\end{equation}
where $\tilde{p}_{C}$ and $\tilde{p}$ are defined as in
Eq.~(\ref{pcset}).

See Table~I of the main text for a schematic presentation of these
analytical formulas.

\section{Supplementary Note 5: Results for multi-path routing\label{secMULTI}}

\subsection{Converse part (upper bound)}

In order to write a single-letter upper bound for the multi-path
capacity of a quantum network, we need to introduce the concept of
\textit{multi-edge} flow of REE through a cut, under some
simulation of the network. Consider an arbitrary quantum network
$\mathcal{N}=(P,E)$ whose simulation has an
associate resource representation $\sigma(\mathcal{N})=\{\sigma_{\mathbf{xy}%
}\}_{(\mathbf{x},\mathbf{y})\in E}$. Then, consider an arbitrary
entanglement cut $C$ with corresponding cut-set $\tilde{C}$. Under
the simulation considered, we define the multi-edge flow of REE
through the cut as the
following quantity%
\begin{equation}
E_{\mathrm{R}}^{\text{m}}(C):=\sum_{(\mathbf{x},\mathbf{y})\in\tilde{C}%
}E_{\mathrm{R}}(\sigma_{\mathbf{xy}})~. \label{multiooo}%
\end{equation}
By minimizing $E_{\mathrm{R}}^{\text{m}}(C)$ over all possible
entanglement cuts of the network, we build our upper bound for the
multi-path capacity. In fact, we may prove the following.

\begin{theorem}
[Converse for multi-path capacity]\label{TheoMP1}Consider an
arbitrary quantum network $\mathcal{N}=(P,E)$ with some resource
representation $\sigma
(\mathcal{N})=\{\sigma_{\mathbf{xy}}\}_{(\mathbf{x},\mathbf{y})\in
E}$. In particular, $\sigma(\mathcal{N})$ may be a
Choi-representation for a teleportation-covariant network. Then,
the multi-path capacity of $\mathcal{N}$
must satisfy the single-letter bound%
\begin{equation}
\mathcal{C}^{\text{m}}(\mathcal{N})\leq\min_{C}E_{\mathrm{R}}^{\text{m}}(C)~,
\label{theommll}%
\end{equation}
where the multi-edge flow of REE in Eq.~(\ref{multiooo}) is
minimized across all cuts of the network. Formulations may be
asymptotic for networks of bosonic channels.
\end{theorem}

\textbf{Proof.}~~Let us start from the general weak converse upper
bound proven in the Methods section of the main manuscript. In
terms of the REE and for flooding protocols
$\mathcal{P}_{\text{\textrm{flood}}}$, it takes the
following form%
\begin{equation}
\mathcal{C}^{\text{m}}(\mathcal{N})\leq E_{\mathrm{R}}^{\bigstar}%
(\mathcal{N}):=\sup_{\mathcal{P}_{\text{\textrm{flood}}}}\underset{n}{\lim
}\frac{E_{\mathrm{R}}(\rho_{\mathbf{ab}}^{n})}{n}.\label{m2m2}%
\end{equation}
According to previous Lemma~\ref{reduceCHOI}, for any flooding
protocol $\mathcal{P}_{\text{\textrm{flood}}}$ and entanglement
cut $C$, we may write Eq.~(\ref{cutEQ}) with $n_{\mathbf{xy}}=n$.
Computing the REE on this
decomposition and exploiting basic properties of the REE, we derive%
\begin{equation}
E_{\mathrm{R}}\left[  \rho_{\mathbf{ab}}^{n}(C)\right]  \leq n\sum
_{(\mathbf{x},\mathbf{y})\in\tilde{C}}E_{\mathrm{R}}(\sigma_{\mathbf{xy}%
})=nE_{\mathrm{R}}^{\text{m}}(C).\label{ddd3}%
\end{equation}
By using Eq.~(\ref{ddd3}) in Eq.~(\ref{m2m2}), both the supremum
and the limit
disappear, and we are left with the bound%
\begin{equation}
\mathcal{C}^{\text{m}}(\mathcal{N})\leq E_{\mathrm{R}}^{\text{m}%
}(C),~~\text{for any~}C\text{.}\label{preRESS22}%
\end{equation}
By minimizing over all cuts, we therefore prove
Eq.~(\ref{theommll}). The extension to asymptotic simulations
follows the same derivation in the proof of
Theorem~\ref{theoUBnet} but setting $n_{\mathbf{xy}}=n$. We find
again Eq.~(\ref{theommll}) but where the REE takes the weaker
formulation for asymptotic states of
Eq.~(\ref{REE_weaker}).~$\blacksquare$

\subsection{Direct part (achievable rate)}

We now provide a general lower bound to the multi-path capacity.
Consider an
arbitrary quantum network $\mathcal{N}=(P,E)$ where edge $(\mathbf{x}%
,\mathbf{y})\in E$ is connected by channel
$\mathcal{E}_{\mathbf{xy}}$ with
two-way capacity $\mathcal{C}_{\mathbf{xy}}=\mathcal{C}(\mathcal{E}%
_{\mathbf{xy}})$. Given an arbitrary entanglement cut $C$ of the
network, we define its multi-edge capacity as the total number of
target bits distributed
by all the edges across the cut, i.e.,%
\begin{equation}
\mathcal{C}^{\text{m}}(C):=\sum_{(\mathbf{x},\mathbf{y})\in\tilde{C}%
}\mathcal{C}_{\mathbf{xy}}~. \label{Cmpform}%
\end{equation}
In this setting, a minimum cut $C_{\text{min}}$ is such that
\begin{equation}
\mathcal{C}^{\text{m}}(C_{\text{min}})=\min_{C}\mathcal{C}^{\text{m}}(C).
\label{achievTARGET}%
\end{equation}
We now prove that the later is an achievable rate for multi-path
quantum/private communication.

\begin{theorem}
[Lower bound]\label{TheoMP2}Consider an arbitrary quantum network
$\mathcal{N}=(P,E)$ where two end-points may be disconnected by an
entanglement cut $C$. The multi-path capacity of the network satisfies%
\begin{equation}
\mathcal{C}^{\text{m}}(\mathcal{N})\geq\min_{C}\mathcal{C}^{\text{m}}(C).
\label{cut2}%
\end{equation}
In other words, the minimum multi-edge capacity of the
entanglement cuts is an achievable rate. This rate is achieved by
a flooding protocol whose\ multi-path routing can be found in
$O(|P|\times|E|)$ time by solving the classical maximum flow
problem.
\end{theorem}

\textbf{Proof.}~~To show the achievability of the rate in
Eq.~(\ref{achievTARGET}), we resort to the classical max-flow
min-cut theorem~\cite{Fords}. In the literature, this theorem has
been widely adopted for the study of directed graphs. In general,
it can also be applied to directed multi-graphs as well as
undirected graphs/multi-graphs (e.g.,
see~\cite[Sec.~6]{netflows}). The latter cases can be treated by
splitting the undirected edges into directed ones (e.g.,
see~\cite[Sec.~2.4]{netflows}).

\begin{figure}[th]
\begin{center}
\vspace{-2.5cm}
\includegraphics[width=0.48\textwidth]{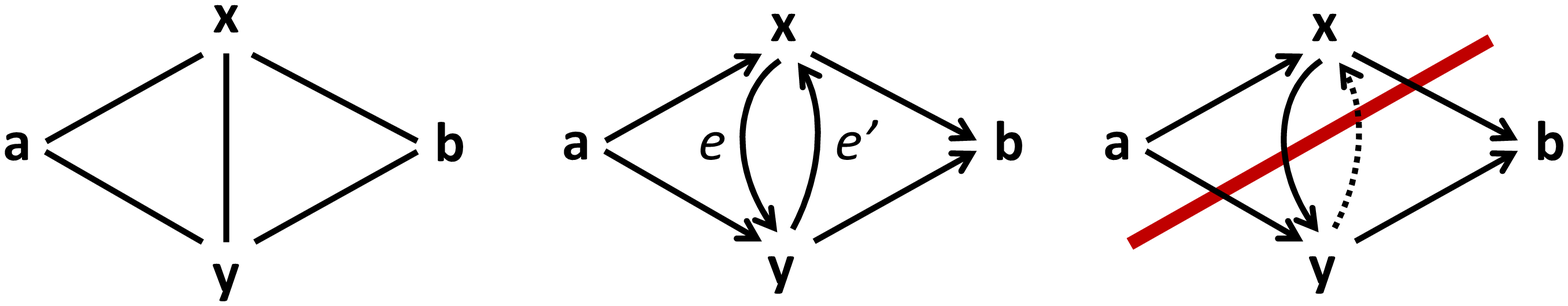}
\vspace{-3.0cm}
\end{center}
\caption{Manipulations of the undirected diamond network.
(Left)~Original undirected quantum network
$\mathcal{N}^{\diamond}$. (Middle)~Flow network
$\mathcal{N}_{\text{flow}}^{\diamond}$ with Alice $\mathbf{a}$ as
source and
Bob $\mathbf{b}$ as sink, where the middle undirected edge $(\mathbf{x}%
,\mathbf{y})$ has been split in two directed edges $e$ and
$e^{\prime}$ with the same capacity. (Right)~Assuming the
displayed Alice-Bob cut, the dotted
edge does not belong to the directed cut-set $\tilde{C}_{D}$.}%
\label{diamondsPLOT}
\end{figure}

Our first step is therefore the transformation of the undirected
graph of the quantum network $\mathcal{N}=(P,E)$ into a suitable
directed graph (in general, these may be multi-graphs, in which
case the following derivation still holds but with more technical
notation). Starting from $(P,E)$, we consider the directed graph
where Alice's edges are all out-going (so that she is a source),
while Bob's edges are all in-going (so that he is a sink). Then,
for any pair $\mathbf{x}$ and $\mathbf{y}$ of intermediate points
$P\backslash\{\mathbf{a},\mathbf{b}\}$, we split the undirected
edge
$(\mathbf{x},\mathbf{y})\in E$ into two directed edges $e:=(\mathbf{x}%
,\mathbf{y})\in E_{D}$ and $e^{\prime}:=(\mathbf{y},\mathbf{x})\in
E_{D}$, having capacities equal to the capacity
$\mathcal{C}_{\mathbf{xy}}$ of the original undirected edge. (Note
that one may always enforce a single direction between
$\mathbf{x}$ and $\mathbf{y}$ by introducing an artificial point
$\mathbf{z}$\ in one of the two directed edges. For instance, we
may keep $(\mathbf{x},\mathbf{y})$ as is, while replacing
$(\mathbf{y},\mathbf{x})$ with $(\mathbf{y},\mathbf{z})$ and
$(\mathbf{z},\mathbf{x})$, both having the same capacity of
$(\mathbf{y},\mathbf{x})$. This further modification does not
affect the maximum flow value and the minimum cut capacity, but
increases the complexity of the network.) These manipulations
generate our flow network $\mathcal{N}_{\text{flow}}=(P,E_{D})$.
See Fig.~\ref{diamondsPLOT} for a simple example.

We then adopt the standard definition of cut-set for flow
networks, here called \textquotedblleft directed
cut-set\textquotedblright. Given an
Alice-Bob cut $C$ of the flow network, with bipartition $(\mathbf{A}%
,\mathbf{B})$ of the points $P$, its directed cut-set is defined
as $\tilde
{C}_{D}=\{(\mathbf{x},\mathbf{y})\in E_{D}:\mathbf{x}\in\mathbf{A}%
,\mathbf{y}\in\mathbf{B}\}$. This means that directed edges of the
type $(\mathbf{y}\in\mathbf{B},\mathbf{x}\in\mathbf{A})$ do not
belong to this set (see Fig.~\ref{diamondsPLOT}). Using this
definition, the cut-properties of the flow network
$\mathcal{N}_{\text{flow}}$ are exactly the same as those of the
original undirected graph $\mathcal{N}$, for which we used the
\textquotedblleft undirected\textquotedblright\ definition of
cut-set. For this reason, we have
\begin{equation}
\left[  \min_{C}\sum_{(\mathbf{x},\mathbf{y})\in\tilde{C}}\mathcal{C}%
_{\mathbf{xy}}\right]  _{\mathcal{N}}=\left[  \min_{C}\sum_{(\mathbf{x}%
,\mathbf{y})\in\tilde{C}_{D}}\mathcal{C}_{\mathbf{xy}}\right]  _{\mathcal{N}%
_{\text{flow}}}, \label{bbACH0}%
\end{equation}
where the first quantity is computed on $\mathcal{N}$, while the
second one is computed on the flow network
$\mathcal{N}_{\text{flow}}$. We aim to show that the latter is an
achievable rate.

Let us now define the \textquotedblleft flow\textquotedblright\ in
the network $\mathcal{N}_{\text{flow}}$ as the number of qubits
per use which are reliably transmitted from $\mathbf{x}$ to
$\mathbf{y}$ along the directed edge $e=(\mathbf{x},\mathbf{y})\in
E_{D}$, denoted by $R_{\mathbf{xy}}^{e}\geq0$. This quantum
transmission is performed by means of a point-to-point protocol
where $\mathbf{x}$ and $\mathbf{y}$ exploit adaptive LOCCs, i.e.,
unlimited two-way CCs and adaptive LOs, without the help of the
other points of the network. It is therefore bounded by the
two-way quantum capacity of the associated channel
$\mathcal{E}_{\mathbf{xy}}$, i.e., $R_{\mathbf{xy}}^{e}\leq
Q_{2}(\mathcal{E}_{\mathbf{xy}})$. The actual physical direction
of the quantum channel does not matter since it is used with
two-way CCs, so that the two points $\mathbf{x}$ and $\mathbf{y}$
first distill entanglement and then they teleport qubits in the
\textquotedblleft logical direction\textquotedblright\ specified
by the directed edge.

Since every directed edge $e=(\mathbf{x},\mathbf{y})$ between two
intermediate points $\mathbf{x},\mathbf{y}\in
P\backslash\{\mathbf{a},\mathbf{b}\}$ has an opposite counterpart\
$e^{\prime}:=(\mathbf{y},\mathbf{x})$, we may simultaneously
consider an opposite flow of qubits from $\mathbf{y}$ to
$\mathbf{x}$ with rate $0\leq R_{\mathbf{yx}}^{e^{\prime}}\leq Q_{2}%
(\mathcal{E}_{\mathbf{xy}})$. As a result, there will be an
\textquotedblleft effective\textquotedblright\ point-to-point\
rate between $\mathbf{x}$ and $\mathbf{y}$ which is defined by the
difference of the two \textquotedblleft
directed\textquotedblright\ rates
\begin{equation}
R_{\mathbf{xy}}:=R_{\mathbf{xy}}^{e}-R_{\mathbf{yx}}^{e^{\prime}}.
\end{equation}
Its absolute value $|R_{\mathbf{xy}}|$ provides the effective
number of qubits transmitted between $\mathbf{x}$ to $\mathbf{y}$
per use of the undirected edge. For $R_{\mathbf{xy}}\geq0$,
effective qubits flow from $\mathbf{x}$ to $\mathbf{y}$, while
$R_{\mathbf{xy}}\leq0$ means that effective qubits flow from
$\mathbf{y}$ to\textbf{\ }$\mathbf{x}$. The effective rate is
correctly bounded $|R_{\mathbf{xy}}|\leq
Q_{2}(\mathcal{E}_{\mathbf{xy}})$ and we set $R_{\mathbf{xy}}=0$
if two points are not connected. The ensemble of positive directed
rates $\{R_{\mathbf{xy}}^{e}\}_{e\in E_{D}}$ represents a flow
vector in $\mathcal{N}_{\text{flow}}$. For any choice of this
vector, there is a
corresponding ensemble of effective rates $\{R_{\mathbf{xy}}\}_{(\mathbf{x}%
,\mathbf{y})\in E}$ for the original network $\mathcal{N}$. The
signs
$\{\mathrm{sgn}(R_{\mathbf{xy}})\}_{(\mathbf{x},\mathbf{y})\in E}$
specify an orientation $\mathcal{N}_{D}=(P,E_{D}^{\prime})$ for
$\mathcal{N}$, and the absolute values
$\{|R_{\mathbf{xy}}|\}_{(\mathbf{x},\mathbf{y})\in E}$ provide
point-to-point quantum communication rates for the associated
protocol.

It is important to note that $\{R_{\mathbf{xy}}^{e}\}_{e\in
E_{D}}$ represents
a \textquotedblleft legal\textquotedblright\ flow vector in $\mathcal{N}%
_{\text{flow}}$\ only if we impose the property of flow
conservation~\cite{netflows}. This property can be stated for $\{R_{\mathbf{xy}%
}^{e}\}_{e\in E_{D}}$\ or, equivalently, for the effective vector
$\{R_{\mathbf{xy}}\}_{(\mathbf{x},\mathbf{y})\in E}$. At any
intermediate point, the number of qubits simultaneously received
must be equal to the number of qubits simultaneously transmitted
through all the point-to-point communications with neighbor
points. In other words, for any $\mathbf{x}\in
P\backslash\{\mathbf{a},\mathbf{b}\}$, we must impose
\begin{equation}
\sum_{\mathbf{y}\in P}R_{\mathbf{xy}}=0.
\end{equation}

This property does not hold for Alice $\mathbf{a}$ (source) and
Bob
$\mathbf{b}$ (sink), for which we impose%
\[
\sum_{\mathbf{y}\in P}R_{\mathbf{ay}}=-\sum_{\mathbf{y}\in P}R_{\mathbf{by}%
}:=|R|,
\]
where $|R|$ is known as the value of the flow. This is an
achievable end-to-end rate since it represents the total number of
qubits per network use which are transmitted by Alice and
correspondingly received by Bob via all the end-to-end routes,
where the intermediate points quantum-communicate at the rates
$\{R_{\mathbf{xy}}\}_{(\mathbf{x},\mathbf{y})\in E}$.

Now, from the classical max-flow min-cut theorem, we know that the
maximum value of the flow in the network $|R|_{\max}$ is equal to
the capacity of the minimum cut~\cite{Fords,netflows}, i.e., we
may write
\begin{equation}
|R|_{\max}=\min_{C}\sum_{(\mathbf{x},\mathbf{y})\in\tilde{C}_{D}}%
Q_{2}(\mathcal{E}_{\mathbf{xy}})~. \label{maxPROO}%
\end{equation}
Thus, by construction, we have that $|R|_{\max}$ is an achievable
rate for quantum communication. The previous reasoning can be
repeated for private bits by defining a corresponding flow of
private information through the network. Thus, in general, we may
write that
\begin{equation}
|R|_{\max}=\min_{C}\sum_{(\mathbf{x},\mathbf{y})\in\tilde{C}_{D}}%
\mathcal{C}_{\mathbf{xy}}%
\end{equation}
is an achievable rate for any of the quantum tasks. This proves
that Eq.~(\ref{bbACH0}) is an achievable rate.

In order to better understand the flooding protocol that achieves $|R|_{\max}%
$, call $\{\tilde{R}_{\mathbf{xy}}^{e}\}_{e\in E_{D}}$ the optimal
flow vector in $\mathcal{N}_{\text{flow}}$. There is a
corresponding vector $\{\tilde
{R}_{\mathbf{xy}}\}_{(\mathbf{x},\mathbf{y})\in E}$ which
determines an optimal orientation
$\mathcal{N}_{D}=(P,E_{D}^{\prime})$ for the quantum network
$\mathcal{N}=(P,E)$, besides providing the optimal rates
$\{|\tilde {R}_{\mathbf{xy}}|\}_{(\mathbf{x},\mathbf{y})\in E}$ to
be reached by the point-to-point connections. In other words,
starting from the capacities $\mathcal{C}_{\mathbf{xy}}$, the
points solve the maximum flow problem and establish an optimal
multi-path routing $\mathcal{R}_{\text{opt}}^{\text{m}}$. After
this, each point $\mathbf{x}\in P$ communicates with its
out-neighborhood $N^{\text{out}}(\mathbf{x})$, according to the
optimal rates and the optimal orientation.

Finally, let us discuss the complexity of finding the optimal
multi-path routing $\mathcal{R}_{\text{opt}}^{\text{m}}$. By
construction, the flow network
$\mathcal{N}_{\text{flow}}=\{P,E_{D}\}$ has only a small overhead
with respect to the original network $\mathcal{N}=\{P,E\}$. In
fact, we just have $|E_{D}|\leq2|E|$. Within
$\mathcal{N}_{\text{flow}}$, the maximum flow can be found with
classical algorithms. If the capacities are rational, we can apply
the Ford-Fulkerson algorithm~\cite{Fords} or the Edmonds--Karp
algorithm~\cite{Karps}, the latter running in
$O(|P|\times|E_{D}|^{2})$ time.
An alternative is Dinic's algorithm~\cite{Dinics}, which runs in $O(|P|^{2}%
\times|E_{D}|)$ time. More powerful algorithms are
available~\cite{Alons,Ahujas,Cheriyans} and the best running
performance is currently $O(|P|\times|E_{D}|)$
time~\cite{Kings,Orlins}. Thus, adopting Orlin's
algorithm~\cite{Orlins}, we find the solution in $O(|P|\times|E_{D}%
|)=O(|P|\times|E|)$ time.~$\blacksquare$

\subsection{Formulas for teleportation-covariant and distillable networks}

Consider a teleportation-covariant quantum network
$\mathcal{N}=(P,E)$ whose teleportation simulation has an
associated Choi-representation $\sigma
(\mathcal{N})=\{\sigma_{\mathcal{E}_{\mathbf{xy}}}\}_{(\mathbf{x}%
,\mathbf{y})\in E}$. Then, from Theorems~\ref{TheoMP1}
and~\ref{TheoMP2}, we
may write the following sandwich for the multi-path capacity%
\begin{equation}
\min_{C}\mathcal{C}^{\text{m}}(C)\leq\mathcal{C}^{\text{m}}(\mathcal{N}%
)\leq\min_{C}E_{\mathrm{R}}^{\text{m}}(C)~, \label{sandmp}%
\end{equation}
with $\mathcal{C}^{\text{m}}(C)$ being defined in Eq.~(\ref{Cmpform}), and%
\begin{equation}
E_{\mathrm{R}}^{\text{m}}(C)=\sum_{(\mathbf{x},\mathbf{y})\in\tilde{C}%
}E_{\mathrm{R}}(\sigma_{\mathcal{E}_{\mathbf{xy}}})~.
\end{equation}
As usual, the latter may have an asymptotic formulation for
networks of bosonic channels, with the REE\ taking the form as in
Eq.~(\ref{REE_weaker})
over $\sigma_{\mathcal{E}_{\mathbf{xy}}}:=\lim_{\mu}\sigma_{\mathcal{E}%
_{\mathbf{xy}}}^{\mu}$, where
$\sigma_{\mathcal{E}_{\mathbf{xy}}}^{\mu}$ is a sequence of states
with finite energy.

In particular, consider now a distillable network. This means
that, for any edge $(\mathbf{x},\mathbf{y})\in E$, we may write
Eq.~(\ref{dueEELL}), exactly or asymptotically. By imposing this
condition in Eq.~(\ref{sandmp}), we find that upper and lower
bounds coincide. We have therefore the following result
which establishes the multi-path capacity $\mathcal{C}^{\text{m}}(\mathcal{N}%
)$\ of a distillable network and fully extends the max-flow
min-cut theorem~\cite{Harriss,Fords,ShannonFLOWs} to quantum
communications.

\begin{corollary}
[multi-path capacities]\label{coroNETmp}Consider a distillable
network $\mathcal{N}=(P,E)$, whose arbitrary edge
$(\mathbf{x},\mathbf{y})\in E$\ is connected by a distillable
channel $\mathcal{E}_{\mathbf{xy}}$\ with two-way
capacity $\mathcal{C}_{\mathbf{xy}}$\ and Choi matrix $\sigma_{\mathcal{E}%
_{\mathbf{xy}}}$. Then, the multi-path capacity of the network is
equal to
\begin{equation}
\mathcal{C}^{\text{m}}(\mathcal{N})=\min_{C}E_{\mathrm{R}}^{\text{m}}%
(C)=\min_{C}\sum_{(\mathbf{x},\mathbf{y})\in\tilde{C}}E_{\mathrm{R}}%
(\sigma_{\mathcal{E}_{\mathbf{xy}}}),
\end{equation}
with an implicit asymptotic formulation for bosonic channels.
Equivalently, $\mathcal{C}^{\text{m}}(\mathcal{N})$ is also equal
to the minimum
(multi-edge) capacity of the entanglement cuts%
\begin{equation}
\mathcal{C}^{\text{m}}(\mathcal{N})=\min_{C}\mathcal{C}^{\text{m}}(C).
\label{coroMAINkk}%
\end{equation}
The optimal multi-path routing can be found in $O(|P|\times|E|)$
time by solving the classical maximum flow problem. A
capacity-achieving flooding protocol corresponds to performing
one-way entanglement distillation between neighbor points,
followed by multiple sessions of teleportation in the direction of
the optimal network orientation.
\end{corollary}

The proof is a direct application of the previous reasonings. In
particular, from Theorem~\ref{TheoMP2}, we have that the routing
problem is reduced to the solution of a classical optimization
problem, i.e., finding the maximum flow in a flow network. This
solution provides an optimal orientation $\mathcal{N}_{D}$ of the
quantum network and also the point-to-point rates
$|\tilde{R}_{\mathbf{xy}}|$ to be used in the various multipoint
communications. Under this optimal routing, the multi-path
capacity is achieved by a non-adaptive flooding protocol based on
one-way CCs. In fact, because the channels are distillable, each
pair of points $\mathbf{x}$ and $\mathbf{y}$ may distill $n|\tilde
{R}_{\mathbf{xy}}|$ ebits. By using the distilled ebits, Alice's
qubits are teleported to Bob along the multi-path routes
associated with the maximum flow. Since Alice's qubits can be part
of ebits and, therefore, private bits, this protocol can also
distill entanglement and keys at the same end-to-end rate.

Thus, Corollary~\ref{coroNETmp} reduces the computation of the
multi-path capacity of a distillable quantum network to the
determination of the maximum flow in a classical network. In this
sense the max-flow min-cut theorem is extended from classical to
quantum communications. In particular, the distillable network can
always be transformed in a teleportation network, where quantum
information is teleported as a flow from Alice to Bob.

\subsection{Multi-path capacities of fundamental networks}

Consider the practical scenario of quantum optical communications
affected by loss, e.g., free-space or fiber-based. A specific
distillable network is a bosonic network connected by lossy
channels $\mathcal{N}_{\text{loss}}$, so that each undirected edge
$(\mathbf{x},\mathbf{y})$ has an associated lossy channel
$\mathcal{E}_{\mathbf{xy}}$ with transmissivity
$\eta_{\mathbf{xy}}$
or equivalent \textquotedblleft loss parameter\textquotedblright%
\ $1-\eta_{\mathbf{xy}}$. We may then apply
Corollary~\ref{coroNETmp} and
express the multi-path capacity $\mathcal{C}^{\text{m}}(\mathcal{N}%
_{\text{loss}})$\ in terms of the loss parameters of the network.

Let us define the loss of an Alice-Bob entanglement cut $C$ as the
product of
the loss parameters of the channels in the cut-set, i.e., we set%
\begin{equation}
l(C):=%
{\textstyle\prod\limits_{(\mathbf{x},\mathbf{y})\in\tilde{C}}}
(1-\eta_{\mathbf{xy}}).
\end{equation}
This quantity determines the multi-edge capacity of the cut, since
we have $\mathcal{C}^{\text{m}}(C)=-\log_{2}l(C)$. By applying
Eq.~(\ref{coroMAINkk}),
we find that the multi-path capacity of the lossy network is given by%
\begin{equation}
\mathcal{C}^{\text{m}}(\mathcal{N}_{\text{loss}})=\min_{C}\left[
-\log _{2}l(C)\right]  =-\log_{2}\left[  \max_{C}l(C)\right]  .
\end{equation}
Thus, we may define the total loss of the network as the
maximization of
$l(C)$ over all cuts, i.e.,%
\begin{equation}
l(\mathcal{N}_{\text{loss}}):=\max_{C}l(C),
\end{equation}
and write the simple formula%
\begin{equation}
\mathcal{C}^{\text{m}}(\mathcal{N}_{\text{loss}})=-\log_{2}l(\mathcal{N}%
_{\text{loss}}). \label{toGENmulti}%
\end{equation}

In general, we may consider a multiband lossy network $\mathcal{N}%
_{\text{loss}}^{\text{band}}$, where each edge
$(\mathbf{x},\mathbf{y})$ represents a multiband lossy channel
$\mathcal{E}_{\mathbf{xy}}^{\text{band}}$ with bandwidth
$M_{\mathbf{xy}}$\ and constant transmissivity $\eta
_{\mathbf{xy}}$. In other words, each single edge
$(\mathbf{x},\mathbf{y})$ corresponds to $M_{\mathbf{xy}}$
independent lossy channels with the same
transmissivity $\eta_{\mathbf{xy}}$. In this case, we have $\mathcal{C}%
(\mathcal{E}_{\mathbf{xy}}^{\text{band}})=-M_{\mathbf{xy}}\log_{2}%
(1-\eta_{\mathbf{xy}})$ and we write
\begin{equation}
\mathcal{C}^{\text{m}}(\mathcal{N}_{\text{loss}}^{\text{band}})=-\log
_{2}\left[  \max_{C}%
{\textstyle\prod\limits_{(\mathbf{x},\mathbf{y})\in\tilde{C}}}
(1-\eta_{\mathbf{xy}})^{M_{\mathbf{xy}}}\right]  ,
\end{equation}
which directly generalizes Eq.~(\ref{toGENmulti}).

In particular, suppose that we have the same loss in each edge of
the
multiband network, i.e., $\eta_{\mathbf{xy}}:=\eta$ for any $(\mathbf{x}%
,\mathbf{y})\in E$, which may occur when points $\mathbf{x}$ and
$\mathbf{y}$
are equidistant. Then, we may simply write%
\begin{align}
\mathcal{C}^{\text{m}}(\mathcal{N}_{\text{loss}}^{\text{band}})  &
=-M_{\min
}\log_{2}(1-\eta),\\
M_{\min}  &  :=\min_{C}\sum_{(\mathbf{x},\mathbf{y})\in\tilde{C}%
}M_{\mathbf{xy}},
\end{align}
where $M_{\min}$ is the effective bandwidth of the network.

Consider now other types of distillable networks. Start with a
bosonic network of quantum-limited amplifiers
$\mathcal{N}_{\text{amp}}$, where the generic edge
$(\mathbf{x},\mathbf{y})$\ has an associated gain
$g_{\mathbf{xy}}$. Its multi-path capacity is given by
\begin{equation}
\mathcal{C}^{\text{m}}(\mathcal{N}_{\text{amp}})=-\log_{2}\left[  \max_{C}%
{\textstyle\prod\limits_{(\mathbf{x},\mathbf{y})\in\tilde{C}}}
(1-g_{\mathbf{xy}}^{-1})\right]  .
\end{equation}
For a qubit network of dephasing channels
$\mathcal{N}_{\text{deph}}$, where the generic edge
$(\mathbf{x},\mathbf{y})$ has dephasing probability
$p_{\mathbf{xy}}$, we may write the multi-path capacity%
\begin{equation}
\mathcal{C}^{\text{m}}(\mathcal{N}_{\text{deph}})=\min_{C}%
{\textstyle\sum\limits_{(\mathbf{x},\mathbf{y})\in\tilde{C}}}
\left[  1-H_{2}(p_{\mathbf{xy}})\right]  .
\end{equation}
Finally, for a qubit network of erasure channels
$\mathcal{N}_{\text{erase}}$
with erasure probabilities $p_{\mathbf{xy}}$, we simply have%
\begin{equation}
\mathcal{C}^{\text{m}}(\mathcal{N}_{\text{erase}})=\min_{C}%
{\textstyle\sum\limits_{(\mathbf{x},\mathbf{y})\in\tilde{C}}}
(1-p_{\mathbf{xy}}).
\end{equation}
Similar expressions may be derived for qudit networks of dephasing
and erasure channels in arbitrary dimension. See Table~I of the
main text for a list of formulas.

\section{Supplementary Note 6: Related literature}

Independently from this work, and simultaneously with its first
appearance on the arXiv in 2016~\cite{longVersion}, Azuma et
al.~\cite{ref1s} also studied upper bounds for private
communication over quantum networks in the single-path
configuration. They specifically employed the squashed
entanglement and adopted different techniques (not based on the
simplification of an entanglement measure via channel simulation).
Because of these choices, they derived completely different
single-path upper bounds. In particular, their bounds are not as
tight as ours for networks connected by teleportation-covariant
channels, such as Pauli, erasure or Gaussian channels. The
methodology of Ref.~\cite{ref1s} cannot identify the single-path
capacities for networks connected by distillable channels (such as
lossy channels, quantum-limited amplifiers, dephasing or erasure
channels). These capacities were instead established in our work
thanks to the use of the REE as entanglement measure and the
network generalization of the channel simulation techniques
introduced by PLOB~\cite{QKDpapers}.

Later, in another work, Azuma and Kato~\cite{ref2s} studied upper
bounds for multi-path routing, mainly using the squashed
entanglement but also resorting to the REE as a consequence of the
results in PLOB~\cite{QKDpapers}. Differently from our work, they
did not consider flooding protocols, where each edge is used
exactly once in each parallel use of the quantum network. The
imposition of this flooding condition is essential for finding our
general upper bound for the multi-path capacity. Flooding is also
essential for extending the max-flow/min-cut theorem to quantum
communications and therefore establishing the formulas of the
multi-path capacities for distillable networks, all results which
have been found here in our work.

\subsection{Followup works and recent developments}

The methods and results of this work~\cite{longVersion} have been
already exploited in a number of recent studies. Rigovacca et
al.~\cite{ref3s} combined the REE approach of this work (based on
channel simulation) and the squashed-entanglement approach of
Refs.~\cite{ref1s,ref2s} (\textit{not} based on channel
simulation) to provide versatile bounds. On the other hand, the
present author~\cite{TomKenns} investigated the end-to-end
capacities of networks composed of Holevo-Werner channels by
considering both the REE\ and the squashed entanglement
\textit{while using} channel simulation and teleportation
stretching (these network results of Ref.~\cite{TomKenns} were
directly based on the techniques devised here).

In another study, Pant et al.~\cite{ref4s} further explored one of
the results of the present work~\cite{longVersion}: the
superiority of multi-path versus single-path protocols for
distributing entanglement and secret keys between end-users.
Differently from here, where this advantage is shown in an
information-theoretic sense with ideal quantum repeaters, Pant et
al.~\cite{ref4s} studies this advantage by considering
realistic/practical models of repeater nodes.

Among other recent developments, let us also mention the recent
work by B\"{a}uml et al.~\cite{ref5s}, which has defined different
types of network capacities. In B\"{a}uml et al.~\cite{ref5s}, the
network capacities are not defined \textit{per network use} but
rather \textit{per total number of channel uses} (which is based
on counting the number of channels that are sequentially used in a
route between the end-parties).

Finally, the limits established by this work for the optimal
performance of quantum repeaters have been already considered in
works of quantum key distribution (QKD), including the
relay-assisted protocols of twin-field QKD~\cite{Marcos} and
Phase-Matching QKD~\cite{Mas}.


\end{document}